\documentclass[aps,prb,twocolumn,superscriptaddress,showpacs]{revtex4-1}

\usepackage{graphicx}
\usepackage{amsmath}
\usepackage{graphics}
\usepackage{exscale}
\usepackage{epsfig}
\usepackage{color}
\newcommand{\ev}[1]{\langle{#1}\rangle}
\newcommand{\ketbra}[1]{|{#1}\rangle\langle{#1}|}
\begin{document}

\preprint{APS/123-QED}

\title{Quantum entanglement in a pure state of strongly correlated quantum impurity systems}

\author{Yunori Nishikawa}
\email{nishikaway@omu.ac.jp}
\thanks{Nambu Yoichiro Institute of Theoretical and Experimental Physics, Osaka Metropolitan University, Sumiyoshi-ku, Osaka 558-8585 Japan}
\author{Tomoki Yoshioka}
%
\affiliation{Dept. of Physics, Graduate School of Science, Osaka
 Metropolitan University, Sumiyoshi-ku, Osaka 558-8585 Japan}
%


\date{\today}

\begin{abstract}
We consider quantum entanglement in strongly correlated quantum impurity
 systems for states manifesting interesting properties such as multi-level Kondo effect and
 dual nature between itineracy and localization etc..
For this purpose, we set up a system consisting of one or two quantum
 impurities arbitrarily selected from the system as a subsystem, and
 investigate quantum entanglement with its environmental system.
%
We reduce the pure state of interest as described above to the subsystem, 
and formulate quantum informative quantities such as entanglement
 entropy, mutual information and relative entropy.
%
To demonstrate the potential of the method proposed here, we apply them
 to a dimer with several types of internal interaction, short Hubbard chains
 and  the single impurity Anderson model 
to study the relationship between 
their states and the behaviors of quantum informative quantities.
%
The obtained results suggest that the method proposed here is promising  for elucidating the quantum
 entanglement of pure states in various quantum impurity systems.
\end{abstract}

\pacs{72.10.F,72.10.A,73.61,11.10.G}

\maketitle


\section{Introduction}
In solid-state electron systems, interesting ground states emerge at low
temperatures as a result of the interaction of many electrons.
Understanding the ground state itself is one of the central issues,
which is especially difficult in strongly correlated electron systems
where correlation effects between electrons cannot be ignored.
In addition, the question of the quantum entanglement transitions
 from high temperature to low temperature to reach the ground state 
is interesting and important from the 
viewpoint of understanding the ground state from a new perspective.
In recent years, quantum entanglement for interesting phenomena in
crystalline systems has been studied\cite{ent-solid}, including area laws and their
deviations etc..

In this paper, we primarily focus on strongly correlated quantum impurity
systems and discuss quantum entanglement in such systems\cite{costi,mil,mil1,stat,YT_shim,YT_shim2,YT_rams,YT_hur,YT_wagn,YT_stoc,YT_affl,YT_sore,YT_erik,YT_sale}.
A system in which multiple quantum dots are connected to electron
reservoirs such as conduction electron systems is a typical strongly
correlated quantum impurity system. In such a system, electron correlation
effects such as the Kondo effect can be studied under controlled
conditions\cite{Natn,goldhaber1998kondo,PhysRevLett.81.5225,potok2007observation,cronenwett1998tunable,jeong2001kondo,iftikhar2015two,sasaki2000kondo,PhysRevLett.104.036804}.

It has been theoretically pointed out that various interesting ground
state such as Nagaoka ferromagnetism and Lieb ferrimagnetism can be
realized on multiple quantum dots\cite{lieb1,lieb2,nagaoka,nagaoka1,tasaki,PhysRevB.100.224421,tnlieb}, and experimentally, Nagaoka
ferromagnetism has been realized on quartet quantum dots\cite{Natn}.
These states are generally entangled states among multiple quantum dots\cite{tnlieb,kondo-naga}.
When electron reservoirs are connected to these systems, the degrees of
freedom of these states are progressively screened from high to low
temperatures by the Kondo effect, and the whole system becomes an entangled
ground state between the quantum dots and the electron reservoirs.
This situation can be understood to some extent by calculating the
contribution of thermodynamic entropy from multiple quantum dots using
Numerical Renormalization Group(NRG) calculations\cite{kww}.
From the results of such studies\cite{tnlieb,kondo-naga,no,onh,PhysRevB.102.085418,PhysRevB.77.035120,PhysRevB.71.075305,PhysRevB.72.165309,PhysRevB.82.161411}, it is known that these Kondo screenings
of multiple degrees of freedom from high temperature to low temperature
may not simply be a process in which independent magnetic moments on
the quantum dots at each location are individually screened.
This is because the states on multiple quantum dots screened by the
Kondo effect are entangled states among multiple quantum dots.
However, this method is not sufficient to obtain a more detailed picture
of screening, and other methods that complement it
 are desired.

In a certain parameter region of the two single-impurity Anderson model
coupled by antiferromagnetic interaction, the  electrons on each
impurity 
are strongly antiferromagnetically correlated with each other, but can
itinerate to the conduction electron systems, which is the dual nature of
localization and itinerancy. 
Here, a new Fermi liquid state is brought about, where the quasi-particle picture holds
but the adiabatic continuity does not hold.\cite{cn,nc,aligia}
The new Fermi liquid is characterized by a topological invariant derived from electron correlation.
In this phase, it would be interesting to compare 
the quantum entanglement between two impurities due to antiferromagnetic coupling 
 and the quantum entanglement between each impurity and the conduction electron system due to the
Kondo effect.
Recently, it has been pointed out that such new Fermi liquids may be realized in
the real systems\cite{expaligia, expaligia1,expaligia2}.
%

In a multiple quantum impurity system as described above, unlike a
crystalline system, each quantum impurity in the system is not
necessarily equivalent, so the quantum entanglement between the selected quantum
impurity in the system and its environmental system generally depends on
the selection of that quantum impurity.
Hence, one can consider the distribution of entanglement on quantum
impurities in the system.
Further investigation of the entanglement in quantum impurity pairs in
the system would provide a deeper understanding of, for example,  the evolution of
quantum entanglement in multi-level Kondo screening, and the dual nature of
localization and itinerancy in the topological Fermi liquid.

%
Therefore, in this paper, quantum impurities are modeled by Anderson
impurities, and in section \ref{forma},
we consider the density operator describing the pure state of interest
reduced to the subsystem consisting of one or two selected Anderson
impurities in the system.
We then formulate quantum informative quantities such as entanglement
entropy, mutual information, and relative entropy.
In sections \ref{app0} and \ref{app}, 
we apply the formulations in the previous section
to
a dimer with several types of internal interaction, short Hubbard chains
 and 
 the Single Impurity Anderson Model(SIAM) 
to study the relationship between 
their states and the behaviors of quantum informative quantities.
The discussion and conclusions are provided in the last section
\ref{last}.

\section{Formulation and Systems}\label{forma}

In this paper, systems such as quantum impurity systems are described by
the Hilbert space ${\mathcal H}$, and are regarded as a composite system of subsystem $A$ and $B$, 
described by the Hilbert spaces ${\mathcal H}_{A}$ and ${\mathcal H}_{B}$, respectively.
Here we restrict the considered state to pure state as
\begin{equation}
\psi\in{\mathcal H}={\mathcal H}_{A}\otimes{\mathcal H}_{B}.\label{pstate}
\end{equation}

The state $\psi$ is expanded as follows;
\begin{equation}
\psi=\sum_{a,b}C_{a,b}e^{(A)}_{a}\otimes e^{(B)}_{b}
\end{equation}
, where $\{e^{(A)}_{a}\}_{a}$ and $\{e^{(B)}_{b}\}_{b}$ are respectively 
Complete OrthoNormal Systems (C.O.N.S.) of ${\mathcal H}_{A}$ and
${\mathcal H}_{B}$.

We reduce the state $\psi$ to subsystem $A$.
The density operator describing the reduced state is 
\begin{equation}
\rho^{(A)}_{\psi}
\equiv
{\rm Tr}_{{\mathcal H}_{B}}|\psi\rangle\langle\psi|
=
\sum_{a,a^{\prime}}
\sum_{b}C_{a,b}C^{\ast}_{a^{\prime},b}
|e^{(A)}_{a}\rangle\langle e^{(A)}_{a^{\prime}} |
\end{equation}
, and the expansion coefficients can be expressed as 
the expected values of operators on ${\mathcal H}_{A}$ in the state $\psi$ 
as follows;
\begin{equation}
\sum_{b}C_{a,b}C^{\ast}_{a^{\prime},b}
=
\langle 
\psi
|
\left(
|e^{(A)}_{a^{\prime}}\rangle
\langle e^{(A)}_{a}|
\otimes
I^{(B)}
\right)
|
\psi
\rangle .
\end{equation}

Henceforth, we write $O^{(A)}\otimes I^{(B)}$ as $O^{(A)}$, and also
denote the expected value $\langle\psi,O^{(A)}\psi\rangle$ in state
$\psi$ by $\langle O^{(A)}\rangle=\langle
O^{(A)}\rangle_{\psi}$($O^{(A)}$ is a linear operator on ${\mathcal H}_{A}$).

\subsection{Subsystem $A$ consisting of one selected quantum impurity}\label{oneI}
In this subsection, we consider the total system including at least one
Anderson impurity, and the system $A$ shall be one 
Anderson impurity system selected from the total system.
The C.O.N.S. of ${\mathcal H}_{A}$ is as follows;
\begin{equation}
\left\{
e^{(A)}_{\alpha}
\right\}_{\alpha}
=
\left\{
|0\rangle_{A},
d^{\dagger}_{\uparrow}|0\rangle_{A},
d^{\dagger}_{\downarrow}|0\rangle_{A},
d^{\dagger}_{\uparrow}d^{\dagger}_{\downarrow}|0\rangle_{A}
\right\}\label{consA}
\end{equation}
, where $d^{\dagger}_{\sigma}$ is the creation operator of the electron
in the selected Anderson impurity with spin $\sigma$
and 
$|0\rangle_{A}$ is the vacuum state.
Using this C.O.N.S.(\ref{consA}), we obtain the representation matrix of the reduced density operator
$\rho_{\psi}^{(A)}$ as follows;
\begin{equation}
\tilde{\rho}^{(A)}_{\psi}
\equiv\left(
\begin{array}{cccc}
\langle h_{\uparrow}h_{\downarrow}\rangle&\langle
 d_{\uparrow}^{\dagger}h_{\uparrow}\rangle&\langle d^{\dagger}_{\downarrow}h_{\downarrow}\rangle&\langle d^{\dagger}_{\downarrow}d^{\dagger}_{\uparrow}\rangle\\
\langle d_{\uparrow}h_{\uparrow}\rangle&\langle
 n_{\uparrow}h_{\uparrow}\rangle&\langle s_{-}\rangle&-\langle n_{\uparrow}d^{\dagger}_{\downarrow}\rangle\\
\langle
 d_{\downarrow}h_{\downarrow}\rangle&\langle
 s_{+}\rangle&\langle n_{\downarrow}h_{\downarrow}\rangle&\langle n_{\downarrow}d^{\dagger}_{\uparrow}\rangle\\
\langle d_{\uparrow}d_{\downarrow}\rangle&-\langle
n_{\uparrow}d_{\downarrow}\rangle&\langle
n_{\downarrow}d_{\uparrow}\rangle&\langle n_{\uparrow}n_{\downarrow}\rangle\\
\end{array}
\right)\label{rho-a}
\end{equation}
, where $s_{+}\equiv d^{\dagger}_{\uparrow}d_{\downarrow}$, $s_{-}\equiv
d^{\dagger}_{\downarrow}d_{\uparrow}$, 
$h_{\sigma}\equiv I-n_{-\sigma}$, and using the relation
$|0\rangle_{A}{}_{A}\langle 0|=h_{\uparrow} h_{\downarrow}$.

The entanglement entropy $S_{\rm E.E.}$ for $\psi$ between system $A$ and $B$ is the von Neumann entropy 
of the reduced state. To wit,
\begin{equation}
S_{\rm E.E.}(=S_{\rm E.E.}(\rho^{(A)}_{\psi}))
\equiv
-{\rm Tr}_{{\mathcal H}_{A}}\left(\rho^{(A)}_{\psi}\log(\rho^{(A)}_{\psi})\right).
\end{equation}

For the general pure state $\psi$, we need to evaluate 9 independent matrix
elements in the right hand of Eq.(\ref{rho-a}) to obtain $\tilde{\rho}^{(A)}_{\psi}$. 
In condensed matter physics, especially in multiplet quantum impurity
systems that manifest the Kondo effects, etc., the states of interest
are often simultaneous eigenstates of $S_{z}$ and $Q$.
Here $S_{z}$ is the spin operator of the $z$-component for the
total system, and $Q$ is the electron number operator of the total
system counted from the half-filling number.
In this case, all off-diagonal matrix elements of
 $\tilde{\rho}^{(A)}_{\psi}$
 vanish and then
$\tilde{\rho}^{(A)}_{\psi}$ becomes a diagonal matrix.
Therefore, $S_{\rm E.E.}$ has a simple expression as follows;
\begin{eqnarray}
S_{\rm E.E.}
&=&
-\langle n_{\uparrow}n_{\downarrow}\rangle\log\langle
n_{\uparrow}n_{\downarrow}\rangle
-
\langle h_{\uparrow}h_{\downarrow}\rangle\log\langle
h_{\uparrow}h_{\downarrow}\rangle\nonumber\\
& &\mbox{}
-\langle n_{\uparrow}h_{\uparrow}\rangle\log\langle
n_{\uparrow}h_{\uparrow}\rangle
-
\langle n_{\downarrow}h_{\downarrow}\rangle\log\langle
n_{\downarrow}h_{\downarrow}\rangle.\label{see-a}
\end{eqnarray}

Hereafter, we will mainly deal with the case where $\psi$ is a
simultaneous eigenstate of $Q$ and $S_{z}$.

From Eq.(\ref{see-a}), it is found that
$S_{\rm E.E.}$ is a function of three variables
$\langle n_{\uparrow}n_{\downarrow}\rangle$,
$\langle n\rangle$ and
$\langle n_{\uparrow}-n_{\downarrow}\rangle (n\equiv n_{\uparrow}+n_{\downarrow})$.
To investigate the stationary value problem of $S_{\rm E.E.}$ and the linear
response of $S_{\rm E.E.}$ to these three variables, differentiating $S_{\rm
E.E.}$ by the three variables and setting them equal to zero yields the following results;
\begin{eqnarray}
\frac{\partial S_{\rm E.E.}}{\partial \langle
 n_{\uparrow}n_{\downarrow}\rangle}
&=&\log\left(\frac{\langle n_{\uparrow}h_{\uparrow}\rangle\langle
	n_{\downarrow}h_{\downarrow}\rangle}
{\langle n_{\uparrow}n_{\downarrow}\rangle\langle h_{\uparrow}h_{\downarrow}\rangle}
\right)=0,\\
\frac{\partial S_{\rm E.E.}}{\partial \langle
 n\rangle}
&=&\log\left(
\frac
{\langle h_{\uparrow}h_{\downarrow}\rangle}
{\sqrt{
\langle n_{\uparrow}h_{\uparrow}\rangle
\langle n_{\downarrow}h_{\downarrow}\rangle
}}
\right)=0,\\
\frac{\partial S_{\rm E.E.}}{\partial \langle
 n_{\uparrow}-n_{\downarrow}\rangle}
&=&\log\left(
\sqrt{
\frac
{\langle n_{\downarrow}h_{\downarrow}\rangle}
{\langle n_{\uparrow}h_{\uparrow}\rangle}
}
\right)=0.
\end{eqnarray}
From the above results, we find that 
the stationary value condition of $S_{\rm E.E.}$ for each variable
gives the relationship among several correlation functions.
The logarithm of the displacement of the relation gives the linear
response of $S_{\rm E.E.}$ to each variable.
Combining the stationary conditions of $S_{\rm E.E.}$ for all
variables,
the well-known equal probability condition is derived, yielding a
maximum value of $S_{\rm E.E.}$, $\log(4)$.
The equal probability condition is expressed as
$\langle n_{\uparrow}n_{\downarrow}\rangle=\langle
n_{\uparrow}\rangle\langle n_{\downarrow}\rangle(=\frac{1}{4})$,
$\langle n\rangle=1$ and
$\langle n_{\uparrow}- n_{\downarrow}\rangle=0$ in the three variables. 
From this result, for maximum entanglement entropy, the up- and
down-spin electrons must be uncorrelated, which is a type of 
monogamy of quantum entanglement\cite{monog,monog1}.

To quantitatively consider the entanglement between the up- and down- spin electrons
, let us consider the mutual information.
For this purpose, we consider system $A$ as a composite system of up-
and down-spin electron as follows;
${\mathcal H}_{A}={\mathcal H}_{\uparrow}\otimes{\mathcal H}_{\downarrow}$
where
${\mathcal H}_{\sigma}\equiv{\rm span}\left\{|0\rangle_{\sigma}, \
|\sigma\rangle_{\sigma}\right\}$ ($\sigma=\uparrow, \ \downarrow$, $|\sigma\rangle_{\sigma}\equiv d^{\dagger}_{\sigma}|0\rangle_{\sigma}$).
The reduced state $\rho^{(\sigma)}_{\psi}$of $\rho^{(A)}_{\psi}$ to
${\mathcal H}_{\sigma}$ is defined by $\rho^{(\sigma)}_{\psi}\equiv 
{\rm Tr}_{{\mathcal H}_{-\sigma}}\rho^{(A)}_{\psi}$
and is expressed as follows; 
\begin{equation}
\rho^{(\sigma)}_{\psi}
=
\langle n_{\sigma}\rangle
|\sigma\rangle_{\sigma}{}_{\sigma}\langle\sigma|
+
\left(1-\langle n_{\sigma}\rangle\right)
|0\rangle_{\sigma}{}_{\sigma}\langle 0|.
\end{equation}

The mutual information $I$ which describes the entanglement between the
up- and down- spin electron is defined by 
\begin{equation}
I(=I(\rho^{(A)}_{\psi}))
\equiv
D\left(
\rho^{(A)}_{\psi}
||
\rho^{(\uparrow)}_{\psi}
\otimes
\rho^{(\downarrow)}_{\psi}
\right),\label{I}
\end{equation}
where 
\begin{equation}
D(\rho||\rho^{\prime})
\equiv
{\rm Tr}_{{\mathcal H}_{A}}
\rho
\left(
\log\rho
-
\log\rho^{\prime}
\right)
\end{equation}
is the Umegaki's relative entropy of $\rho$ with respect to $\rho^{\prime}$\cite{ume}.

We can easily derive the following equation;
\begin{equation}
I=S_{\uparrow}+S_{\downarrow}-S_{\rm E.E.}\label{relIS},
\end{equation}
where
\begin{equation}
S_{\sigma}\equiv -\langle n_{\sigma}\rangle\log\left(\langle
						n_{\sigma}\rangle\right)
-\left(1-\langle n_{\sigma}\rangle\right)\log\left(1-\langle n_{\sigma}\rangle\right)\label{Ssigma}
\end{equation}
is the von Neumann entropy of $\rho^{(\sigma)}_{\psi}$.

The direct measure of the difference between 
$\rho^{(A)}_{\psi}$ and $\rho^{(\uparrow)}_{\psi}
\otimes
\rho^{(\downarrow)}_{\psi}$
is the trace distance.
We easily find that
\begin{equation}
{\rm Tr}\left|\rho^{(\uparrow)}_{\psi}\otimes\rho^{(\downarrow)}_{\psi}-\rho^{(A)}_{\psi}\right|
=
4\left(
-
\langle
n_{\uparrow}n_{\downarrow}
\rangle
+
\langle
n_{\uparrow}
\rangle
\langle
n_{\downarrow}
\rangle
\right).
\end{equation}
The above result shows that the trace distance is proportional to the correlation
between the up- and down-spin electrons.
Note that while $I$ contains a logarithmic function, the trace distance
does not, so these functions are generally independent functions.

Now we consider capturing the Kondo effect from the 
perspective of quantum information theory.
Here, we further assume that $\psi$ is  an eigenstate of the spin magnitude
operator ${\bf S}^{2}$ of the total system,
This assumption applies, for example, to the energy eigenstates of
systems with spin SU(2) symmetry and many interesting quantum impurity
systems have spin SU(2) symmetry.
Here we write $\psi$ as $\psi_{\mathcal Q}({\mathcal S},{\mathcal S}_{z})$ and
${\bf S}^{2}\psi_{\mathcal Q}({\mathcal S},{\mathcal S}_{z})={\mathcal
S}({\mathcal S}+1)\psi_{\mathcal Q}({\mathcal S},{\mathcal S}_{z})$,
$S_{z}\psi_{\mathcal Q}({\mathcal S},{\mathcal S}_{z})={\mathcal
S}_{z}\psi_{\mathcal Q}({\mathcal S},{\mathcal S}_{z})$
and
$Q\psi_{\mathcal Q}({\mathcal S},{\mathcal S}_{z})={\mathcal
Q}\psi_{\mathcal Q}({\mathcal S},{\mathcal S}_{z})$
hold.
We assume that $\psi_{\mathcal Q}({\mathcal S},{\mathcal S}_{z})$ and
$\psi_{\mathcal Q}({\mathcal S},{\mathcal S}_{z}-1)$ are related typically by the following equation;
$S_{-}\psi_{\mathcal Q}({\mathcal S},{\mathcal S}_{z})=\sqrt{({\mathcal
S}+{\mathcal S}_{z})({\mathcal S}-{\mathcal S}_{z}+1)}\psi_{\mathcal
Q}({\mathcal S},{\mathcal S}_{z}-1) \
({\mathcal S}\ge {\mathcal S}_{z}>-{\mathcal S})$, where $S_{-}\equiv S_{x}-iS_{y}$.

Under the above assumptions, we find that
$\langle n \rangle_{\psi_{\mathcal Q}({\mathcal S},{\mathcal S}_{z})}$
,
$\langle n_{\uparrow}n_{\downarrow} \rangle_{\psi_{\mathcal Q}({\mathcal
S},{\mathcal S}_{z})}$
are independent of ${\mathcal S}_{z}$ and 
$\langle s_{z}\rangle_{\psi_{\mathcal Q}({\mathcal S},-{\mathcal S})}=
-\langle s_{z}\rangle_{\psi_{\mathcal Q}({\mathcal S},{\mathcal S})}$
holds, 
where
\begin{equation}
s_{z}\equiv \frac{1}{2}\left(n_{\uparrow}-n_{\downarrow}\right).
\end{equation}

Hence, we obtain
\begin{equation}
D\left(\rho^{(A)}_{\psi_{\mathcal Q}({\mathcal S},{\mathcal
  S})}||\rho^{(A)}_{\psi_{\mathcal Q}({\mathcal S},-{\mathcal S})}\right)
=
2
\langle
s_{z}
\rangle
\log
\left(
\frac
{\langle n \rangle/2-\langle n_{\uparrow}n_{\downarrow}\rangle+\langle s_{z}\rangle}
{\langle n \rangle/2-\langle n_{\uparrow}n_{\downarrow}\rangle-\langle s_{z}\rangle}
\right),
\end{equation}
where $\langle\cdots\rangle=\langle\cdots\rangle_{\psi_{\mathcal
Q}({\mathcal S},{\mathcal S})}$.
The Kondo screening of the selected impurity can be captured through
the relative entropy of $\rho^{(A)}_{\psi_{\mathcal Q}({\mathcal S},{\mathcal S})}$ with respect to $\rho^{(A)}_{\psi_{\mathcal Q}({\mathcal S},-{\mathcal S})}$
, as shown in subsection \ref{flow}.

The trace distance between 
$\rho^{(A)}_{\psi_{\mathcal Q}({\mathcal S},{\mathcal S})}$ and 
$\rho^{(A)}_{\psi_{\mathcal Q}({\mathcal S},-{\mathcal S})}$ 
is 
\begin{equation}
{\rm Tr}
\left|
\rho^{(A)}_{\psi_{\mathcal Q}({\mathcal S},{\mathcal S})}
-
\rho^{(A)}_{\psi_{\mathcal Q}({\mathcal S},-{\mathcal S})}
\right|
=
4\left|
\langle
s_{z}
\rangle
\right|,
\end{equation}
exactly proportional to the  magnetization of the selected Anderson impurity.
Since $D$ is a function of $\langle s_{z}\rangle$, $\langle n \rangle$, and $\langle
n_{\uparrow}n_{\downarrow}\rangle$
and the trace distance is a function of $\langle s_{z}\rangle$ only,
they are independent functions in general.
However, when $\langle n \rangle$, and $\langle
n_{\uparrow}n_{\downarrow}\rangle$ are constant and the value of 
$\langle s_{z}\rangle$ is small compared to those values, 
they are essentially the same functions since
$D\propto \langle s_{z}\rangle^{2}$
.

At the end of this subsection, 
we discuss the entanglement entropy for three special cases.

The first case is the case where further
half-filling state 
$\langle n\rangle=1$
and magnetic isotropic state
$\langle n_{\uparrow}\rangle=\langle n_{\downarrow}\rangle$.
In this case, the expression of $S_{\rm E.E.}$ (\ref{see-a})
is as follows;
\begin{equation}
S_{\rm E.E.}
=\log(2)
-p\log p
-(1-p)\log(1-p)\label{s_hf},
\end{equation}
where
$p=1-2\langle n_{\uparrow}n_{\downarrow}\rangle$.

From this expression, the Anderson impurity and its environment are always
entangled because the inequality $S_{\rm E.E.}\ge \log(2)$ holds.
This is because there is no separable state $\psi$ that satisfies
$\langle n_{\uparrow}\rangle=\langle n_{\downarrow}\rangle$ and 
$\langle n\rangle=1$ under the assumptions that led to Eq.(\ref{see-a}), as will be shown next.
Assume that there exists a separable state
$\psi=\psi^{(A)}\otimes\psi^{(B)}$
that satisfies 
$\langle n_{\uparrow}\rangle=\langle n_{\downarrow}\rangle$ and 
$\langle n\rangle=1$.
In the discussion where we derived Eq.(\ref{s_hf}),
the state $\psi$ is the eigenstate of $S_{z}$ and $Q$;
$S_{z}\psi={\mathcal S}_{z}\psi$,
$Q\psi={\mathcal Q}\psi$.
From the assumption, the state $\psi^{(A)}$
is an eigenstate of $S_{z}^{(A)}\equiv s_{z}$ and $Q^{(A)}\equiv n-1$;
$S_{z}^{(A)}\psi^{(A)}={\mathcal S}_{z}^{(A)}\psi^{(A)}$,
$Q^{(A)}\psi^{(A)}={\mathcal Q}^{(A)}\psi^{(A)}$.
Then we obtain ${\mathcal S}_{z}^{(A)}=0$ and ${\mathcal Q}^{(A)}=0$ because
$0=\langle
n_{\uparrow}-n_{\downarrow}\rangle=2\langle\psi^{(A)}|S_{z}^{(A)}|\psi^{(A)}\rangle=2{\mathcal
S}_{z}^{(A)}$, $1=\langle
n\rangle=\langle\psi^{(A)}|Q^{(A)}+1|\psi^{(A)}\rangle={\mathcal Q}^{(A)}+1$.
However, simultaneous eigenstate for ${\mathcal S}_{z}^{(A)}=0$ and
${\mathcal Q}^{(A)}=0$ does not
exist in the Anderson impurity system.

Next, we consider the case where $\psi$ is an eigenstate of $S_{z}$ but
not necessarily an eigenstate of $Q$.
This may be the case, for example, in systems containing superconducting
leads etc..
In this case, the off-diagonal matrix elements of the right hand of
Eq.(\ref{rho-a}) vanish except for (1-4) and (4-1) elements.
So we obtain the relatively simple expression of $S_{\rm E.E.}$ as follows;
\begin{eqnarray}
S_{\rm E.E.}
&=&
-\langle n_{\uparrow}h_{\uparrow}\rangle\log\langle
n_{\uparrow}h_{\uparrow}\rangle
-\langle n_{\downarrow}h_{\downarrow}\rangle\log\langle
n_{\downarrow}h_{\downarrow}\rangle\nonumber\\
\mbox{}&&
-\lambda_{+}\log\lambda_{+}
-\lambda_{-}\log\lambda_{-},
\end{eqnarray}
where
\begin{equation}
\lambda_{\pm}
\equiv
\frac{1}{2}
\left(
1
-
\langle n\rangle
+
2\langle n_{\uparrow}n_{\downarrow}\rangle
\pm
\sqrt{
(1-\langle n\rangle)^{2}
+
4
|
\langle d^{\dagger}_{\uparrow}d^{\dagger}_{\downarrow}\rangle
|^{2}
}
\right).
\end{equation}
.
The last case is when $\psi$ is an eigenstate of $Q$ but not necessarily
an eigenstate of $S_{z}$.
In this case, the off-diagonal matrix elements of the right hand of
Eq.(\ref{rho-a}) vanish except for (2-3) and (3-2) elements.
So we also obtain the relatively simple expression of $S_{\rm E.E.}$ as follows;
\begin{eqnarray}
S_{\rm E.E.}
&=&
-\langle n_{\uparrow}n_{\downarrow}\rangle\log\langle
n_{\uparrow}n_{\downarrow}\rangle
-\langle h_{\uparrow}h_{\downarrow}\rangle\log\langle
h_{\uparrow}h_{\downarrow}\rangle\nonumber\\
\mbox{}&&
-\mu_{+}\log\mu_{+}
-\mu_{-}\log\mu_{-}.
\end{eqnarray}
where 
\begin{equation}
\mu_{\pm}
\equiv
\frac{1}{2}
\left(
\langle n\rangle
-
2\langle n_{\uparrow}n_{\downarrow}\rangle
\pm
\sqrt{
\langle n_{\uparrow}-n_{\downarrow}\rangle^{2}
+
4
|
\langle s_{+}\rangle
|^{2}
}
\right).
\end{equation}

\subsection{Subsystem $A$ consisting of two selected quantum impurities}\label{twoI}
In this subsection, we consider the total system including at least two Anderson impurities.
We can select two Anderson impurities from the total system, which are
called $a$, $b$ and described by
creations operators $d_{a\sigma}^{\dagger}(=a^{\dagger}_{\sigma})$ and $d_{b\sigma}^{\dagger}(=b^{\dagger}_{\sigma})$,
respectively($\sigma=\uparrow, \downarrow$ represents spin index).
Here we define the subsystem $A$ consisting of two Anderson impurities $a$ and $b$.
The 16-dimensional Hilbert space ${\mathcal H}_{A}$ of system $A$ is spanned by the following basis;
\begin{equation}
\left\{
|\{N_{i\sigma}\}\rangle\equiv
\prod_{i=a,b,\sigma=\uparrow,\downarrow}
\left(d_{i\sigma}^{\dagger}\right)^{N_{i\sigma}}
|0\rangle_{ab}
|
N_{i\sigma}=0,1 
\right\}.\label{cons-ab}
\end{equation}


We consider the reduced state of $\psi$ defined by
the Eq.(\ref{pstate}), to ${\mathcal H}_{A}$  as follows;
\begin{equation}
\rho^{(ab)}_{\psi}
=
{\rm Tr}_{{\mathcal H}_{B}}
|\psi\rangle
\langle\psi|,
\end{equation}
where ${\mathcal H}_{B}$ is the Hilbert space describing the environmental
system $B$ of $A$.

The reduced state $\rho^{(ab)}_{\psi}$ can be represented as a
 16$\times$16
 matrix having 135 independent matrix elements
, which are shown in the appendix.

%
 In the main text here, it is useful to discuss special cases where
there are at least specific applications.

Because of the following equation,
\begin{equation}
\left[Q,|\{N_{i\sigma}\}\rangle\langle\{N^{\prime}_{i\sigma}\}|\right]=\left(N_{Q}-N_{Q}^{\prime}\right)|\{N_{i\sigma}\}\rangle\langle\{N^{\prime}_{i\sigma}\}|,
\end{equation}
 where$N_{Q}=\sum_{i\sigma}N_{i\sigma}$ and$N_{Q}^{\prime}=\sum_{i\sigma}N^{\prime}_{i\sigma}$,
when we assume that $\psi$ is an eigenstate of $Q$
, the representation matrix of $\rho^{(ab)}_{\psi}$
becomes a block diagonal matrix consisting of
$1\times 1$ matrix for $N_{Q}=0$,
$4\times 4$ matrix for $N_{Q}=1$, 
$6\times 6$ matrix for $N_{Q}=2$, 
$4\times 4$ matrix for $N_{Q}=3$ and 
$1\times 1$ matrix for $N_{Q}=4$. 

\begin{figure}[h]
\includegraphics[width=0.5\textwidth]{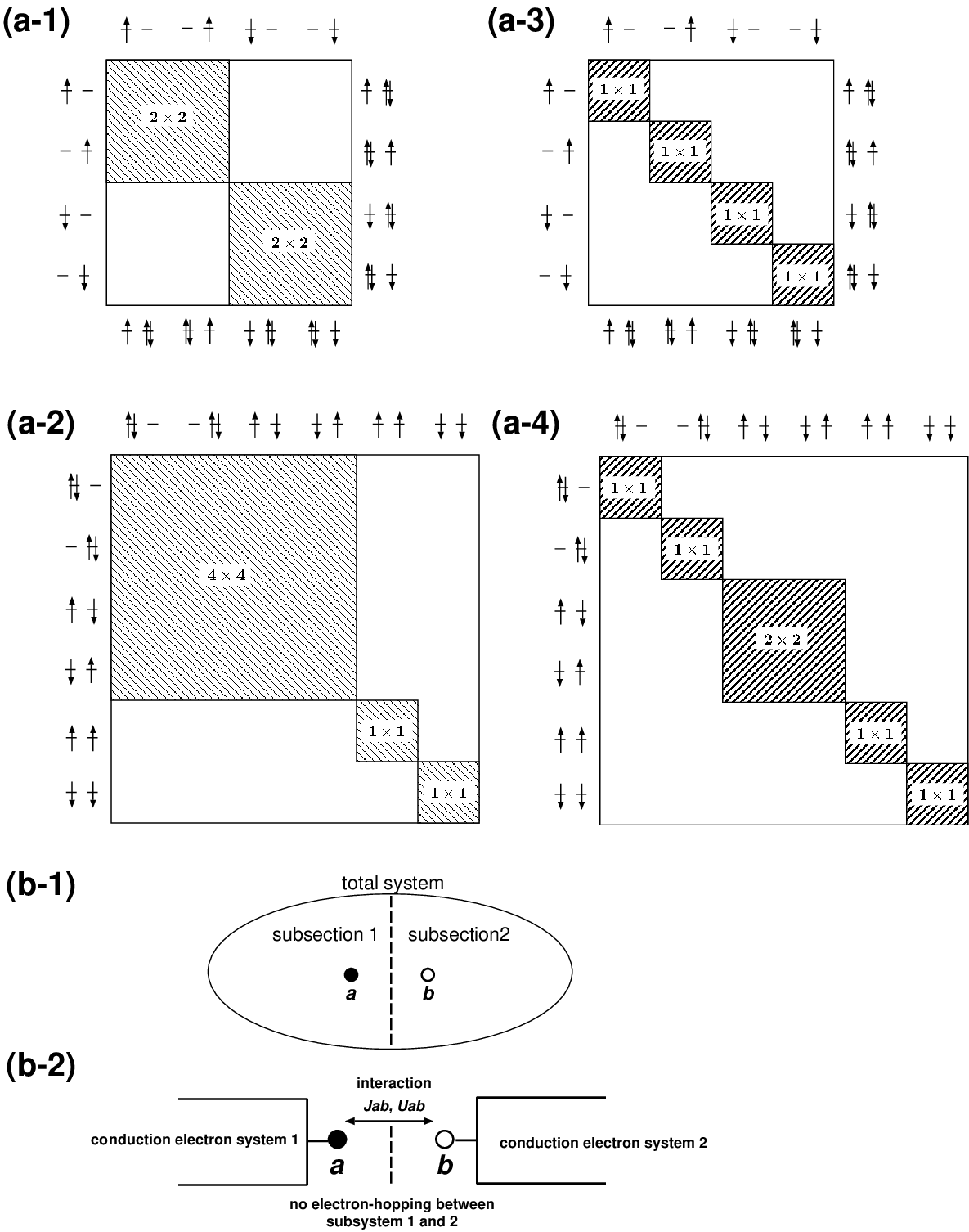}
\caption{(a)Block matrices of sectors for $N_{Q}=1,3$(a-1) and $N_{Q}=2$(a-2), 
composing the representation block diagonal matrix of
 $\rho^{(A)}_{\psi}$ when $\psi$ is an eigenstate of $Q$.
Each block matrix is further block diagonalized if $\psi$ is an
 eigenstate of $S_{z}$ (a-1,2),$S_{z}$ and $\Delta Q$(a-3,4).
The pictures placed on the upper and left sides of each matrix show the electron
 occupancy of the Anderson impurities $a$ and $b$ for $N_{Q}=1$(a-1,3)
 and $N_{Q}=2$(a-2,4).  
The electron occupancy of the  Anderson impurities $a$ and $b$
 for $N_{Q}=3$(a-1,3) is also shown in the pictures placed on the bottom and
 right sides of each matrix.
(b-1) Bisection of all orbitals in total system, where Anderson
 impurities $a$ and $b$ belong to different subsections.
(b-2) Two single
impurity Anderson models coupled by magnetic interaction $J_{ab}$ and
repulsion $U_{ab}$ between the two impurities without electron
 transfer, as an example where the bisection shown in (b-1) is useful and
 natural.
}\label{blockzu}
\end{figure}

We further assume that $\psi$ is an eigenstate of
$S_{z}$.
Because the following equation holds; 
\begin{equation}
\left[S_{z},|\{N_{i\sigma}\}\rangle\langle\{N^{\prime}_{i\sigma}\}|\right]=\left(N_{S}-N_{S}^{\prime}\right)|\{N_{i\sigma}\}\rangle\langle\{N^{\prime}_{i\sigma}\}|,
\end{equation}
 where
$N_{S}=\sum_{i\sigma}\sigma N_{i\sigma}$ and
$N_{S}^{\prime}=\sum_{i\sigma}\sigma N^{\prime}_{i\sigma}$, 
the $4\times 4$ matrices for $N_{Q}=1, 3$ 
become the block diagonal matrices 
consisting of
$2\times2$  matrix for $N_{S}=\frac{1}{2}$, 
$2\times 2$ matrix for $N_{S}=-\frac{1}{2}$
and 
the $6\times 6$ matrix for $N_{Q}=2$
becomes the block diagonal matrix consisting of
$4\times 4$ matrix for $N_{S}=0$, 
$1\times 1$ matrix for $N_{S}=1$, 
$1\times 1$ matrix for $N_{S}=-1 $,
as shown in the Fig.\ref{blockzu}(a-1) and (a-2), respectively.


From the above discussion, 
the representation matrix of
$\rho^{(ab)}_{\psi}$
is consist of following block matrices;

the $1\times 1$ matrix for the sector $(N_{Q},N_{S})=(0,0)$
\begin{equation}
\tilde{\rho}^{(ab)}_{\psi,N_{Q}=0,N_{S}=0}\equiv\langle
 h_{a\uparrow}h_{a\downarrow}h_{b\uparrow}h_{b\downarrow}\rangle,
\end{equation}
the $2\times 2$ matrix for the sector $(N_{Q},N_{S})=\left(1,\frac{1}{2}\right)$
\begin{eqnarray}
& &\mbox{}\tilde{\rho}^{(ab)}_{\psi,N_{Q}=1,N_{S}=\frac{1}{2}}\nonumber\\
&\equiv&\left(
\begin{array}{cc}
\langle n_{a\uparrow}h_{a\uparrow}h_{b\uparrow}h_{b\downarrow}\rangle&
\langle b^{\dagger}_{\uparrow}h_{a\uparrow}h_{b\uparrow}a_{\uparrow}\rangle\\
\langle a^{\dagger}_{\uparrow}h_{a\uparrow}h_{b\uparrow}b_{\uparrow}\rangle&
\langle h_{a\uparrow}h_{a\downarrow}h_{b\uparrow}n_{b\uparrow}\rangle
\end{array}
\right),\label{rho-a2}
\end{eqnarray}
the $2\times 2$ matrix for the sector $(N_{Q},N_{S})=\left(1,-\frac{1}{2}\right)$
\begin{eqnarray}
& &\mbox{}
\tilde{\rho}^{(ab)}_{\psi,N_{Q}=1,N_{S}=-\frac{1}{2}}\nonumber\\
&\equiv&\left(
\begin{array}{cc}
\langle
 n_{a\downarrow}h_{a\downarrow}h_{b\uparrow}h_{b\downarrow}\rangle&
\langle b^{\dagger}_{\downarrow}h_{a\downarrow}h_{b\downarrow}a_{\downarrow}\rangle\\
\langle a^{\dagger}_{\downarrow}h_{a\downarrow}h_{b\downarrow}b_{\downarrow}\rangle
&
\langle
 h_{a\uparrow}h_{a\downarrow}h_{b\downarrow}n_{b\downarrow}\rangle
\end{array}
\right),\label{rho-a2}
\\
\end{eqnarray}
the $4\times 4$ matrix for the sector $(N_{Q},N_{S})=(2,0)$
\begin{eqnarray}
\left(\tilde{\rho}^{(ab)}_{\psi,N_{Q}=2,N_{S}=0}\right)_{11}
&\equiv&
\langle
n_{a\uparrow}n_{a\downarrow}h_{b\uparrow}h_{b\downarrow}
\rangle,\\
\left(\tilde{\rho}^{(ab)}_{\psi,N_{Q}=2,N_{S}=0}\right)_{22}
&\equiv&
\langle
h_{a\uparrow}h_{a\downarrow}n_{b\uparrow}n_{b\downarrow}
\rangle,\\
\left(\tilde{\rho}^{(ab)}_{\psi,N_{Q}=2,N_{S}=0}\right)_{33}
&\equiv&
\langle
n_{a\uparrow}h_{a\uparrow}h_{b\downarrow}n_{b\downarrow}
\rangle,\\
\left(\tilde{\rho}^{(ab)}_{\psi,N_{Q}=2,N_{S}=0}\right)_{44}
&\equiv&
\langle
n_{a\downarrow}h_{a\downarrow}h_{b\uparrow}n_{b\uparrow}
\rangle,\\
\left(\tilde{\rho}^{(ab)}_{\psi,N_{Q}=2,N_{S}=0}\right)_{21}
&\equiv&
\langle
a_{\uparrow}^{\dagger}a_{\downarrow}^{\dagger}b_{\downarrow}b_{\uparrow}
\rangle,\\
\left(\tilde{\rho}^{(ab)}_{\psi,N_{Q}=2,N_{S}=0}\right)_{31}
&\equiv&
\langle
a_{\downarrow}^{\dagger}n_{a\uparrow}h_{b\downarrow}b_{\downarrow}
\rangle,\\
\left(\tilde{\rho}^{(ab)}_{\psi,N_{Q}=2,N_{S}=0}\right)_{41}
&\equiv&
-
\langle
a_{\uparrow}^{\dagger}n_{a\downarrow}h_{b\uparrow}b_{\uparrow}
\rangle,\\
\left(\tilde{\rho}^{(ab)}_{\psi,N_{Q}=2,N_{S}=0}\right)_{32}
&\equiv&
\langle
b_{\uparrow}^{\dagger}h_{a\uparrow}n_{b\downarrow}a_{\uparrow}
\rangle,\\
\left(\tilde{\rho}^{(ab)}_{\psi,N_{Q}=2,N_{S}=0}\right)_{42}
&\equiv&
-
\langle
b_{\downarrow}^{\dagger}h_{a\downarrow}n_{b\uparrow}a_{\downarrow}
\rangle,\\
\left(\tilde{\rho}^{(ab)}_{\psi,N_{Q}=2,N_{S}=0}\right)_{43}
&\equiv&
\langle
s_{+}^{(a)}s_{-}^{(b)}
\rangle,
\end{eqnarray}

the $1\times 1$ matrix for the sector $(N_{Q},N_{S})=(2,1)$
\begin{equation}
\tilde{\rho}^{(ab)}_{\psi,N_{Q}=2,N_{S}=1}
\equiv
\langle
n_{a\uparrow}n_{b\uparrow}h_{a\uparrow}h_{b\uparrow}
\rangle,
\end{equation}
the $1\times 1$ matrix for the sector $(N_{Q},N_{S})=(2,-1)$
\begin{equation}
\tilde{\rho}^{(ab)}_{\psi,N_{Q}=2,N_{S}=-1}
\equiv
\langle
n_{a\downarrow}n_{b\downarrow}h_{a\downarrow}h_{b\downarrow}
\rangle,
\end{equation}
the $2\times 2$ matrix for the sector $(N_{Q},N_{S})=\left(3,\frac{1}{2}\right)$
\begin{eqnarray}
& &\mbox{}
\tilde{\rho}^{(ab)}_{\psi,N_{Q}=3,N_{S}=\frac{1}{2}}\nonumber\\
&\equiv&\left(
\begin{array}{cc}
\langle n_{a\uparrow}h_{a\uparrow}n_{b\uparrow}n_{b\downarrow}\rangle&
-\langle a^{\dagger}_{\downarrow}n_{a\uparrow}n_{b\uparrow}b_{\downarrow}\rangle\\
-\langle b^{\dagger}_{\downarrow}n_{a\uparrow}n_{b\uparrow}a_{\downarrow}\rangle&
\langle n_{a\uparrow}n_{a\downarrow}n_{b\uparrow}h_{b\uparrow}\rangle
\end{array}
\right),\label{rho-a2}
\end{eqnarray}
the $2\times 2$ matrix for the sector $(N_{Q},N_{S})=\left(3,-\frac{1}{2}\right)$
\begin{eqnarray}
& &\mbox{}
\tilde{\rho}^{(ab)}_{\psi,N_{Q}=3,N_{S}=-\frac{1}{2}}\nonumber\\
&\equiv&\left(
\begin{array}{cc}
\langle n_{a\downarrow}h_{a\downarrow}n_{b\uparrow}n_{b\downarrow}\rangle&
-\langle a^{\dagger}_{\uparrow}n_{a\downarrow}n_{b\downarrow}b_{\uparrow}\rangle\\
-\langle b^{\dagger}_{\uparrow}n_{a\downarrow}n_{b\downarrow}a_{\uparrow}\rangle&
\langle n_{a\uparrow}n_{a\downarrow}n_{b\downarrow}h_{b\downarrow}\rangle
\end{array}
\right),\label{rho-a2}
\end{eqnarray}
and 
the $1\times 1$ matrix for the sector $(N_{Q},N_{S})=(4,0)$
\begin{equation}
\tilde{\rho}^{(ab)}_{\psi,N_{Q}=4,N_{S}=0}
\equiv
\langle
n_{a\uparrow}n_{a\downarrow}n_{b\uparrow}n_{b\downarrow}
\rangle,
\end{equation}
where, $n_{i\sigma}\equiv
d_{i\sigma}^{\dagger}d_{i\sigma}$, $h_{i\sigma}\equiv I-n_{i,-\sigma}$,
 $s_{\pm}^{(i)}\equiv d_{i\sigma}^{\dagger}d_{i,-\sigma}$ ($i=a,b,
\sigma=\uparrow (+),\downarrow (-)$).

The entanglement entropy between system $A$ and $B$ is represented as follows;
\begin{eqnarray}
S_{\rm E.E.}&=&
-\lambda^{(0)}\log\lambda^{(0)}
-\sum_{i=\pm,\sigma=\uparrow,\downarrow }\mu^{(1)}_{i,\sigma}\log\mu^{(1)}_{i,\sigma}\nonumber\\
& &\mbox{}
-\sum_{i=1}^{4}\mu^{(2)}_{i}\log\mu^{(2)}_{i}
-\sum_{\sigma=\uparrow,\downarrow}\lambda^{(2)}_{a\sigma,b\sigma}\log\lambda^{(2)}_{a\sigma,b\sigma}\nonumber\\
& &\mbox{}
-\sum_{i=\pm,\sigma=\uparrow,\downarrow}\mu^{(3)}_{i,\sigma}\log\mu^{(3)}_{i,\sigma}
-
\lambda^{(4)}\log\lambda^{(4)},
\end{eqnarray}
where $\mu^{(2)}_{i} \ (i=1,\cdots,4)$ are the eigenvalues of the
$4\times 4$ matrix
$\tilde{\rho}^{(ab)}_{\psi,N_{Q}=2,N_{S}=0}$, 
$\mu^{(n_{Q})}_{i,\sigma} \ (i=\pm, n_{Q}=1,3, \ \
\sigma=\uparrow(=\frac{1}{2}),\downarrow(=-\frac{1}{2}))$
are eigenvalues of the $2\times 2$ matrix  
$\tilde{\rho}^{(ab)}_{\psi,N_{Q}=n_{Q},N_{S}=\sigma}$
and
\begin{eqnarray}
\lambda^{(0)}&\equiv&\langle
 h_{a\uparrow}h_{a\downarrow}h_{b\uparrow}h_{b\downarrow}\rangle,\\
%
\lambda^{(2)}_{a\sigma,b\sigma}&\equiv&\langle n_{a\sigma}n_{b\sigma}h_{a\sigma}h_{b\sigma}\rangle,\\
\lambda^{(4)}
&\equiv&
\langle
n_{a\uparrow}n_{a\downarrow}n_{b\uparrow}n_{b\downarrow}
\rangle,
\end{eqnarray}
are  the eigenvalues of the $1\times 1$ matrices
$\tilde{\rho}^{(ab)}_{\psi,N_{Q}=0,N_{S}=0}$,
$\tilde{\rho}^{(ab)}_{\psi,N_{Q}=2,N_{S}=2\sigma}$($\sigma=\pm\frac{1}{2}$) and
$\tilde{\rho}^{(ab)}_{\psi,N_{Q}=4,N_{S}=0}$, respectively.

$S_{\rm E.E.}$ is a function of the following correlation functions;
$c^{(1)}_{i\sigma}\equiv\langle n_{i\sigma}\rangle$
,
$c^{(2)}_{i\uparrow,i\downarrow}\equiv\langle n_{i\uparrow}n_{i\downarrow}\rangle$
,
$c^{(2)}_{a\sigma,b\sigma}\equiv\langle n_{a\sigma}n_{b\sigma}\rangle$
,
$c^{(2)}_{a\sigma,b-\sigma}\equiv\langle n_{a\sigma}n_{b-\sigma}\rangle$
,
$c^{(3)}_{a\uparrow,a\downarrow,b\sigma}\equiv\langle n_{a\uparrow}n_{a\downarrow}n_{b\sigma}\rangle$
,
$c^{(3)}_{a\sigma,b\uparrow,b\downarrow}\equiv\langle n_{a\sigma}n_{b\uparrow}n_{b\downarrow}\rangle$
,
$c^{(4)}\equiv\langle n_{a\uparrow}n_{a\downarrow}n_{b\uparrow}n_{b\downarrow}\rangle$
,
$C\equiv \langle s^{(a)}_{+}s^{(b)}_{-}\rangle$,
$\langle a^{\dagger}_{\sigma}b_{\sigma}\rangle$,
$\langle a^{\dagger}_{\sigma}n_{i,-\sigma}b_{\sigma}\rangle$,
$\langle
a^{\dagger}_{\uparrow}a^{\dagger}_{\downarrow}b_{\downarrow}b_{\uparrow}\rangle$,
and 
$\langle a^{\dagger}_{\sigma}n_{a,-\sigma}n_{b,-\sigma}b_{\sigma}\rangle$
($i=a,b, \ \sigma=\uparrow, \downarrow$) 
.

To evaluate $S_{\rm E.E.}^{(ab)}$, we need to calculate these 25
correlation functions.
This is not an easy calculation, but fortunately the 25 quantities are
independent and can be calculated separately, if necessary, in numerical
calculations.

The degree of entanglement between Anderson impurity $a$ and $b$ 
can be determined by the 
mutual information $I$ defined by the
following equation;
\begin{eqnarray}
I&\equiv&D\left(\rho^{(ab)}_{\psi}||\rho^{(a)}_{\psi}\otimes\rho^{(b)}_{\psi}\right)\\
&=&S^{(a)}_{\rm E.E.}+S_{\rm E.E.}^{(b)}-S_{\rm E.E.}^{(ab)},
\end{eqnarray}

where 
\begin{eqnarray}
\rho^{(i)}_{\psi}
&\equiv&
\langle h_{i\uparrow}h_{i\downarrow}\rangle
|0\rangle_{i}{}_{i}\langle 0|
+
\langle n_{i\uparrow}h_{i\uparrow}\rangle
|\uparrow\rangle_{i}{}_{i}\langle \uparrow|\nonumber\\
& &\mbox{}
+
\langle n_{i\downarrow}h_{i\downarrow}\rangle
|\downarrow\rangle_{i}{}_{i}\langle \downarrow|
+
\langle n_{i\uparrow}n_{i\downarrow}\rangle
|{\rm F}\rangle_{i}{}_{i}\langle {\rm F}|,
\end{eqnarray}
$|\sigma\rangle_{i}\equiv d_{i\sigma}^{\dagger}|0\rangle_{i}$,
$|{\rm F}\rangle_{i}\equiv d_{i\uparrow}{\dagger}d_{i\downarrow}^{\dagger}|0\rangle_{i}$
and
\begin{eqnarray}
S_{\rm E.E.}^{(i)}
&=&
-\langle n_{i\uparrow}n_{i\downarrow}\rangle\log\langle
n_{i\uparrow}n_{i\downarrow}\rangle
-
\langle h_{i\uparrow}h_{i\downarrow}\rangle\log\langle
h_{i\uparrow}h_{i\downarrow}\rangle\nonumber\\
& &\mbox{}
-\langle n_{i\uparrow}h_{i\uparrow}\rangle\log\langle
n_{i\uparrow}h_{i\uparrow}\rangle
-
\langle n_{i\downarrow}h_{i\downarrow}\rangle\log\langle
n_{i\downarrow}h_{i\downarrow}\rangle\nonumber\\
\end{eqnarray}
($i=a,b$).

For our purpose, as stated in the Introduction,
we will further consider the case where $\psi$ is an eigenstate of
$\Delta Q$. 
Here, $\Delta Q\equiv Q_{1}-Q_{2}$ and $Q_{i} ( i=1, 2)$ is the electron
number operator of the subsection $i$ counted from the half-filling
number.
In the above, all orbitals in the system are divided into two parts and
called subsection 1 and 2.
Subsection 1 contains Anderson impurity $a$ and subsection 2 contains
Anderson impurity $b$ (See also Fig.\ref{blockzu}(b-1)).
Note that this bisection is not to be confused with
the previously mentioned division of the system into $A$ and $B$.
Such bisection of orbitals and eigenstates of $\Delta Q$ are natural bisection
and states for a model and its energy eigenstates in which, for example, two single
impurity Anderson models are coupled by magnetic interaction $J_{ab}$ and
repulsion $U_{ab}$ between the two impurities without electron transfer, as shown
in Fig.\ref{blockzu}(b-2).

In this case, the above mentioned two $2\times 2$ matrices are
diagonalized and the above mentioned $4\times 4$ matrix become the block
diagonal matrix consisting of $1\times 1$, $1\times 1$ and $2\times 2$ matrices,
as show in Fig.\ref{blockzu} (a-3) and (a-4).
This is because the following equation holds;   
$\left[\Delta Q,|\{N_{i\sigma}\}\rangle\langle\{N^{\prime}_{i\sigma}\}|\right]=\left(N_{\Delta}-N_{\Delta}^{\prime}\right)|\{N_{i\sigma}\}\rangle\langle\{N^{\prime}_{i\sigma}\}|$
where
$N_{\Delta}=\sum_{\sigma}(N_{a\sigma}-N_{b\sigma})$ and
$N_{\Delta}^{\prime}=\sum_{\sigma}(N^{\prime}_{a\sigma}-N^{\prime}_{b\sigma})$.

The expression of the entanglement entropy between system $A$ and $B$ becomes simpler as follows;
\begin{eqnarray}
S_{\rm E.E.}^{(ab)}
&=&
-\lambda^{(0)}\log\lambda^{(0)}
-\sum_{i=a,b,\sigma=\uparrow,\downarrow}\lambda^{(1)}_{i\sigma}\log\lambda^{(1)}_{i\sigma}\nonumber\\
& &\mbox{}
-\sum_{i=a,b}\lambda^{(2)}_{i\uparrow,i\downarrow}\log\lambda^{(2)}_{i\uparrow,i\downarrow}
-\sum_{\sigma=\uparrow,\downarrow}\lambda^{(2)}_{a\sigma,b\sigma}\log\lambda^{(2)}_{a\sigma,b\sigma}\nonumber\\
& &\mbox{}
-\sum_{j=\pm}\lambda^{(2)}_{j}\log\lambda^{(2)}_{j}
-\sum_{i=a,b,\sigma=\uparrow,\downarrow}\lambda^{(3)}_{i\sigma}\log\lambda^{(3)}_{i\sigma}\nonumber\\
& &\mbox{}
-
\lambda^{(4)}\log\lambda^{(4)},
\end{eqnarray}
where,
\begin{eqnarray}
\lambda^{(1)}_{i\sigma}&\equiv&
\langle
n_{i\sigma}h_{i\sigma}h_{i^{c}\uparrow}h_{i^{c}\downarrow}
\rangle,\\
\lambda^{(2)}_{i\uparrow,i\downarrow}
&\equiv&
\langle
n_{i\uparrow}n_{i\downarrow}h_{i^{c}\uparrow}h_{i^{c}\downarrow}
\rangle,\\
\lambda^{(2)}_{a\sigma,b-\sigma}
&\equiv&
\langle
n_{a\sigma}n_{b-\sigma}h_{a\sigma}h_{b-\sigma}
\rangle,\\
\lambda^{(3)}_{i\sigma}
&\equiv&
\langle
n_{i^{c}\uparrow}n_{i^{c}\downarrow}n_{i\sigma}h_{i\sigma}
\rangle,\\
\end{eqnarray}

and
\begin{equation}
\lambda^{(2)}_{\pm}
\equiv
\frac{
\lambda^{(2)}_{a\uparrow,b\downarrow}
+
\lambda^{(2)}_{a\downarrow,b\uparrow}
\pm
\sqrt{
(
\lambda^{(2)}_{a\uparrow,b\downarrow}
-
\lambda^{(2)}_{a\downarrow,b\uparrow}
)^{2}
+
4|\langle s^{(a)}_{+}s^{(b)}_{-}\rangle|^{2}
}
}
{2}.
\end{equation}

\section{Results of application to a dimer with several types of internal interaction, short Hubbard chains}\label{app0}
\subsection{Dimer with several types of internal interaction}

At first, we consider isolated two quantum impurity system describing
the following Hamiltonian;
\begin{eqnarray}
H_{\rm two}&=&H_{\rm two}^{(0)}+H_{\rm two}^{(1)},\\
H_{\rm
 two}^{(0)}&=&\sum_{i=a}^{b}\varepsilon_{i}n_{i}+t\sum_{\sigma}\left(a_{\sigma}^{\dagger}b_{\sigma}+{\rm h.c.}\right),\\
H_{\rm
 two}^{(1)}&=&\sum_{i=a}^{b}U_{i}n_{i\uparrow}n_{i\downarrow}+Vn_{a}n_{b}+J{\bf s}^{(a)}\cdot{\bf s}^{(b)},
\end{eqnarray}
where $i_{\sigma}(i=a,b) $ annihilates an electron with
spin $\sigma$ on the quantum impurity $i$, characterized by the onsite
energy $\varepsilon_{i}$ and the intra-impurity repulsion $U_{i}$
($n_{i\sigma}=i^{\dagger}_{\sigma}i_{\sigma}$, $n_{i}=\sum_{\sigma}n_{i\sigma}$).
The quantum impurities $a$ and $b$ are coupled by the hybridization
$t(\neq 0)$,
inter-impurity repulsion $V$ and the inter-impurity spin interaction
$J$(${\bf s}^{(i)}$ is the spin operator of the quantum impurity $i$).
We calculate the entanglement entropy $S_{\rm
E.E.}\left(\rho^{(A=\{a\})}_{\psi}\right)=S_{\rm
E.E.}\left(\rho^{(A=\{b\})}_{\psi}\right)$ for the energy eigen state
$\psi$ of $H_{\rm two}$.
The energy eigen states can be classified by using the eigen values
$({\mathcal S},{\mathcal S}_{z},{\mathcal Q})$ of
the operators
${\bf S}^{2}, S_{z}, Q$ because $[H_{\rm two},S_{\alpha}]=0 \
(\alpha=1,2,3)$ and $[H_{\rm two},Q]=0$.
The classified eigenstates and the calculated entanglement entropy  are presented in Table \ref{tab_tqi}.
\begin{table}
\caption{}\label{tab_tqi}
\begin{tabular}{|l|}\hline
\multicolumn{1}{|l|}{$({\mathcal S},{\mathcal S}_{z},{\mathcal Q})=(0,0,-2)$}\\ \hline
$\psi_{0,0,-2}=|0\rangle$\\ 
$S_{\rm E.E.}\left(\rho^{(A=\{a\})}_{\psi_{0,0,-2}}\right)=0$ \\
 \hline
\multicolumn{1}{|l|}{$({\mathcal S},{\mathcal S}_{z},{\mathcal Q})=(0,0,2)$}\\ \hline
$\psi_{0,0,2}=a_{\uparrow}^{\dagger}b_{\uparrow}^{\dagger}a_{\downarrow}^{\dagger}b_{\downarrow}^{\dagger}|0\rangle$\\ 
$S_{\rm E.E.}\left(\rho^{(A=\{a\})}_{\psi_{0,0,2}}\right)=0$ \\
 \hline
\multicolumn{1}{|l|}{$({\mathcal S},{\mathcal S}_{z},{\mathcal Q})=(\frac{1}{2},\sigma,-1)$, $\sigma=\pm\frac{1}{2}(\uparrow,\downarrow)$}\\ \hline
$\psi_{\frac{1}{2},\sigma,-1}=c_{a}a^{\dagger}_{\sigma}|0\rangle+c_{b}
 b^{\dagger}_{\sigma}|0\rangle$\\
$\phantom{\psi_{\frac{1}{2},\sigma,-1}}( p_{a}=|c_{a}|^{2},
 p_{b}=|c_{b}|^{2}, p_{a}+p_{b}=1 )$\\ 
$S_{\rm
 E.E.}\left(\rho^{(A=\{a\})}_{\psi_{\frac{1}{2},\sigma,-1}}\right)=-p_{a}\log(p_{a})-(1-p_{a})\log(1-p_{a})$\\
 \hline
\multicolumn{1}{|l|}{$({\mathcal S},{\mathcal S}_{z},{\mathcal Q})=(\frac{1}{2},\sigma,1)$, $\sigma=\pm\frac{1}{2}(\uparrow,\downarrow)$}\\ \hline
$\psi_{\frac{1}{2},\sigma,1}=c_{a}b^{\dagger}_{\sigma}a^{\dagger}_{\uparrow}a^{\dagger}_{\downarrow}|0\rangle+c_{b}a^{\dagger}_{\sigma}b^{\dagger}_{\uparrow}b^{\dagger}_{\downarrow}|0\rangle$\\
$\phantom{\psi_{\frac{1}{2},\sigma,1}} ( p_{a}=|c_{a}|^{2},
 p_{b}=|c_{b}|^{2}, p_{a}+p_{b}=1 )$\\ 
$S_{\rm
 E.E.}\left(\rho^{(A=\{a\})}_{\psi_{\frac{1}{2},\sigma,1}}\right)=-p_{a}\log(p_{a})-(1-p_{a})\log(1-p_{a})$\\
 \hline
\multicolumn{1}{|l|}{$({\mathcal S},{\mathcal S}_{z},{\mathcal Q})=(1,m,0) , m=\pm 1$}\\ \hline
$\psi_{1,m,0}=a^{\dagger}_{\sigma}b^{\dagger}_{\sigma}|0\rangle,
 \sigma=\uparrow (m=1), \downarrow (m=-1)$\\ 
$S_{\rm E.E.}\left(\rho^{(A=\{a\})}_{\psi_{1,m,0}}\right)=0$ \\
 \hline
\multicolumn{1}{|l|}{$({\mathcal S},{\mathcal S}_{z},{\mathcal Q})=(1,0,0)$}\\ \hline
$\psi_{1,0,0}=\frac{1}{\sqrt{2}}\left(a^{\dagger}_{\uparrow}b^{\dagger}_{\downarrow}+a^{\dagger}_{\downarrow}b^{\dagger}_{\uparrow}\right)|0\rangle$\\ 
$S_{\rm E.E.}\left(\rho^{(A=\{a\})}_{\psi_{1,0,0}}\right)=\log(2)$ \\
 \hline
\multicolumn{1}{|l|}{$({\mathcal S},{\mathcal S}_{z},{\mathcal Q})=(0,0,0)$}\\ \hline
$\psi_{0,0,0}=c_{Da}a^{\dagger}_{\uparrow}a^{\dagger}_{\downarrow}|0\rangle+c_{Db}b^{\dagger}_{\uparrow}b^{\dagger}_{\downarrow}|0\rangle$\\
$\phantom{psi_{0,0,0}}+c_{S}\frac{1}{\sqrt{2}}\left(a^{\dagger}_{\uparrow}b^{\dagger}_{\downarrow}-a^{\dagger}_{\downarrow}b^{\dagger}_{\uparrow}\right)|0\rangle$\\ 
$ \phantom{psi_{0,0,0}}( p_{Da}=|c_{Da}|^{2},
 p_{Db}=|c_{Db}|^{2},p_{S}=|c_{S}|^{2},$\\
$\phantom{psi_{0,0,0}} p_{Da}+p_{Db}+p_{S}=1 )$\\ 
$S_{\rm
 E.E.}\left(\rho^{(A=\{a\})}_{\psi_{0,0,0}}\right)=p_{S}\log(2)-p_{Da}\log(p_{Da})$\\
$\phantom{S_{\rm E.E.}\left(\rho^{(A=\{a\})}_{\psi_{}}\right)}-p_{Db}\log(p_{Db})-p_{S}\log(p_{S})$ \\
 \hline
\end{tabular}
\end{table}

In Table \ref{tab_tqi},
the vector $(c_{a},c_{b})^{t}$ is an eigenvector of the $2\times 2$ representation
matrix of $H$ for the $({\mathcal S},{\mathcal S}_{z},{\mathcal Q})=(\frac{1}{2},\sigma,\pm 1)$.
For the eigenvalue $E_{\pm} \ ( E_{+}> E_{-} )$, 
$(c_{a},c_{b})^{t}=\frac{1}{\sqrt{1+(g_{\pm})^{2}}}(1,g_{\pm})^{t}$, 
$g_{\pm}\equiv -x\pm {\rm sgn}(c)\sqrt{x^{2}+1}$
and 
$x=\frac{\varepsilon_{a}-\varepsilon_{b}}{2t}, \ c=t$ 
for the sector $({\mathcal S},{\mathcal S}_{z},{\mathcal Q})=(\frac{1}{2},\sigma,-1)$, 
$x=\frac{\varepsilon_{a}-\varepsilon_{b}+U_{a}-U_{b}}{2t}, \ c=-t$
for the sector $({\mathcal S},{\mathcal S}_{z},{\mathcal Q})=(\frac{1}{2},\sigma,1)$.
The corresponding entanglement entropy is independent from the double
sign of $E_{\pm}$ because of $\frac{1}{1+(g_{+})^{2}}+\frac{1}{1+(g_{-})^{2}}=1$.

The vector $(c_{Da},c_{Db},c_{S})^{t}$ is an eigen vector of the
representation matrix of $H$ for the sector $({\mathcal S},{\mathcal S}_{z},{\mathcal Q})=(0,0,0)$.
%
The condition for maximum vales of $S_{\rm
E.E.}\left(\rho^{(A=\{a\})}_{\psi_{0,0,0}}\right)$ is
$(p_{Da},p_{Db},p_{S})=\left(\frac{1}{4},\frac{1}{4},\frac{1}{2}\right)$.

We further consider the case of geometrically symmetric system, i.e.,
$\varepsilon_{a}=\varepsilon_{b}, \ U_{a}=U_{b}$.
In this case, the Hamiltonian $H_{\rm two }$  and the permutation
operator $P$ are commutative, where $Pa_{\sigma}P^{\dagger}=b_{\sigma},P^{\dagger}=P
, P^{2}=I$ and $P|0\rangle =|0\rangle$.
The eigenvalue $p$ of $P$ is $p=\pm 1 (p=1={\rm even}, \ p=-1={\rm odd})$. 
The permutation operator $P$ and ${\bf S}^{2}$, $S_{z}$, $Q$ are also commutative.
Therefore, we can classify the energy eigenstates by using the
eigenvalues $({\mathcal S},{\mathcal S}_{z},{\mathcal Q},p)$ of the
operators ${\bf S}^{2}, S_{z}, Q$ and $P$.

The eigenstates classified by $({\mathcal S},{\mathcal S}_{z},{\mathcal Q})$ in Table \ref{tab_tqi} are 
already eigenstates of $P$ except for 
$({\mathcal S},{\mathcal S}_{z},{\mathcal Q})= (\frac{1}{2},\sigma,\pm
1),\  (0,0,0)$.
%
The two-dimensional subspace labeled with 
$({\mathcal S},{\mathcal S}_{z},{\mathcal Q})= (\frac{1}{2},\sigma,\pm 1)$
is split into two one-dimensional subspaces labeled with
$({\mathcal S},{\mathcal S}_{z},{\mathcal Q},p)= (\frac{1}{2},\sigma,\pm
1,p) \ (p={\rm even}, {\rm odd})$.

The eigenstate labeled with $({\mathcal S},{\mathcal S}_{z},{\mathcal Q},p)= (\frac{1}{2},\sigma,\pm
1,p)$ is
\begin{equation}
\psi_{\frac{1}{2},\pm\frac{1}{2},\pm1,p}
=
\frac{1}{\sqrt{2}}
\left(
A+pB
\right)|0\rangle,
\end{equation}
where $A=a_{\sigma}^{\dagger}, \ B=b^{\dagger}_{\sigma}$ for 
$({\mathcal S},{\mathcal S}_{z},{\mathcal Q},p)=(\frac{1}{2},\pm\frac{1}{2},-1,p)$
,
$A=a_{\sigma}^{\dagger}b^{\dagger}_{\uparrow}b^{\dagger}_{\downarrow}, \
B=b^{\dagger}_{\sigma}a^{\dagger}_{\uparrow}a^{\dagger}_{\downarrow}$
for $({\mathcal S},{\mathcal S}_{z},{\mathcal Q},p)=(\frac{1}{2},\pm\frac{1}{2},1,p)$.

The entanglement entropy is given by
$S_{\rm E.E.}\left(\rho^{(A=\{a\})}_{\psi_{\frac{1}{2},\pm\frac{1}{2},\pm
1,p}}\right)=\log(2)$.

The three-dimensional subspace labeled with 
$({\mathcal S},{\mathcal S}_{z},{\mathcal Q})= (0,0,0)$
is split into the one-dimensional subspace labeled with
$({\mathcal S},{\mathcal S}_{z},{\mathcal Q},p)= (0,0,
0,{\rm odd})$ and 
the two-dimensional subspace labeled with
$({\mathcal S},{\mathcal S}_{z},{\mathcal Q},p)= (0,0,
0,{\rm even})$ 
.
The eigenstate labeled with $({\mathcal S},{\mathcal S}_{z},{\mathcal Q},p)= (0,0,
0,{\rm odd})$  is 
$\psi_{0,0,0,{\rm odd}}\equiv \frac{1}{\sqrt{2}}\left(a^{\dagger}_{\uparrow}a^{\dagger}_{\downarrow}-b^{\dagger}_{\downarrow}b^{\dagger}_{\uparrow}\right)|0\rangle$
and the entanglement entropy is given by 
$S_{\rm E.E.}\left(\rho^{(A=\{a\})}_{\psi_{0,0,0,{\rm odd}}}\right)=\log(2)$.

The eigenstate labeled with 
$({\mathcal S},{\mathcal S}_{z},{\mathcal Q},p)=(0,0,0,{\rm even})$
is
\begin{eqnarray}
\psi_{0,0,0,{\rm
 even}}&=&c_{D}\frac{1}{\sqrt{2}}\left(a^{\dagger}_{\uparrow}a^{\dagger}_{\downarrow}+b^{\dagger}_{\downarrow}b^{\dagger}_{\uparrow}\right)|0\rangle\nonumber\\
& &\mbox{}+c_{S}\frac{1}{\sqrt{2}}\left(a^{\dagger}_{\uparrow}b^{\dagger}_{\downarrow}-a^{\dagger}_{\downarrow}b^{\dagger}_{\uparrow}\right)|0\rangle
\end{eqnarray}
, where the vector $(c_{D},c_{S})^{t}$ is the eigenvector of the $2\times 2$ representation
matrix of $H$ for the $({\mathcal S},{\mathcal S}_{z},{\mathcal Q},p)=(0,0,0,{\rm even})$.
For the eigenvalue $E_{\pm} \ ( E_{+}> E_{-} )$, 
$(c_{D},c_{S})^{t}=\frac{1}{\sqrt{1+(g_{\pm})^{2}}}(1,g_{\pm})^{t}$, where
\begin{equation}
g_{\pm}=-\frac{U-V+\frac{3}{4}J}{4t}\pm {\rm sgn}(t) \sqrt{1+\left(\frac{U-V+\frac{3}{4}J}{4t}\right)^{2}}.
\end{equation}

\begin{figure}[h]
\includegraphics[width=0.5\textwidth]{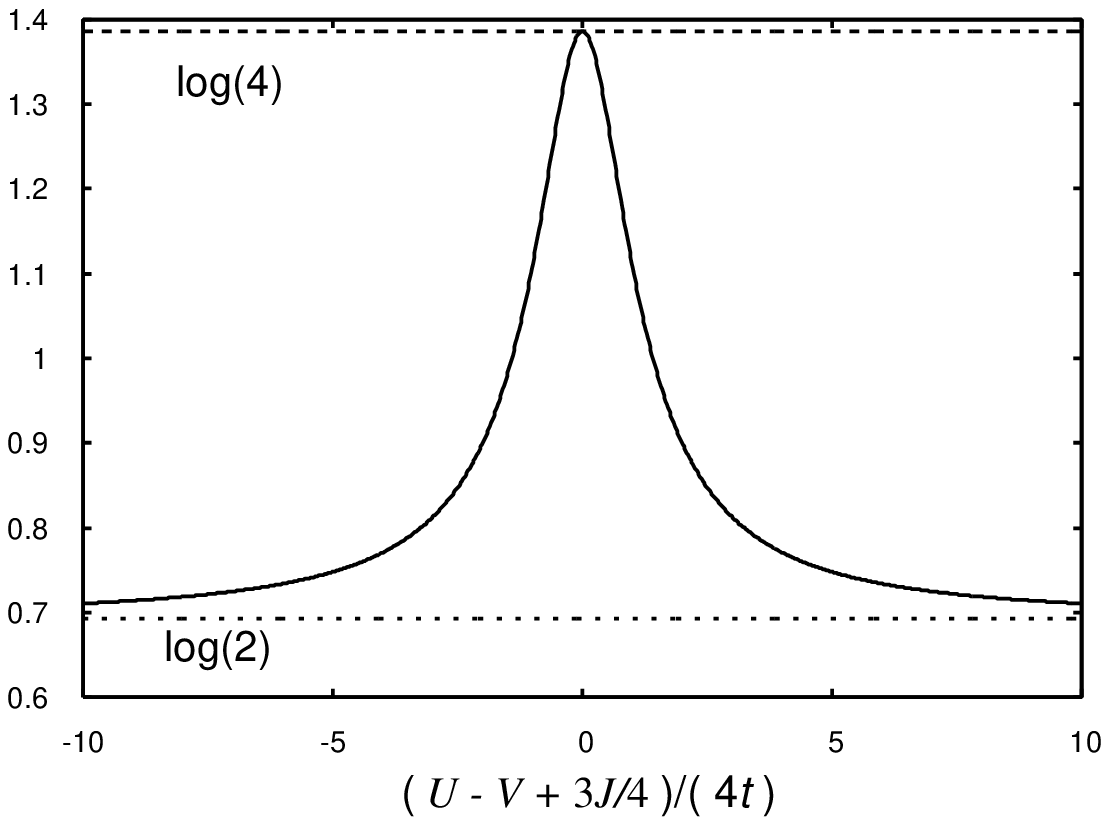}
\caption{The entanglement entropy 
$S_{\rm E.E.}\left(\rho^{(A=\{a\})}_{\psi_{0,0,0,{\rm even}}}\right)$ as a function of $\frac{U-V+\frac{3J}{4}}{4t}$}\label{see_tqi}
\end{figure}

The entanglement entropy $S_{\rm E.E.}\left(\rho^{(A=\{a\})}_{\psi_{0,0,0,{\rm
even}}}\right)$ is represented as follows;
\begin{eqnarray}
S_{\rm E.E.}\left(\rho^{(A=\{a\})}_{\psi_{0,0,0,{\rm
	     even}}}\right)&=&\log(2)-p_{D}\log(p_{D})-p_{S}\log(p_{S}),\nonumber\\
\end{eqnarray}
where 
$p_{D}=|c_{D}|^{2}\left(=\frac{1}{\sqrt{1+g_{\pm}^{2}}}\right)$
 and $p_{S}=|c_{S}|^{2}(=1-|c_{D}|^{2})$.
Therefore,  $S_{\rm E.E.}\left(\rho^{(A=\{a\})}_{\psi_{0,0,0,{\rm
even}}}\right)$ is a function of $\frac{U-V+\frac{3J}{4}}{4t}$.
Its graph is shown in Fig.\ref{see_tqi}.
The condition for maximum value of $S_{\rm E.E.}\left(\rho^{(A=\{a\})}_{\psi_{0,0,0,{\rm
even}}}\right)$ is
\begin{equation}
U-V+\frac{3}{4}J=0.\label{max_tqi}
\end{equation}
This condition also gives the Hund's rule, i.e.
$H_{\rm two}^{(1)}\psi_{1,m,0,{\rm
odd}}=\left(V+\frac{1}{4}J\right)\psi_{1,m,0,{\rm odd}} \
(m=0,\pm 1)$,
$H_{\rm
two}^{(1)}\psi_{0,0,0,p}=\left(V-\frac{3}{4}J\right)\psi_{0,0,0,p}$ when
$J<0$,
and the condition for
the {\rm SU(4)} symmetry of $H_{\rm two}^{(1)}$ when $J=0$.
It can be seen that the presence of only one interaction does not
achieve the entanglement entropy maximization, but the presence of two
or three interactions can achieve that maximization by satisfying Eq.(\ref{max_tqi}).
\subsection{Short Hubbard chains}
In this subsection, we apply the ideas of 
the distribution of entanglement on quantum
impurities formulated in \ref{oneI} and 
the entanglement in quantum impurity pairs
formulated in \ref{twoI} to several half-filled Hubbard chains.
The Hubbard chain consider here consists of $M$ equivalent sites with
on-site energy $\varepsilon$ and on-site repulsion $U$ and  
is described by the following Hamiltonian;
\begin{eqnarray}
H_{{\rm Hub},M}
&\equiv&
\varepsilon\sum_{i=1}^{M}n_{i}+t\sum_{i=1}^{M-1}\sum_{\sigma}\left(d^{\dagger}_{i,\sigma}d_{i+1,\sigma}+{\rm
							    h.c.}\right)\nonumber\\
& &\mbox{}
+U\sum_{i=1}^{M}n_{i,\uparrow}n_{i,\downarrow},
\end{eqnarray}
where $t(\neq 0)$ is the hopping integral between $i$ and $i+1$ sites, $d_{i,\sigma}$ is the annihilation operator for the electron of
spin $\sigma$ at the site $i$  and
$n_{i,\sigma}=d_{i,\sigma}^{\dagger}d_{i,\sigma}$, $n_{i}=\sum_{\sigma}n_{i,\sigma}$.
We set $\varepsilon=-\frac{U}{2}$ to achieve a half-filled state.

The set of the quantum numbers $({\mathcal S}_{z},{\mathcal S},{\mathcal
Q})$ of the ground state $\psi^{({\rm G})}_{{\mathcal S}_{z},{\mathcal S},{\mathcal Q}}$
 is 
$({\mathcal S}_{z},{\mathcal S},{\mathcal Q})=\left(0,0,0\right)$ if $M$
is even and
$({\mathcal S}_{z},{\mathcal S},{\mathcal Q})=\left(\pm\frac{1}{2},\frac{1}{2},0\right)$
if $M$ is odd.
This is because in all cases there are antiferromagnetic correlations in
the ground state, but the magnetic moments cancel each other when $M$ is
even, and there is an unpaired magnetic moment when $M$ is odd.

At first, we consider 
the distribution of entanglement entropy on quantum
impurities $\left\{S_{i}\right\}_{i=1,\cdots,M}$,
where $S^{(i)}= S_{\rm
E.E.}\left(\rho^{(A=\{i\})}_{\psi^{({\rm G})}_{{\mathcal S_{z}},{\mathcal
S},{\mathcal Q}}}\right)$.

\begin{figure}
\includegraphics[width=0.4\textwidth]{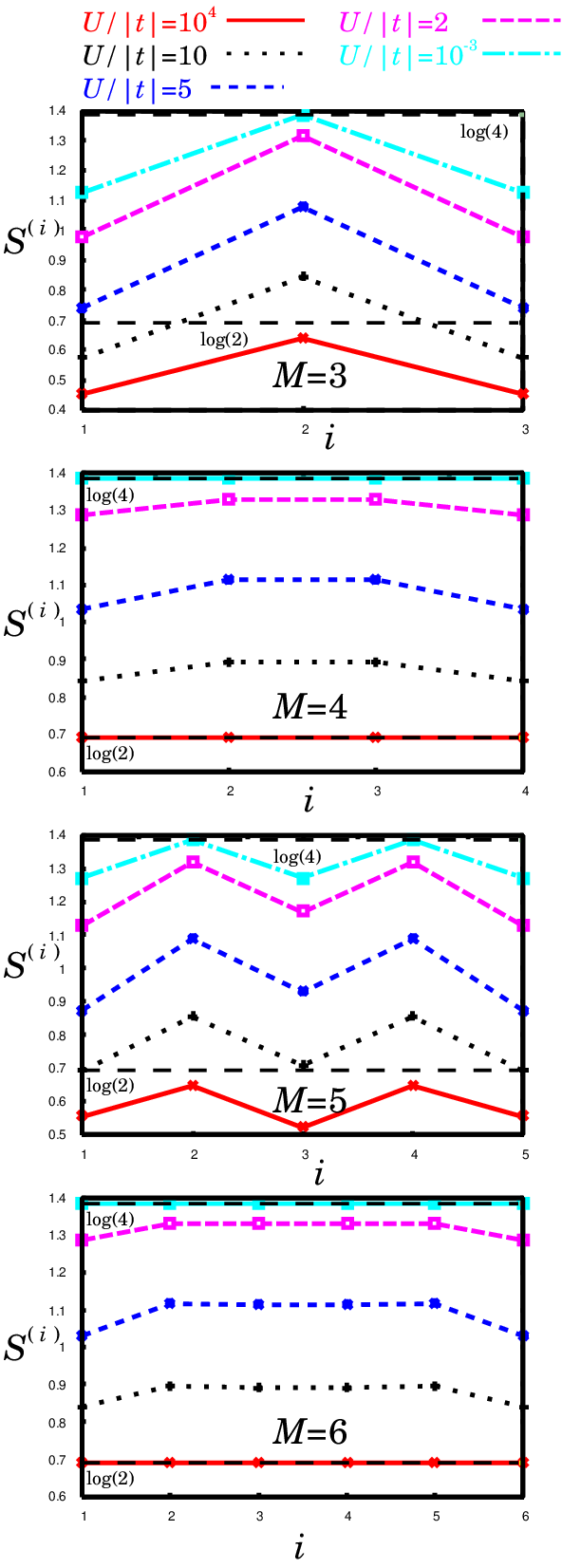}
\caption{Entanglement entropy distribution on Hubbard chains ( $M=3 \
 ,4, \ 5, \ 6$)}\label{hub_S}
\end{figure}

The results for $M=3,\ 4, \ 5$ and $6$ are shown in Fig.\ref{hub_S}.
The $U$-dependence of the distributions is also shown here. 

In all cases considered here, increasing the value of $U$ decreases the value of entanglement entropy.
As the value of $U$ increases, the Hubbard chain approaches the
Heisenberg chain, so the distribution of entanglement entropy also
approaches its distribution.
For example, in the case $M=3$, at $U\to+\infty$,
$\psi^{({\rm G})}_{\frac{1}{2},0,0}\to
\frac{1}{\sqrt{6}}(d^{\dagger}_{1,\uparrow}d^{\dagger}_{2,\uparrow}d^{\dagger}_{3,\downarrow}
-2d^{\dagger}_{1,\uparrow}d^{\dagger}_{2,\downarrow}d^{\dagger}_{3,\uparrow}
+
d^{\dagger}_{1,\downarrow}d^{\dagger}_{2,\uparrow}d^{\dagger}_{3,\uparrow}
)|0\rangle$, then
$S^{(1)}=S^{(3)}\to
-\frac{5}{6}\log(\frac{5}{6})-\frac{1}{6}\log(\frac{1}{6})\simeq 0.45$,
$S^{(2)}\to
-\frac{2}{3}\log(\frac{2}{3})-\frac{1}{3}\log(\frac{1}{3})\simeq 0.64$.

For $M=3,5$, the shape of the distribution of entanglement entropy does
not change much with respect to changes in the value of $U$, but changes
in the overall value.
On the other hand, for $M=4,6$, as the value of $U$ is increased from
zero, the distribution changes from a uniform distribution to one in
which the value of entanglement entropy of internal sites (site 2 and 3 for
$M=4$, site 2,3,4, and 5 for $M=6$) increases, and
then back to a uniform distribution again.
We then consider the reasons for the behaviors described above.
First, we consider the cases $M=4$ and $6$.
In this case the ground states are nonmagnetic.
Therefore, $\langle n_{i,\uparrow}\rangle =\langle
n_{i,\downarrow}\rangle$ holds for all site $i$.
We have also confirmed numerically that 
$\langle n_{i,\uparrow}\rangle =\langle
n_{i,\downarrow}\rangle=\frac{1}{2}$ for all site $i$ in the case of
$M=4$ and 6.
Thus, $S^{(i)}$ can be written as follows;
\begin{eqnarray}
S^{(i)}
&=&
\log(2)
-
\left(\frac{1}{2}-2\delta_{i}\right)\log\left(\frac{1}{2}-2\delta_{i}\right)\nonumber\\
& &\mbox{}
-
\left(\frac{1}{2}+2\delta_{i}\right)\log\left(\frac{1}{2}+2\delta_{i}\right)\label{Si-even},
\end{eqnarray}
where $\delta_{i} \equiv \langle n_{i,\uparrow}\rangle \langle
n_{i,\downarrow}\rangle -\langle n_{i,\uparrow}n_{i,\downarrow}\rangle$
is the electron correlation at site $i$.
This result shows that $S^{(i)}$ decreases as $\delta_{i}$ increases, and 
$S^{(i)}$ reflects the electron correlation at each site $i$.

\begin{figure}
\includegraphics[width=0.5\textwidth]{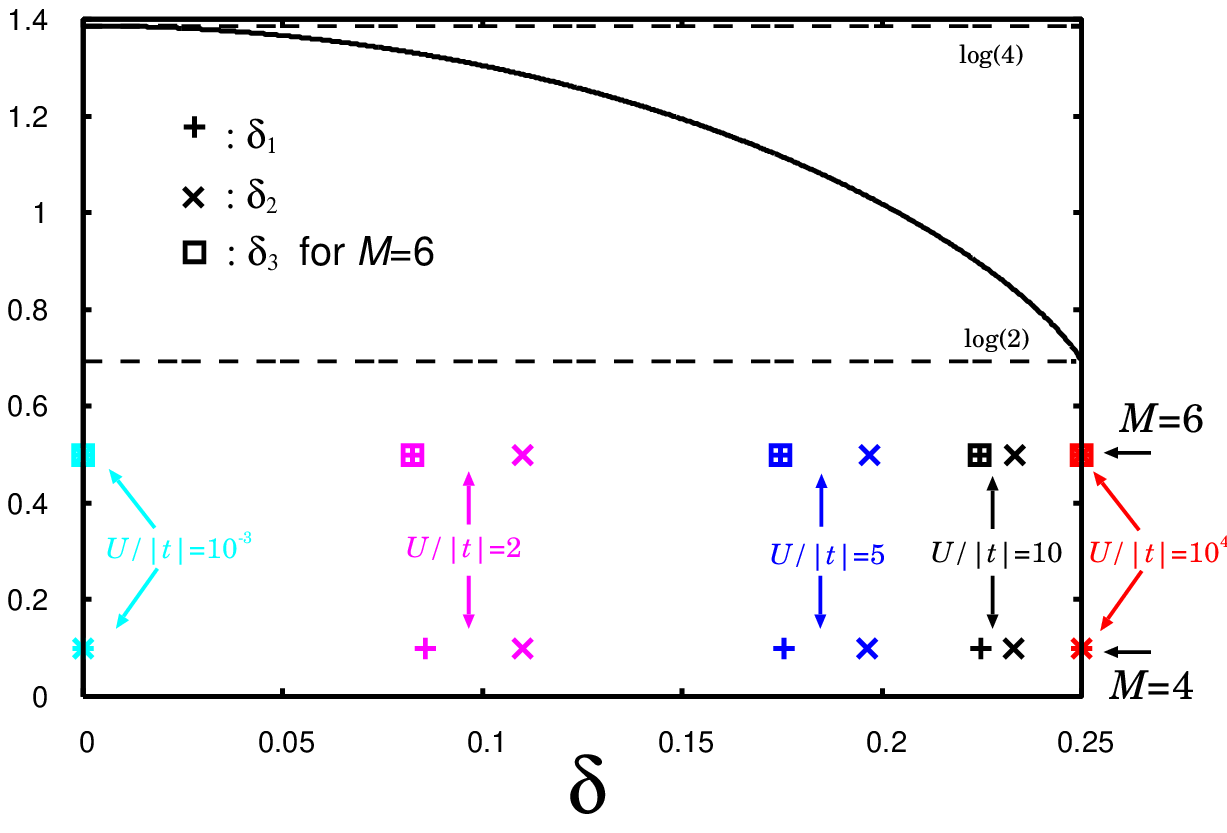}
\caption{(Color on line) Graph of $S^{(i)}$ defined by
 Eq.(\ref{Si-even})(Solid black line). Values of
 $\delta_{i}$ for each site $i$ for $M=4, 6$.}\label{evenhubreason}
\end{figure}

In Fig.\ref{evenhubreason},  we show $S^{(i)}$ as a function of $\delta_{i}$ and ,in the
case $M=4, \ 6$, the value of $\delta_{i}$
at each site for the same several values of $U$ as in 
Fig.\ref{hub_S}.
When $U=0$, $\delta_{i}=0$ at any site $i$, the distribution is uniform.
As the value of $U$ is increased from zero, the electron correlations at
the edge sites become larger than those at the interior sites, and the
distribution changes to a distribution with larger $S^{(i)}$ values at the
interior sites.
In the limit $U\to +\infty$, all sites have the largest electron
correlation $\delta_{i}=\frac{1}{4}$ and the distribution becomes
uniform again.

We next consider the case $M=3$ and $5$.
Because the ground states are magnetic in these cases,
 $\langle n_{i,\uparrow}\rangle\neq \langle
n_{i,\downarrow}\rangle$ in general.
We have numerically confirmed
that $\langle n_{i,\uparrow}\rangle +\langle n_{i,\downarrow}\rangle =1$
for all site $i$.
Thus, $S^{(i)}$ is expressed as a function of the polarization
$p_{i}\equiv \langle n_{i,\uparrow}\rangle -\langle
n_{i,\downarrow}\rangle$ and $\delta_{i}$
as follows;
\begin{eqnarray}
S^{(i)}&=&
-2\left(\frac{1-p_{i}^{2}}{4}-\delta_{i}\right)\log\left(\frac{1-p_{i}^{2}}{4}-\delta_{i}\right)\nonumber\\
& &\mbox{}
-\left(\frac{(1+p_{i})^{2}}{4}+\delta_{i}\right)\log\left(\frac{(1+p_{i})^{2}}{4}+\delta_{i}\right)\nonumber\\
& &\mbox{}
-\left(\frac{(1-p_{i})^{2}}{4}+\delta_{i}\right)\log\left(\frac{(1-p_{i})^{2}}{4}+\delta_{i}\right)\label{Si-odd}.
\end{eqnarray}

\begin{figure}[h]
\includegraphics[width=0.45\textwidth]{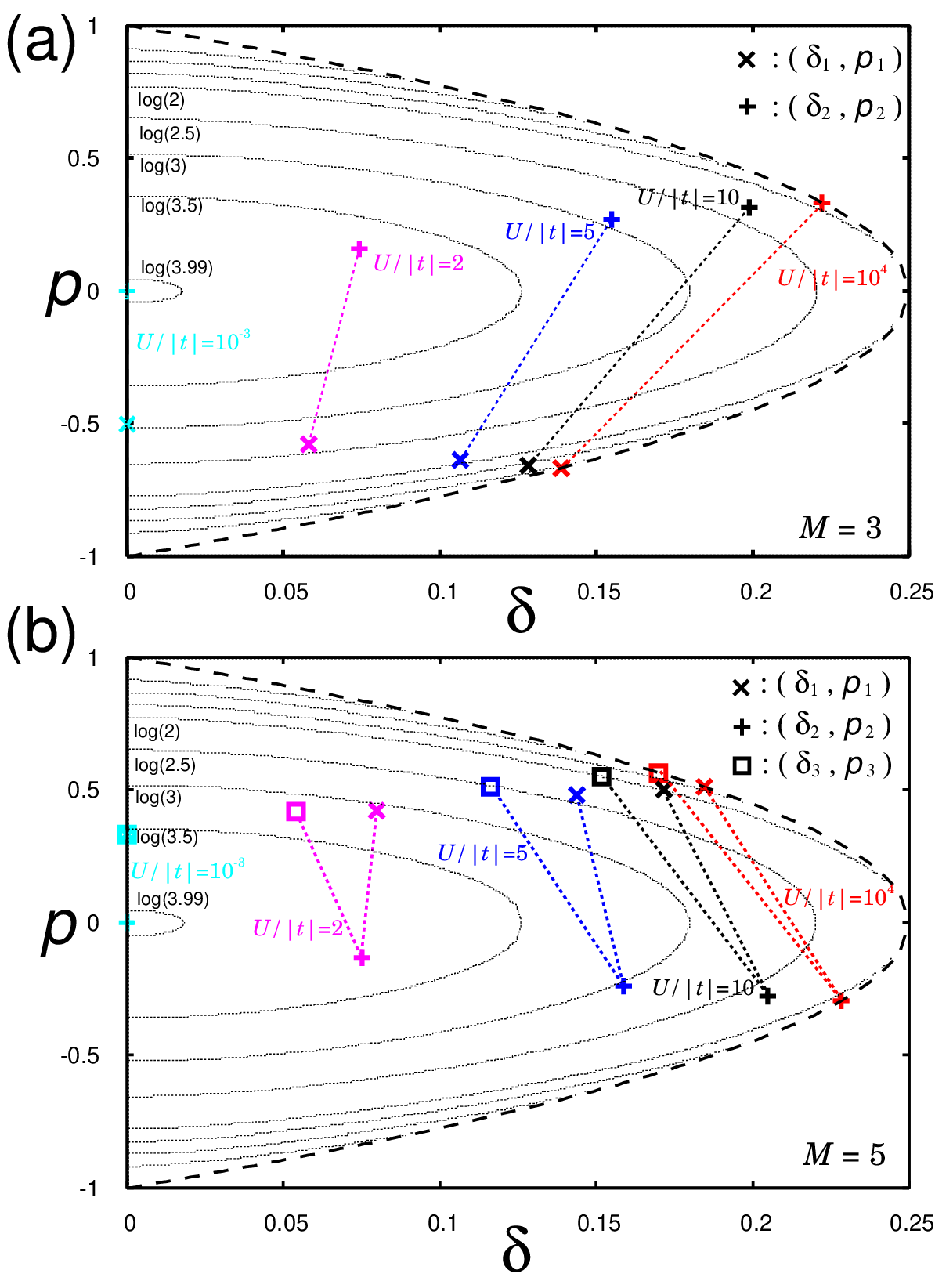}
\caption{(Color on line)  Values of
 $(\delta_{i}, p_{i})$ for each site $i$ for $M=3 (a), 5 (b)$. Some contours of $S^{(i)}$ defined by
 Eq.(\ref{Si-odd})(thin dotted lines).}\label{oddhubreason}
\end{figure}

In the case of $M=3, \ 5$, the values of $p_{i}$ and $\delta_{i}$ for each site are plotted
in Fig.\ref{oddhubreason} for the same several values of $U$ as in Fig.\ref{hub_S} (We used
the ground state $\psi^{({\rm G})}_{-\frac{1}{2},\frac{1}{2},0}$ for
$M=3$ and the ground state $\psi^{({\rm G})}_{\frac{1}{2},\frac{1}{2},0}$ for $M=5$).
Some contours of $S^{(i)}$ defined by Eq.(\ref{Si-odd}) are also shown in Fig.\ref{oddhubreason}.
Fig.\ref{oddhubreason} shows that in the region where the value of $S^{(i)}$ is larger than
$\log(2)$, the value of $S^{(i)}$ increases as the absolute values of $p_{i}$ and
$\delta_{i}$ decreases.
At $(\delta, p)=(0.25,0)$, the $\log(2)$ contour is tangent to the boundary of the definition
region defined by $\delta_{i}=\frac{1-p_{i}^{2}}{4}$(dashed line).
Contours below $\log(2)$ intersect the boundary.
On the boundary,  $S^{(i)}$ is expressed as 
$S^{(i)}=-\frac{1+p_{i}}{2}\log\left(\frac{1+p_{i}}{2}\right)-\frac{1-p_{i}}{2}\log\left(\frac{1-p_{i}}{2}\right)$
, and is a monotonically decreasing function of the absolute value of
$p_{i}$ ($\delta_{i}=\frac{1-p_{i}^{2}}{4}$ is also monotonically decreasing function of the
absolute value of $p_{i}$).
Therefore, increasing the value of $\delta_{i}$
results in an increases in the value of $S^{(i)}$ on the boundary.
In all cases shown in Fig.\ref{oddhubreason}, it can be seen that as the
value of $U$ increases, the values of $\delta_{i}$ and $|p_{i}|$
increases, and the value of $S^{(i)}$ decreases, as shown in Fig.\ref{hub_S}.
When $U=0$, $p_{i}=0$ for even sites $i$ and $p_{j}\neq 0$ for odd sites $j$.
As the value of $U$ increases to some extent, the value of $\delta_{i}$ increases but
the value of $|p_{i}|$ decreases for even sites $i$, and the value of
$\delta_{j}$ decreases but the value of $|p_{j}|$ increases for odd
sites $j$, resulting in a smaller value of $S^{(j)}$ for odd sites $j$
than the value of $S^{(i)}$ for even sites $i$.
The sign of $p_{i}$ alternates between even and odd sites, which is due
to antiferromagnetic correlations in the system.
Therefore, the zigzag structure of the value of $S^{(i)}$ shown in
Fig.\ref{hub_S} reflects the antiferromagnetic correlations in the
system, and the shape of the zigzag structure is not easily affected by
changes in the value of $U$.
For $U/|t|=10^{4}$, $(\delta_{i}, p_{i})$ is located extremely close to
the boundary. However, because of constraint $\sum_{i}p_{i}=1$ or $-1$, 
even in the limit $U\to\infty$, $\delta_{i}\to 0.25$ for all sites $i$
does not hold, as it does when $M$ is even.

Next, we consider the entanglement in quantum impurity pairs $I_{ij} \ (i<j, i,j=1,\cdots M)$
in the system,
where
$I_{ij}\equiv S^{(i)}_{\rm E.E.}+S^{(j)}_{\rm E.E}-S^{(ij)}_{\rm E.E.}$.

We first consider the case $M=3$.
Since we are considering quantum entanglement between subsystems when
the whole system is in a pure state,
$S^{(12)}=S^{(3)}$, $S^{(13)}=S^{(2)}$ and $S^{(23)}=S^{(1)}$ hold.
Hence, $I_{12}=I_{23}=S^{(2)}$ and $I_{13}=2S^{(1)}-S^{(2)}$ hold.

Since we have already obtained $S^{(2)}>S^{(1)}$, 
$I_{12}-I_{13}=2(S^{(2)}-S^{(1)})>0$ holds.

We have already seen that when the value of $U$ is increased, 
the values of $S^{(1)}$ and $S^{(2)}$ decrease while the value of the
difference $S^{(2)}-S^{(1)}$ remains somewhat constant.
Therefore, when the value of $U$ increases, 
the values of $I_{12}$ and $I_{13}$ decrease and the 
relationship $I_{12}> I_{13}$ is maintained.

\begin{figure}
\includegraphics[width=0.51\textwidth]{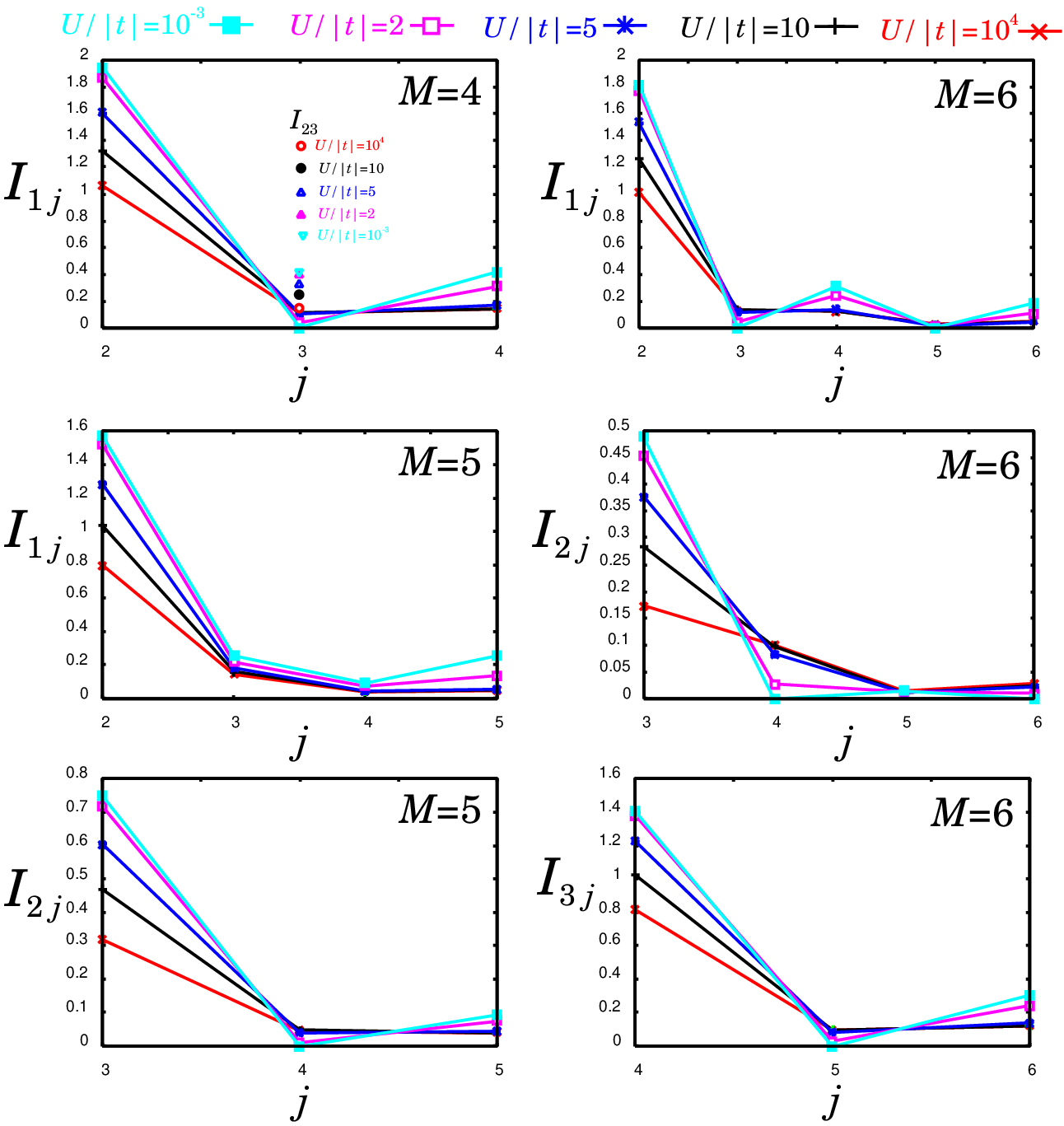}
\caption{$I_{1j} \ (j=2,\cdots, 4)$ and $I_{23}$ for $M=4$,
$I_{1j} \ (j=2,\cdots,5)$ and $I_{2j} \ (j=3,\ 5)$ for $M=5$, 
and
$I_{1j} \ (j=2,\cdots, 6)$, $I_{2j} \ (j=3,\cdots,6)$ and $I_{3j} \
 (j=4,\cdots, 6)$ for
$M=6$}\label{hub_I}
\end{figure}

Next, we will consider the cases $M=4, \ 5, \ 6$.
Because of the symmetry of the Hubbard chain under consideration here, 
$I_{1i} \ (i=2,\cdots, 4)$ and $I_{23}$ for $M=4$,
$I_{1i} \ (i=2,\cdots,5)$ and $I_{2i} \ (i=3,\ 4)$ for $M=5$, 
and
$I_{1i} \ (i=2,\cdots, 6)$, $I_{2i} \ (i=3,\cdots,5)$ and $I_{34}$ for
$M=6$ are independent of each other in each case.
These independent $I_{ij}$ are shown in Fig.\ref{hub_I}
for several different values of $U$ (for graphical clarity, $I_{24}$
for $M=4$
, $I_{25}$ for $M=5$, and
$I_{26}$ for $M=6$ are shown in Fig.\ref{hub_I}
, although $I_{24}=I_{13} (M=4)$, $I_{25}=I_{14} (M=5)$, and 
$I_{26}=I_{15} (M=6)$).
Fig.\ref{hub_I}
shows that $I_{ij}$ generally decreases with increasing values of
$|i-j|$.
However, it is not necessarily a monotonically decreasing function of
$|i-j|$.
Indeed, in the limit $U\to 0$, $I_{13}=0$ for $M=4$, $I_{24}=0$ for
$M=5$, and 
$I_{13}=I_{15}=I_{24}=I_{26}=0$,  for $M=6$, 
and since $I_{ij}$ is a non negative quantity,
$I_{13}<I_{14} \ (M=4)$, 
$I_{24}<I_{25} \ (M=5)$, and 
$I_{13} < I_{14}, \ I_{15} < I_{16}$, $I_{24} < I_{25} \ (M=6)$ necessarily
follow.
As the value of $U$ is increased from zero, some of these inequalities, which
imply non-monotonic decreasing nature of $I_{ij}$, are still valid though
$I_{ij}$ are no longer zero as in the case $U=0$.
For example, the values of $I_{1j}$ for $M=6$, $U/|t|=10^{-3}, \ 5, \ 10$ are shown in Table
\ref{tab_I},
and
it can be understood that 
$I_{ij}$ is not a monotonically decreasing function of $|j-1|$.
\begin{table}
\caption{$I_{1j}$ for $M=6$, $U/|t|=10^{-3}, \ 5$, and $10$}\label{tab_I}
\begin{tabular}{|c|c|c|c|c|c|}\hline
&$I_{12}$&$I_{13}$&$I_{14}$&$I_{15}$&$I_{16}$\\ \hline
$U/|t|=10^{-3}$&1.817 & 0.1120$\times 10^{-7}$ & 0.3086& 0.3098$\times 10^{-8}$ & 0.1846\\ \hline
$U/|t|=5$&1.538 &0.1211 & 0.1421& 0.02068& 0.04741\\ \hline
$U/|t|=10$&1.265 &0.1358 &0.1240 &0.02525  & 0.04642 \\ \hline
\end{tabular}
\end{table}

%
%
This non-monotonic decreasing nature is interesting because it indicates
that quantum entanglement between sites is not necessarily smaller the
further apart they are.

The above property stems from the property that $I_{ij}=0$ at
the inner sites when $U=0$.
$I_{ij}=0$ means there is no correlation between sites $i$ and $j$.
This means that from the results of subsection \ref{twoI}, the correlation functions decouple as
$\langle
n_{i\sigma}n_{i\sigma^{\prime}}n_{j\tau}n_{j\tau^{\prime}}\rangle=
\langle
n_{i\sigma}n_{i\sigma^{\prime}}\rangle\langle n_{j\tau}n_{j\tau^{\prime}}\rangle,
\cdots$.
Furthermore, when $U=0$, $S^{(ij)}_{\rm E.E.}=\log(16)$, and $S^{(i)}_{\rm
E.E.}=S^{(j)}_{\rm E.E.}=\log(4)$ hold when $I_{ij}=0$.

For $U=0$ in the Hubbard chain considered here, the wavefunction
representing one-electron state is described by a sinusoidal wave,
typical of wave.
%
%
%
So the property $I_{ij}=0$ would be brought about by the wave nature of the
system at $U=0$, and the particle nature brought about by the finite
value of $U$ makes $I_{ij}\neq 0$.
We will confirm that the wave nature of the system at $U=0$ yield
$I_{ij}=0$ in a non-interacting two-particle spinless fermion system
moving in  one-dimensional continuous space, avoiding complexity due to both discreteness and many-body
systems.
We then consider the case where the density correlations become
zero at two points in space.
The density operator at position $a$ is defined as
$n(a)\equiv \sum_{i=1}^{2}\delta(a-x_{i})$,
so the product operator of the densities at positions $a$ and $b$ is
$n(a)n(b)=\sum_{i\neq j}\delta(a-x_{i})\delta(b-x_{j})$.
Calculating these expected values for the state 
$\Psi(x_{1},x_{2})\equiv\frac{1}{\sqrt{2}}(\phi_{1}(x_{1})\phi_{2}(x_{2})-\phi_{2}(x_{1})\phi_{1}(x_{2}))$
 yields
$\langle n(a)\rangle\equiv\langle\Psi,n(a)\Psi\rangle=\sum_{i=1}^{2}|\phi_{i}(a)|^{2}$
and
$\langle
n(a)n(b)\rangle=|\phi_{1}(a)|^{2}|\phi_{2}(b)|^{2}+|\phi_{2}(a)|^{2}|\phi_{1}(b)|^{2}-2{\rm
Re}(\overline{\phi_{1}(a)}\phi_{2}(a)\overline{\phi_{2}(b)}\phi_{1}(b))$.
From the nature of sinusoidal functions $\phi_{i}(x)=C\sin\omega_{i}x \
(i=1,2)$, for example, if $a=\frac{n\pi}{\omega_{2}-\omega_{1}}$,
$b=\frac{m\pi}{\omega_{2}-\omega_{1}}$ ($n$:even number, $m$:odd number ),
then $\langle n(a)n(b)\rangle=\langle n(a)\rangle \langle n(b)\rangle$
is obtained and the density correlation disappears(
Another example is $a=\frac{n\pi\omega_{1}-m\pi\omega_{2}}{\omega_{2}^{2}-\omega_{1}^{2}}$,
$b=\frac{n\pi\omega_{2}-m\pi\omega_{1}}{\omega_{2}^{2}-\omega_{1}^{2}}$,
\ $n$:even number, $m$:odd number ).

\section{Results of application to SIAM}\label{app}
In this section, we focus on the simplest non-trivial quantum impurity
system, the Single Impurity Anderson Model (SIAM) defined by the following Hamiltonian
(\ref{siam});
\begin{eqnarray}
H&\equiv&H_{\rm imp}+H_{\rm hyb}+H_{\rm cond},\label{siam}\\
H_{\rm
 imp}&\equiv&\sum_{\sigma}\varepsilon_{d}n_{\sigma}+Un_{\uparrow}n_{\downarrow},\\
H_{\rm hyb}&\equiv&\sum_{\sigma,{\bf k}}V_{\bf
 k}d_{\sigma}c^{\dagger}_{{\bf k}\sigma}+{\rm h.c.},\\
H_{\rm cond}&\equiv&\sum_{{\bf k},\sigma}E_{\bf k}c^{\dagger}_{{\bf
 k}\sigma}c_{{\bf k}\sigma},
\end{eqnarray}
where $d_{\sigma}$ annihilates an electron with spin $\sigma$ at the
Anderson impurity, 
characterized by the onsite
energy $\varepsilon_{d}$ and the intra-impurity repulsion $U$.
Here $n_{\sigma}\equiv d^{\dagger}_{\sigma}d_{\sigma}$ is the number
operator of the electron with spin $\sigma$ in the impurity.
%
In the conduction electron system, $c^{\dagger}_{{\bf k}\sigma}$ creates an electron
with energy $E_{\bf k}$, spin $\sigma$ and momentum ${\bf k}$, and
$V_{\bf k}$ is the hopping integral between the orbital for $({\bf
k},\sigma)$-state of the conduction electron system and the orbital of
the impurity with spin index $\sigma$.
We assume that the hybridization strength $\Delta\equiv\pi
\sum_{\bf k}\left|V_{\bf k}\right|^{2}\delta(\omega-E_{\bf k})$
is a constant independent of the frequency $\omega$, and take
the Fermi energy $\mu$ to be $\mu=0$.
Hence, assuming that the
conduction electron system has a flat band structure with half
bandwidth $D$ and neglecting the ${\bf k}$-dependence of $V_{\bf k}$,
 we have $\Delta=\pi V^{2}/(2D)$.

We then study the behavior of the entanglement entropy, mutual
information, and relative entropy for the energy eigenstates of this
system.
In order to calculate these quantities based on the formulas derived in
 the subsection \ref{oneI}, it is necessary to calculate the physical quantities (correlation
functions) on the Anderson impurity, for which we use the Numerical
Renormalization Group (NRG) calculation.
In the NRG calculation, the logarithmic discretization parameter
$\Lambda$ is set to $\Lambda=2$ and 
$(\pi\Delta)/D=0.05$.
Calculations were performed with several values of $\Lambda$ to confirm
that the choice of $\Lambda$ value does not affect the physical interpretation.

%
Since $[H,Q]=[H,S_{z}]=0$, the discretized Hamiltonian $H_{\rm NRG}$
directly treated in NRG calculation also satisfies $[H_{\rm NRG},Q]=[H_{\rm NRG},S_{z}]=0$.
So, the energy eigenstates can be obtained by iterative
diagonalization while taking into simultaneous eigenstates of $Q$ and
$S_{z}$.
Therefore, the formulas derived in the subsection \ref{oneI}
can be applied to the energy eigenstates of each step $N_{\rm nrg}$ of
the NRG calculation.
%

In the next subsection \ref{fix}, we discuss results for the ground
state in the low-temperature limit (i.e., the ground state of the NRG
fixed point Hamiltonian).
In the following subsection \ref{flow}, we  discuss results for the
states corresponding to the high-temperature regime, including the
excited states (i.e., the states depending on the NRG step(flow) number
$N_{\rm nrg}$).
\subsection{Quantum entanglement in ground state of NRG fixed point Hamiltonian}\label{fix} 
In this subsection, we show the results for the ground state $\psi^{({\rm G})}$
of the NRG fixed point Hamiltonian.

\begin{figure}[h]
\includegraphics[width=0.5\textwidth]{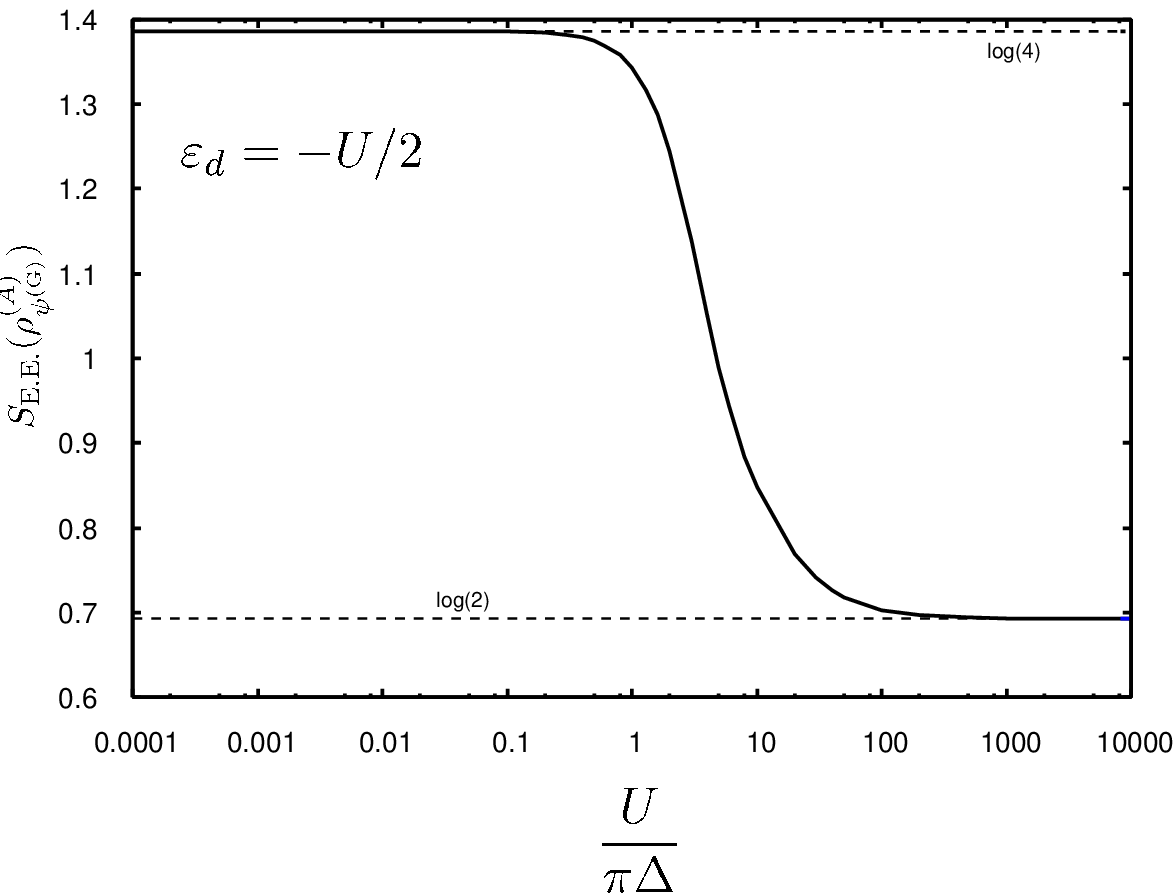}
\caption{Transition of entanglement entropy with change in $U$}\label{u-dep-see-half}
\end{figure}

Fig.\ref{u-dep-see-half} shows the $U$-dependence of the 
entanglement entropy in the presence of electron-hole symmetry.
The horizontal axis is $U/(\pi\Delta)$ on a logarithmic scale and the
vertical axis is the entanglement entropy.
As the value of $U$ is increased,
it can be seen that 
the entanglement entropy monotonically transitions from
$\log(4)$ to $\log(2)$.
In the presence of electron-hole symmetry, the entanglement entropy 
is determined through the value of the correlation function $\langle
n_{\uparrow}n_{\downarrow}\rangle$ as shown in Eq.(\ref{s_hf}).
The exact series representation of $\langle n_{\uparrow}n_{\downarrow}\rangle$
with respect to $U$ has been obtained.
We confirmed that the calculated results of
$\langle n_{\uparrow}n_{\downarrow}\rangle$
using NRG are in good agreement with the exact values for some values of
$U$.
For $U=0$, the up-spin and down-spin correlations disappear, so the
correlation function $\langle n_{\uparrow}n_{\downarrow}\rangle$
takes the maximum value $1/4$ in the case of electron-hole symmetry,
resulting in the entanglement entropy of $\log(4)$ of the maximum value.
The value of $U$ in the transition region from $\log(4)$ to $\log(2)$
belongs to the strongly correlated region.
For typical values of $U$ that can be realized in quantum dot systems etc.,
the entanglement entropy is considerably larger than $\log(2)$,
although it belongs to the transition region.
This indicates that, from the quantum entanglement point of view,
Anderson impurities should not be easily replaced by quantum spins just
because
they are in the strongly correlated regime.

In the remainder of subsection \ref{fix}, we will mainly discuss the
the dependence of onsite energy $\varepsilon_{d}$.
Since SIAM has the electron-hole symmetry, it is obvious that some graphs
of $\varepsilon_{d}$-dependence are symmetric with respect to
$\varepsilon_{d}=-U/2$.
However, we display the results as functions of $\varepsilon_{d}+U/2$
so that its symmetry can be seen at a glance.

\begin{figure}[h]
\includegraphics[width=0.5\textwidth]{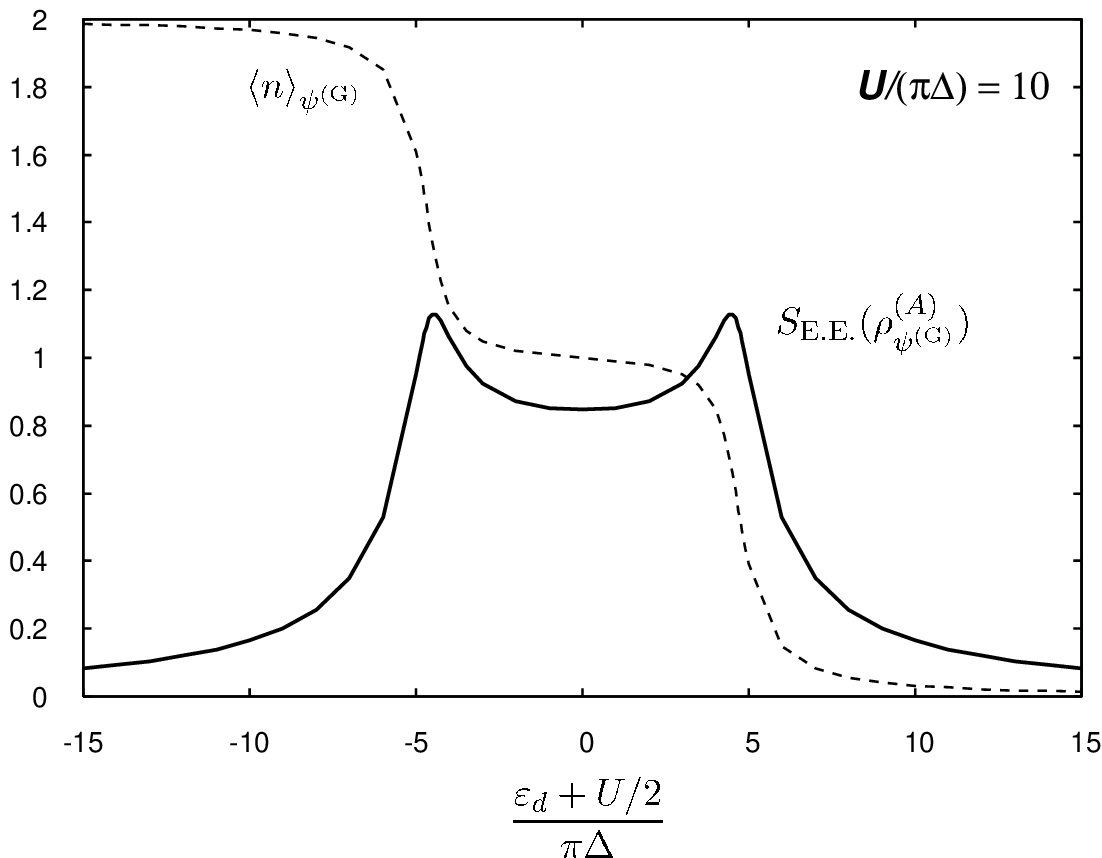}
\caption{Entanglement entropy (solid line) and the average electron
 number in the impurity (dashed line)as functions of $\varepsilon_{d}$}\label{ed-dep-see-half}
\end{figure}

Fig.\ref{ed-dep-see-half} shows the dependence of the entanglement
entropy on $\varepsilon_{d}$
 for $U/(\pi\Delta)=10$.
The horizontal axis is 
the onsite
energy $\varepsilon_{d}$ and the vertical
axis is the entanglement entropy, or
the average electron number $\langle n\rangle_{\psi^{({\rm G})}}$ in the impurity.
It can be seen that 
the entanglement entropy peaks where the average number of electron
changes,
and
that the entanglement entropy takes values greater than  $\log(2)$
around $\langle n\rangle_{\psi^{({\rm G})}}\simeq 1$.
To elucidate this feature, we examine the behavior of the
correlation functions which are the diagonal component of the density operator.

\begin{figure}[h]
\includegraphics[width=0.5\textwidth]{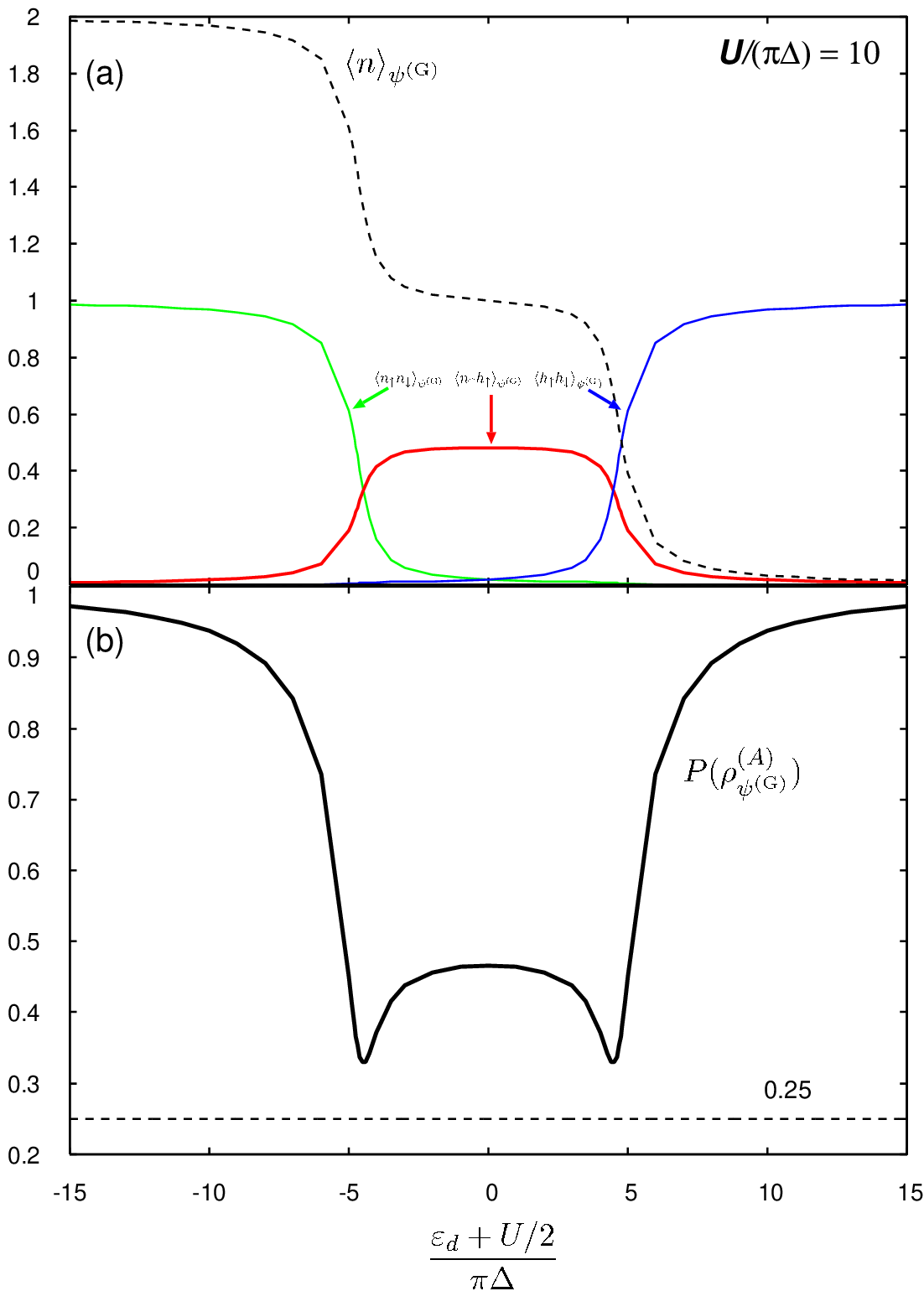}
\caption{(Color on line)(a)The correlation functions $\langle
 n_{\uparrow}n_{\downarrow}\rangle$(green solid line), $\langle
 n_{\uparrow}h_{\downarrow}\rangle$(red solid line), and $\langle h_{\uparrow}h_{\downarrow}\rangle$(blue solid line) as functions of
 $\varepsilon_{d}$. The $\varepsilon_{d}$-dependence of the average
 electron number as a reference(dashed line).
(b)The purity $P$ as a function of $\varepsilon_{d}$.}\label{ed-dep-cf}
\end{figure}

Fig.\ref{ed-dep-cf}(a) shows the $\varepsilon_{d}$-dependence of
the correlation functions $\langle n_{\uparrow}n_{\downarrow}\rangle$, 
$\langle n_{\uparrow}h_{\uparrow}\rangle=\langle n_{\downarrow}h_{\downarrow}\rangle$
and
$\langle h_{\uparrow}h_{\downarrow}\rangle$.
Note that the spin quantum number ${\mathcal S}$ of the ground state we are considering
here is zero, so the ground state is magnetic isotropic and $\langle n_{\uparrow}\rangle=\langle
n_{\downarrow}\rangle$ holds.
The average electron number in the impurity is also shown again in dashed lines for reference
in the figure.
The correlation function $\langle n_{\uparrow}n_{\downarrow}\rangle$
takes values of approximately 1 in the region where $\langle
n\rangle_{\psi^{({\rm G})}}\simeq 2$,
and  
converges monotonically to 0 toward the region where 
$\langle n\rangle_{\psi^{({\rm G})}}\simeq 0$.
Similarly, the correlation function 
$\langle h_{\uparrow}h_{\downarrow}\rangle$
takes values of approximately 1 in the region 
where the average hole number in the impurity  is approximately 2, 
and 
converges monotonically to 0
toward the region where the average hole number in the impurity is approximately
0.
On the other hand, the correlation function 
$\langle n_{\uparrow}h_{\uparrow}\rangle$
has a plateau only in the region where the average particle number
and 
the average hole number in the impurity are approximately 1, and 
converges to 0
in other regions.
The contribution of each correlation function $C$ to 
the entanglement entropy is expressed as
$-C\log(C)$,
which has a unimodal peak at $C=1/e$ and is zero at $C=0$ and $1$.
Thus, the correlation functions 
$\langle n_{\uparrow}n_{\downarrow}\rangle$ and 
$\langle h_{\uparrow}h_{\downarrow}\rangle$
 produce peaks in the entanglement entropy 
in the region where the value of $\langle n\rangle_{\psi^{({\rm G})}}$
changes from 1 to 2 and from 0 to 1, respectively.
The correlation function 
$\langle n_{\uparrow}h_{\downarrow}\rangle$
yields a peak structure in the entanglement entropy in both regions 
where the value of $\langle n\rangle_{\psi^{({\rm G})}}$
changes from 0 to 1 and from 1 to 2,
while even in the region 
where$\langle n\rangle_{\psi^{({\rm G})}}\simeq 1$,
 the  value of this correlation function is only slightly greater than $1/e$, 
which yields the value greater than $\log(2)$ for the entanglement entropy.
The above results indicate that each peak structure of the entanglement
entropy is caused by corresponding two correlation functions ($\langle
n_{\uparrow}h_{\uparrow}\rangle=\langle n_{\downarrow}h_{\downarrow}\rangle$ and $\langle
n_{\uparrow}n_{\downarrow}\rangle$ or $\langle
h_{\uparrow}h_{\downarrow}\rangle$), and 
that the plateau of the entanglement entropy in the region of the
half-filled state is mainly caused by the correlation function 
$\langle n_{\uparrow}h_{\uparrow}\rangle=\langle n_{\downarrow}h_{\downarrow}\rangle$.

We consider the purity $P(=P(\rho^{(A)}_{\psi}))\equiv {\rm Tr}_{{\mathcal
H}_{A}}\left(\rho^{(A)}_{\psi}\right)^{2}$, which measures how close the
reduced state $\rho^{(A)}_{\psi}$ is to the pure state\cite{YT_stoc}.
In the case of the density operator we are considering now, 
it is the sum of the squares of the diagonal components.
The purity $P$ satisfies $1/4\le P \le 1$, where 
$P=1$ is the pure state and 
$P=1/4$ is the maximum entropy state.
Fig.\ref{ed-dep-cf}(b) shows the $\varepsilon_{d}$-dependence of the purity
$P$.
The purity $P$ is close to unity for large absolute values of
$\varepsilon_{d}$, because the state is closer to the vacuum or fully
occupied state, respectively. 
In the valence fluctuation regime, the purity $P$ decreases due to the
contribution of about three states, 
but 
in the vicinity of half-filled state, the purity $P$ increases 
due to the tendency to exclude the vacuum and fully occupied states.
%

\begin{figure}[h]
\includegraphics[width=0.5\textwidth]{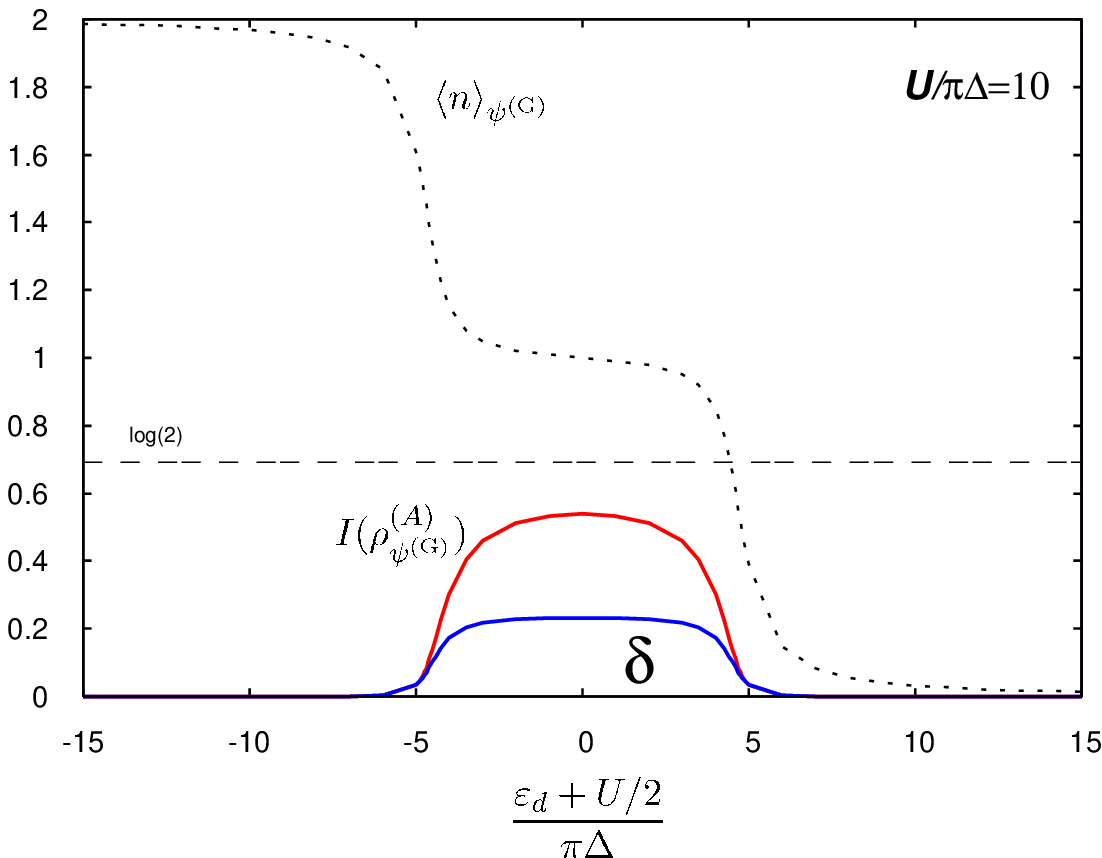}
\caption{(Color on line)The mutual information  $I$(red solid line), and $\delta$(blue solid line) as functions of
 $\varepsilon_{d}$. The $\varepsilon_{d}$-dependence of the average
 electron number in the impurity as a reference (dotted line).}\label{ed-dep-I}
\end{figure}

To examine the entanglement between up- and down-spin electrons, we will
examine the mutual information $I$ defined in Eq.(\ref{I}).
The Fig.\ref{ed-dep-I} shows the $\varepsilon_{d}$-dependence of $I$ and 
also the $\varepsilon_{d}$-dependence of 
$\delta\equiv \langle n_{\uparrow}\rangle\langle n_{\downarrow}\rangle
-\langle n_{\uparrow}n_{\downarrow}\rangle$, one of the quantities
measuring electron correlation(deviation from the mean-field
approximation).
Both $I$ and $\delta$ show rectangle-shaped graphs in the region 
where $\langle n\rangle_{\psi^{({\rm G})}} \simeq 1$.
%
This is because the mean-field approximation breaks down ($\delta\neq 0$) because the
effect of repulsion between up- and down-spin electrons is more 
pronounced in the region where $\langle n\rangle_{\rm G}\simeq 1$, and there
is interdependence where the presence of the electron with one spin in
the impurity makes it harder for the electron with the other spin to be
present in the impurity, 
resulting in non zero value of the mutual information $I$ which indicates
the interdependence of the up- and down-spin electron states.
Although $I$ and $\delta$ are linear independent functions by their
definitions and by their graphs shown here,
the results indicate that the behavior of $I$ and $\delta$ are positively
correlated.
From this result, we can say that the mutual information is one of the
measures of electron correlation in this case.

\begin{figure}[h]
\includegraphics[width=0.5\textwidth]{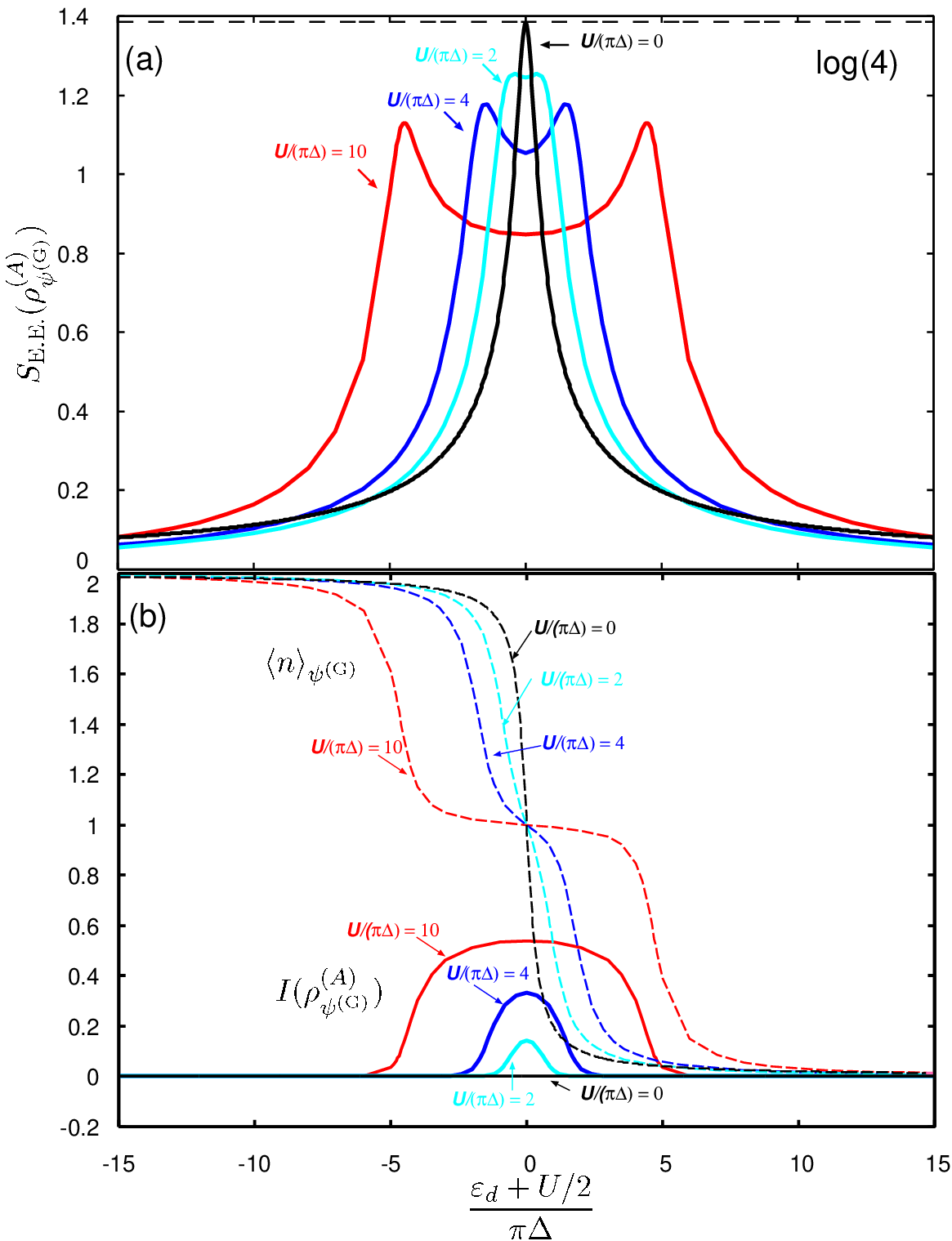}
\caption{(Color online) Entanglement entropy (a), mutual information and average
 electron number in the impurity (b) as functions of $\varepsilon_{d}$ for
 $U/(\pi\Delta)=$10(red), 4(blue), 2(cyan) and 0(black). }\label{u-dep}
\end{figure}

Next, we consider the $\varepsilon_{d}$-dependence of entanglement
entropy $S_{\rm E.E.}$ and mutual information $I$ when 
the value of $U/(\pi\Delta)$ is varied.
Fig.\ref{u-dep} shows the $\varepsilon_{d}$-dependence of $S_{\rm
E.E.}$, $I$ and 
$\langle n\rangle_{\psi^{({\rm G})}}$ 
 for $U/(\pi\Delta)=10, 4, 2$ and
$0$.
As for the electron occupancy of the isolated Anderson impurity,
it is apparent that 
the electron number in the impurity is 1 in the interval $-U/2\le\varepsilon_{d}+U/2\le
U/2$ with interval width $U$.
In the system under consideration, the average electron number
fluctuates around 1 due to the effect of hybridization with the
conduction electron system.
Thus, as shown in Fig.\ref{u-dep}(b), the plateau with $\langle
n\rangle\simeq 1$ becomes shorter and more blurred as the value of $U$
decreases.
As a result, the correlation function 
$\langle n_{\uparrow}n_{\downarrow}\rangle$ with a non-zero value in the
region where $\langle n\rangle_{\psi^{({\rm G})}}\simeq 2$ 
 and 
the correlation function $\langle h_{\uparrow}h_{\downarrow}\rangle$ 
with a non-zero value in the region where $\langle n\rangle_{\psi^{({\rm G})}}\simeq 0$ 
, viewed as functions of $(\varepsilon_{d}+U/2)/(\pi\Delta)$, move in parallel
toward $\varepsilon_{d}+U/2=0$ and overlap each other.
Also, the correlation function $\langle n_{\uparrow}h_{\uparrow}\rangle$
goes from a rectangular graph to a unimodal graph.
(The two facts above are not illustrated here.)
From the above facts, as shown in Fig.\ref{u-dep}, 
the side peaks of the entanglement entropy approach each other while
getting larger, and the height of the plateau in the half-filling region
grows.
Also in the case of $U=0$, the side-peak structure disappears and the graph
becomes unimodal.
Among the three conditions for the maximum entanglement entropy shown in
subsection \ref{oneI},
(A)$\langle n_{\uparrow}\rangle=\langle n_{\downarrow}\rangle$,
(B)$\langle n\rangle=1$, and (C)$\langle
n_{\uparrow}n_{\downarrow}\rangle=\langle
n_{\uparrow}\rangle\langle n_{\downarrow}\rangle$, 
the condition (A) is always satisfied because the quantum number
${\mathcal S}$ of the ground
state considered here is ${\mathcal S}=0$.
The condition (B) is satisfied for $\varepsilon_{d}+U/2=0$ regardless of
the value of $U$.
The condition (C) is satisfied only when $U=0$ under the previous two
conditions.
As a result, as shown in Fig.\ref{u-dep}(a), the entanglement entropy reaches its maximum value
$\log(4)$ only when $\varepsilon_{d}+U/2=0$ and $U=0$.
%
Fig.\ref{u-dep}(b) shows that as the value of $U$ decreases, 
the region of interdependence between up- and down-spin electrons
becomes smaller because the region of average electron number is 1
becomes smaller, as mentioned above, and the degree of interdependence
also decreases because the repulsive interaction becomes smaller,
resulting in a smaller value of mutual information.
Although not shown in the figure, the difference between $\delta$ and $I$
becomes smaller for smaller values of $U$. 
For $U=0$, $\delta=I\equiv 0$  holds.
The above results are consistent with the
interpretation that 
a small value of $U$ leads to low correlation within the Anderson
impurity system and 
a large value of $\Delta$ leads to large hybridization with the
conduction electron system, 
resulting in large entanglement between the Anderson impurity and the
conduction electron system.

\begin{figure}[h]
\includegraphics[width=0.4\textwidth]{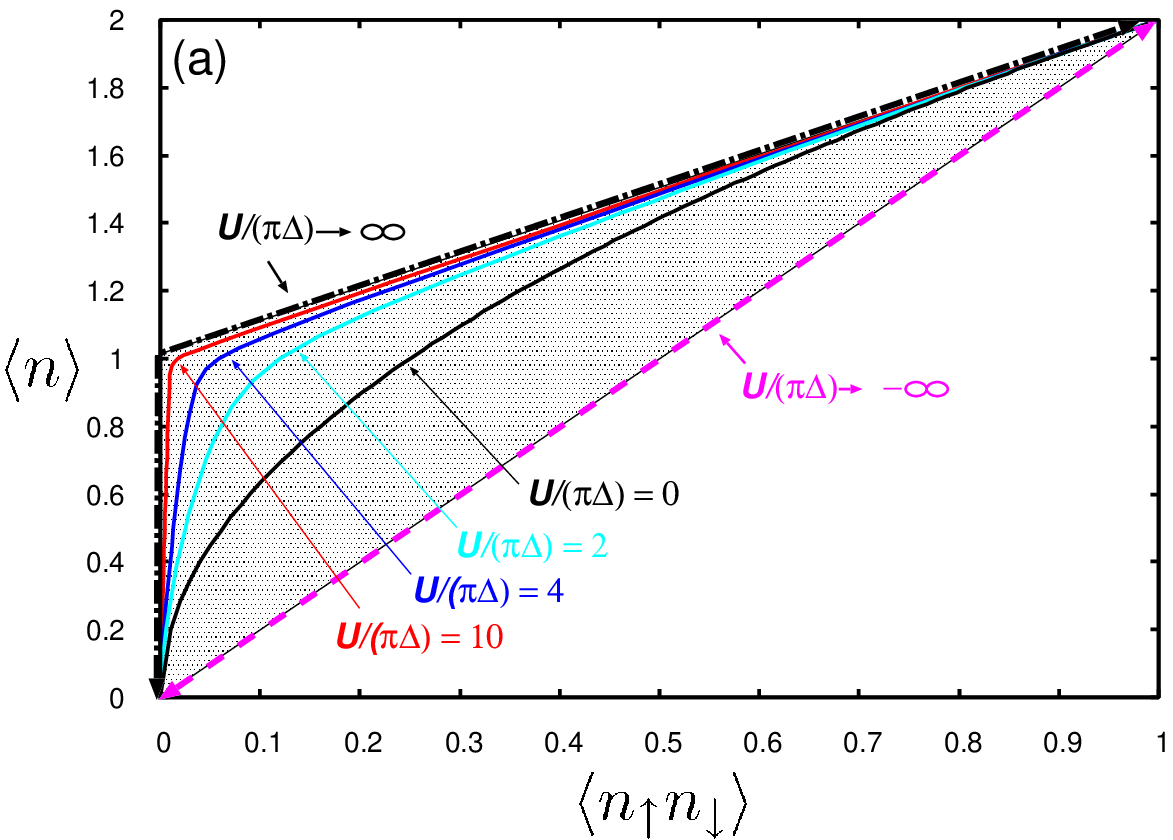}\\
\includegraphics[width=0.4\textwidth]{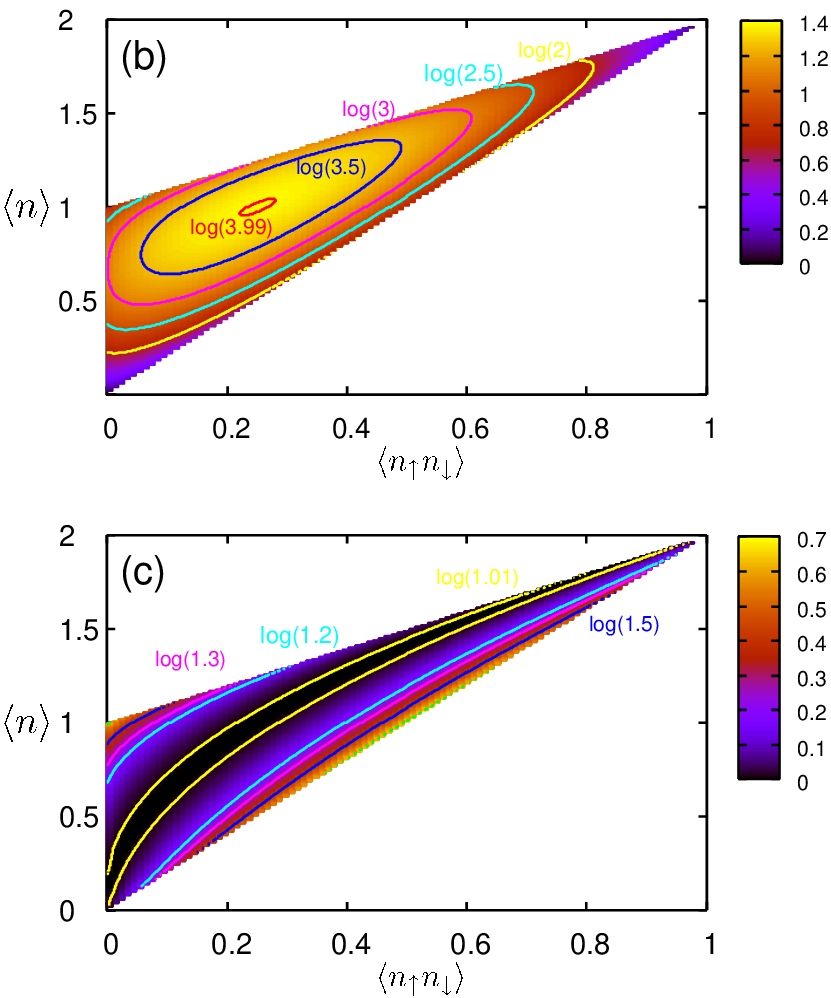}
\caption{(Color on line) (a) Region of existence of $\langle n\rangle$ and 
$\langle n_{\uparrow}n_{\downarrow}\rangle$.Trajectories of $(\langle n\rangle,\langle
n_{\uparrow}n_{\downarrow}\rangle)$ for varying $\varepsilon_{d}$ for
some vales of $U$. (b,c)Contour plots of the entanglement entropy
$S_{\rm E.E.}$(b) and the mutual information $I$(c) as functions of
 $\langle n\rangle$ and $\langle n_{\uparrow}n_{\downarrow}\rangle$.}\label{bv}
\end{figure}

At the end of this subsection, we summarize the results on ground
states $\psi^{({\rm G})}$ investigated so far, considering the limit where $U$ is large.
%
Since the quantum number ${\mathcal S}$ of the ground state considered so far
is ${\mathcal S}=0$, the entanglement entropy $S_{\rm E.E.}$ and 
mutual information $I$ are functions of $\langle n\rangle$ and 
$\langle n_{\uparrow}n_{\downarrow}\rangle$ because $\langle
n_{\uparrow}\rangle =\langle n_{\downarrow}\rangle$.
The region of existence of $\langle n\rangle$ and 
$\langle n_{\uparrow}n_{\downarrow}\rangle$
is the region enclosed by the triangle in $\langle n\rangle$ - 
$\langle n_{\uparrow}n_{\downarrow}\rangle$ plane in Fig.\ref{bv}(a).
The trajectories of $(\langle n\rangle,\langle
n_{\uparrow}n_{\downarrow}\rangle)$ for varying $\varepsilon_{d}$ for
some vales of $U$ are depicted in Fig.\ref{bv}(a).
As $U$ is decreased, the value of the correlation function $\langle
n_{\uparrow}n_{\downarrow}\rangle$
 increases, so the trajectory shifts to the right, and 
the trajectory for $U=0$ is given by $\langle
n_{\uparrow}n_{\downarrow}\rangle=(\langle n\rangle)^{2}/4$.
The region to the the right of this trajectory is the region where the
trajectory exists for the case of attractive $U$, which is not addressed in
this paper.
In the limit $U\to-\infty$, the trajectory asymptotically approaches the
right oblique side of the triangle.
On the other hand, in the limit $U\to +\infty$, the trajectory
asymptotically approaches the left and upper sides of the triangle.
Fig.\ref{bv}(b) and (c) show contour plots of the entanglement entropy
$S_{\rm E.E.}$ and the mutual information $I$ as functions of $\langle
n\rangle$ and $\langle n_{\uparrow}n_{\downarrow}\rangle$, respectively.
Considering these figures together, the results shown so far can be
clearly understood at a glance.
In addition, the following two facts can be understood.
When $U$ is large, the entanglement entropy $S_{\rm E.E.}$ plotted as a
function of $\varepsilon_{d}$ has two sharp side peaks of $\log(3)$, and
has a plateau of $\log(2)$ in the half-filling region.
The mutual information $I$ plotted as a function of $\varepsilon_{d}$
asymptotically approaches a rectangular shape with a maximum value of
$\log(2)$ in the half-filling region, for the limit $U\to +\infty$.
The behavior of the quantum informative quantities formulated in this
paper,
which is calculated for the first time for the ground state of SIAM, 
is shown to be what is expected from the properties of the ground state
of SIAM known so far.

\subsection{Quantum entanglement in energy eigenstates depending on $N_{\rm nrg}$}\label{flow}
In this subsection, we investigate the NRG step $N_{\rm nrg}$ dependence
of entanglement entropy, mutual information and relative entropy for the energy eigenstates including excited
states along the NRG flow.
We show the results using even  $N_{\rm nrg}$.
See \cite{oddN} for results using odd $N_{\rm nrg}$.

%

\begin{figure}[h]
\includegraphics[width=0.5\textwidth]{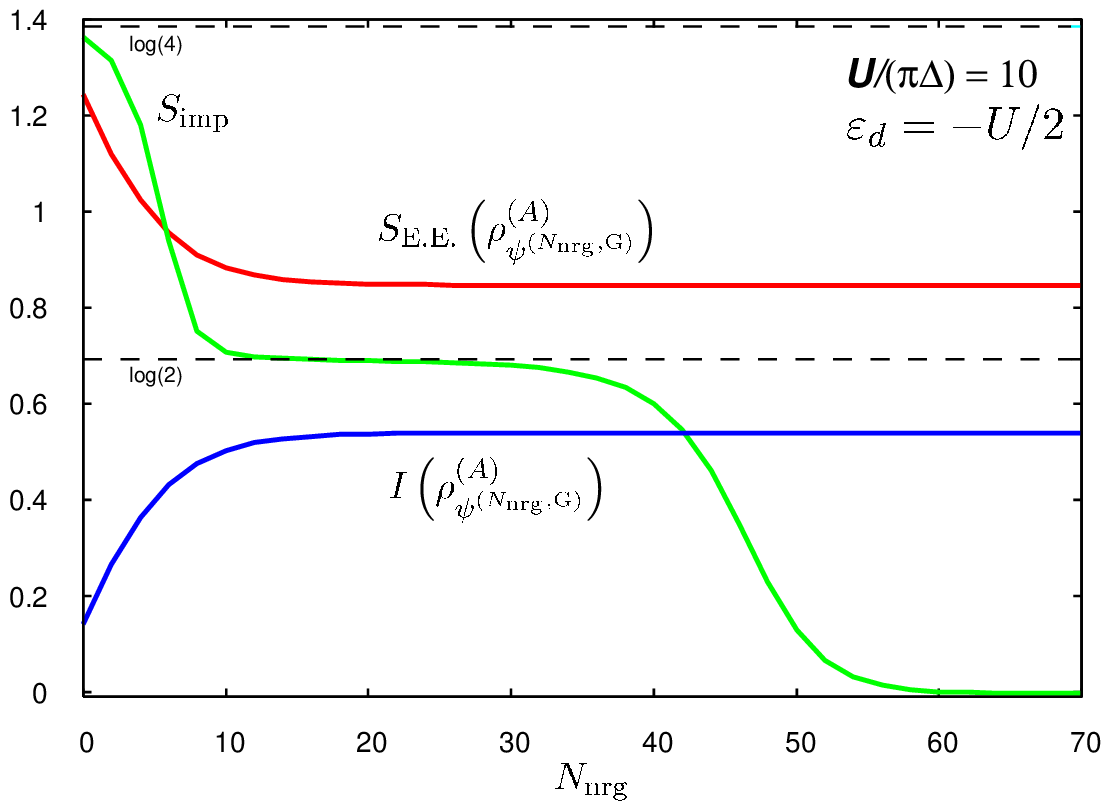}
\caption{(Color on line) $N_{\rm nrg}$-dependence of $S_{\rm imp}$, $S_{\rm
 E.E.}\left(\rho^{(A)}_{\psi^{({\rm G},N_{\rm nrg})}}\right)$, and $I\left(\rho^{(A)}_{\psi^{({\rm G},N_{\rm nrg})}}\right)$.}\label{n-dep-sIsimp-G}
\end{figure}

We first consider the ground state $\psi^{({\rm G},N_{\rm nrg})}$ for
each $N_{\rm nrg}$ in the case of the electron-hole symmetric system.
Fig.\ref{n-dep-sIsimp-G} shows the $N_{\rm nrg}$-dependence of
entanglement entropy $S_{\rm E.E.}$ and mutual information $I$ for 
$U/(\pi\Delta)=10$, along with the $N_{\rm nrg}$-dependence of $S_{\rm
imp}$.
Here, $S_{\rm imp}$ is the thermodynamic entropy of the total system minus the
thermodynamic entropy of the conduction electron systems, generally with
respect to all the quantum impurities in the system (although only one
Anderson impurity is in the system considered here).
The total energy levels, including the excited states obtained in the
NRG calculation contribute to $S_{\rm imp}$ with the weight of the
Boltzmann factor for the temperature $T_{N_{\rm nrg}}\equiv
\tau\Lambda^{-(N_{\rm nrg}-1)/2}$ determined by $N_{\rm nrg}$, 
where $\Lambda$ is the NRG discretization parameter and $\tau=O(1)$ is a
kind of fitting parameter independent from $N_{\rm nrg}$. Note again that the entanglement entropy or
mutual information considered here, on the other hand, is generally a quantity
with selectivity determined by the selected quantum impurity and the
selected eigenstate of the total system.
The graph for $S_{\rm imp}$ in Fig.\ref{n-dep-sIsimp-G} shows that at
high temperatures, the four degrees of freedom, the total degrees of freedom of the Anderson impurity
appears, but as the temperature decreases, the two degrees of freedom
of spin remains, and finally, the Kondo singlet with zero degrees of
freedom is formed.
On the other hand, the entanglement entropy $S_{\rm E.E.}$ 
decreases similarly as $S_{\rm imp}$ decreases from $\log(4)$ to
$\log(2)$, as seen in Fig.\ref{n-dep-sIsimp-G}.
This can be interpreted as reflecting a decrease in the contribution of the vacuum state
and the fully occupied state to entanglement as the temperature
decreases.
Fig.\ref{n-dep-sIsimp-G} shows that the entanglement entropy does not
change as the temperature is further lowered and the temperature enters
the Kondo screening region and becomes even lower.
This means that the surviving degrees of freedom contribute to entanglement with the
conduction electron system even within the low-temperature limit.
The mutual information $I$ increases toward the region where 
$S_{\rm imp}\simeq \log(2)$
, due to its nature of
reflecting the degree of interdependence between the up- and
down-electron states, and constant in the low-temperature region
thereafter, including the Kondo screening region, as shown Fig.\ref{n-dep-sIsimp-G}.
The persistence of the interdependence due to the repulsion between up-
and down-spin electrons in the impurity in the Kondo screening region
and in the lower temperature region thereafter can be understood from
the fact that, due to its persistence, the electron in the impurity and
conduction electrons form a many-body singlet state(Kondo singlet state) 
instead of a singlet state in the impurity in that temperature region.
%
Fig.\ref{n-dep-sIsimp-G} shows that the two graph curves $I$ and $S_{\rm
E.E.}$ are symmetric with respect to $\log(2)$.
This is because $I$ and $S_{\rm E.E.}$ now satisfy the relation
\begin{equation}
I+S_{\rm E.E.}=\log(4)\label{monohalf}
\end{equation}
from Eqs.(\ref{relIS}) and (\ref{Ssigma}),
since $\langle n_{\uparrow}\rangle=\langle
n_{\downarrow}\rangle=\frac{1}{2}$ for all even numbers of $N_{\rm nrg}$
.
Eq.(\ref{monohalf}) also express the monogamy of quantum entanglement as
an equality.
This means that 
the degree of quantum entanglement with the environmental system($S_{\rm
E.E.}$) decreases when the internal quantum entanglement ($I$) is
strong, and vice versa, and the sum of the two is a constant value of $\log(4)$.
%

We can understand the behavior of these quantities being constant in the Kondo screening
region described above from the fact that the spin quantum number ${\mathcal S}$ of the
ground state of each NRG step $N_{\rm nrg}$ is zero, so that $S_{\rm
E.E.}$ and $I$ are represented only by the non-magnetic quantities 
$\langle n\rangle$ and $\langle n_{\uparrow}n_{\downarrow}\rangle$.

Therefore, to capture the Kondo screening region, 
we next consider entanglement entropy $S_{\rm E.E.}$, mutual
information $I$ and relative entropy $D$ for states with ${\mathcal S}\neq
0$(excited states).
Since the state with ${\mathcal S}\neq 0$ has non-zero ${\mathcal S}_{z}$ and the expected
value in these states of
$s_{z}=\frac{1}{2}(n_{\uparrow}-n_{\downarrow})$ is generally non-zero,
$\langle n_{\uparrow}\rangle$ and $\langle n_{\downarrow}\rangle$ in the
expressions of $S_{\rm E.E.}$ and $I$ become 
$\langle n_{\uparrow}\rangle=\langle n\rangle/2+\langle s_{z}\rangle$
and
$\langle n_{\downarrow}\rangle=\langle n\rangle/2-\langle s_{z}\rangle$,
respectively, and include the magnetic quantity $\langle s_{z}\rangle$.
Note that ${\bf S}^{2}$ and $S_{z}$ are operators of the spin magnitude and
the $z$-component of the spin for the total system, respectively, and 
$s_{z}$ is the operator of the $z$-component of the spin of the (selected) Anderson
impurity.

The behaviors of the entanglement entropy $S_{\rm E.E.}$ and the mutual
information $I$ as functions of $N_{\rm nrg}$ 
 for the lowest energy state $\psi_{\mathcal Q}^{({\rm G},N_{\rm
 nrg})}({\mathcal S},{\mathcal S}_{z}={\mathcal S})$ in
states within ${\mathcal S}=1/2$ and ${\mathcal Q}=\pm 1$ are as follows: 
The entanglement entropy is indeed found to change in the Kondo
screening region.
Especially in the case of the electron-hole symmetric systems, it is clearly
distinguishable from changes in other regions.
However, since the non-magnetic quantities $\langle n\rangle$ and 
$\langle n_{\uparrow}n_{\downarrow}\rangle$ included in the expression
of $S_{\rm E.E.}$ also change, especially in case of systems away from
the electron-hole symmetry, they also change outside the Kondo screening region,
making the change in the Kondo screening region unclear.
As for the mutual information, no clear change is observed in the Kondo
screening region because, mathematically,  it is a subtraction of two quantities including
the magnetic quantities, $S_{\uparrow}+S_{\downarrow}$ and $S_{\rm E.E.}$.
The above two results are not illustrated here.

\begin{figure}[h]
\includegraphics[width=0.5\textwidth]{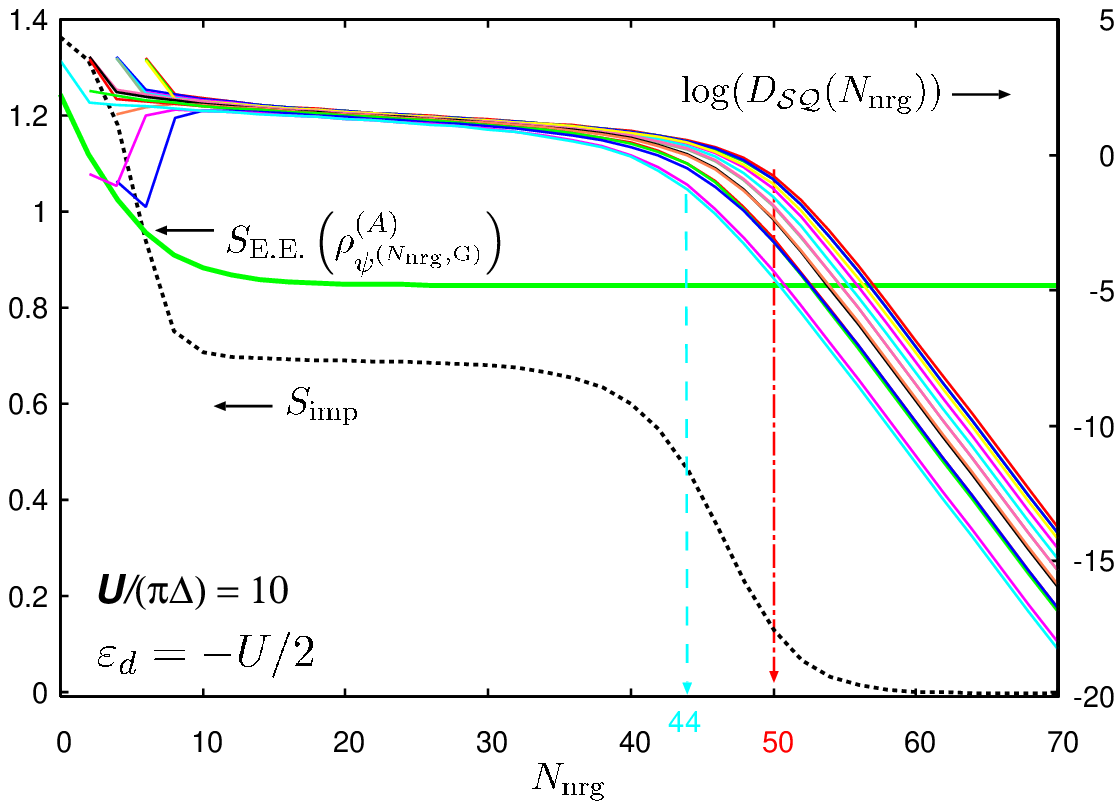}
\caption{(Color on line) $N_{\rm nrg}$-dependence of $S_{\rm imp}$, $S_{\rm
 E.E.}\left(\rho^{(A)}_{\psi^{({\rm G},N_{\rm nrg})}}\right)$, and
 $D_{{\mathcal S}{\mathcal Q}}(N_{\rm nrg})$ for some values of $({\mathcal
 S},{\mathcal Q})$ stated in the main text. See the main text and Table
 \ref{tab_D} for
 the shown curves of $D_{{\mathcal S}{\mathcal Q}}(N_{\rm nrg})$ and values of
 $({\mathcal S},{\mathcal Q})$ correspondences.}\label{n-dep-D}
\end{figure}

 We next consider two states with ${\mathcal S}_{z}={\mathcal S}$
 and ${\mathcal S}_{z}=-{\mathcal S}$ for the
 lowest energy state $\psi^{(N_{\rm nrg},\rm G)}_{\mathcal Q}({\mathcal
 S},{\mathcal S}_{z})$ within ${\mathcal S}\neq 0$ and the corresponding
 appropriate ${\mathcal Q}$ for each $N_{\rm nrg}$, and then reduce these two states to the Anderson
 impurity system and consider the relative entropy
 $D\left(\rho^{(A)}_{\psi^{(N_{\rm nrg},\rm
 G)}_{\mathcal Q}({\mathcal S},{\mathcal S})}||\rho^{(A)}_{\psi^{(N_{\rm nrg},\rm
 G)}_{\mathcal Q}({\mathcal S},-{\mathcal S})}\right)(\equiv
 D_{{\mathcal S}{\mathcal Q}}(N_{\rm nrg}))$.
For ${\mathcal S}\neq 0$, the corresponding appropriate ${\mathcal Q}$ retained in our NRG
calculation are $({\mathcal S},{\mathcal Q})=(n/2,\pm 1)$, $(n/2,\pm 3)
\ (n=1,3,5)$,
$(1/2,\pm 5)$,$(3/2,\pm 5)$, $(m,0), (m,\pm 2)$, $(m,\pm 4) \ (m=1,2)$, $(3,0)$.
For each $({\mathcal
S},{\mathcal Q})$ mentioned above,
the $N_{\rm nrg}$-dependence of relative entropy 
$D_{{\mathcal S}{\mathcal Q}}(N_{\rm nrg})$ 
are shown in Fig.\ref{n-dep-D}.
The $N_{\rm nrg}$-dependence of $S_{\rm imp}$ is also shown in the
figure again as reference.
Here, we have confirmed the following inequalities and equality hold for $N$ in the Kondo screening
region;
$D_{5/2,\pm 3}(N)>D_{3,0}(N)$$>D_{2,\pm 4}(N)>D_{3/2,\pm 5}(N)>D_{5/2,\pm
1}(N)$$>D_{2,\pm 2}(N)>D_{2,0}(N)$ 
$>D_{3/2,\pm 3}(N)>D_{3/2,\pm 1}(N)$ $>D_{1,\pm 4}(N)>D_{1.\pm 2}(N)$ 
$>D_{1,0}(N)>D_{1/2,\pm 5}(N)$ $>D_{1/2,\pm 3}(N)>D_{1/2,\pm 1}(N)$
(double
sign in same order), $D_{SQ}(N)>D_{SQ^{\prime}}(N) \ (|Q|>|Q^{\prime}|)$,
and $D_{S,Q}(N)=D_{S,-Q}(N)$.

The relationship between the quantum number $({\mathcal S,Q})$ and the
order of the $D_{\mathcal S,Q}$ values is summarized in Table \ref{tab_D} for
ease of reading.

\begin{table}
\caption{How to read the table: The label $Q$ of $D_{\mathcal S,Q}$ with
 the $n$-th value reads what is in the table, and ${\mathcal S}$ reads
 the name of the column.}\label{tab_D}
\begin{tabular}{|c|c|c|c|c|c|c|c|}\hline
\phantom{10st}&${\mathcal S}=\frac{1}{2}$&${\mathcal S}=1$&${\mathcal
 S}=\frac{3}{2}$&${\mathcal S}=2$&${\mathcal S}=\frac{5}{2}$&${\mathcal S}=3$\\ \hline
1st&\phantom{1}&\phantom{2}&\phantom{1}&\phantom{1}&${\mathcal Q}=\pm 3$&\phantom{1}\\
 \hline
2nd&\phantom{1}&\phantom{2}&\phantom{1}&\phantom{1}&\phantom{1}&${\mathcal
			 Q=0}$\\ \hline
3rd&\phantom{1}&\phantom{2}&\phantom{1}&${\mathcal Q=\pm 4}$&\phantom{1}&\phantom{1}\\ \hline
4th&\phantom{1}&\phantom{2}&${\mathcal Q}=\pm 5$&\phantom{1}&\phantom{1}&\phantom{1}\\ \hline
5th&\phantom{1}&\phantom{2}&\phantom{1}&\phantom{1}&${\mathcal Q}=\pm 1$&\phantom{1}\\ \hline
6th&\phantom{1}&\phantom{2}&\phantom{1}&${\mathcal Q}=\pm 2$&\phantom{1}&\phantom{1}\\ \hline
7th&\phantom{1}&\phantom{2}&\phantom{1}&${\mathcal Q}=0$&\phantom{1}&\phantom{1}\\ \hline
8th&\phantom{1}&\phantom{2}&${\mathcal Q=\pm 3}$&\phantom{1}&\phantom{1}&\phantom{1}\\ \hline
9th&\phantom{1}&\phantom{2}&${\mathcal Q}=\pm 1$&\phantom{1}&\phantom{1}&\phantom{1}\\ \hline
10th&\phantom{1}&${\mathcal Q}=\pm 4$&\phantom{1}&\phantom{1}&\phantom{1}&\phantom{1}\\ \hline
11th&\phantom{1}&${\mathcal Q=\pm 2}$&\phantom{1}&\phantom{1}&\phantom{1}&\phantom{1}\\ \hline
12th&\phantom{1}&${\mathcal Q}=0$&\phantom{1}&\phantom{1}&\phantom{1}&\phantom{1}\\ \hline
13th&${\mathcal Q}=\pm 5$&\phantom{1}&\phantom{1}&\phantom{1}&\phantom{1}&\phantom{1}\\ \hline
14th&${\mathcal Q}=\pm 3$&\phantom{2}&\phantom{1}&\phantom{1}&\phantom{1}&\phantom{1}\\ \hline
15th&${\mathcal Q}=\pm 1$&\phantom{2}&\phantom{1}&\phantom{1}&\phantom{1}&\phantom{1}\\ \hline
\end{tabular}
\end{table}

%
It can be seen from Fig.\ref{n-dep-D} that $D_{{\mathcal S}{\mathcal Q}}(N)$ 
decreases from $N\simeq 10$ to low temperatures with two continuously
connected power laws.
Here we define $N_{c}({\mathcal S,Q})$, the point of maximum curvature of the graph
curve of $D_{{\mathcal S}{\mathcal Q}}(N)$, as the change-point where
the power law changes.
We find that 
$N_{c}(\frac{1}{2},\pm 1)=44$
 and $N_{c}(\frac{5}{2},\pm 3)=50$,
which give the minimum and maximum change-points in $({\mathcal S,Q})$
considered here, respectively.
The positions of those two points are indicated in Fig.\ref{n-dep-D} by arrows.
It can been seen that these two points are located within the Kondo
screening region indicated by $S_{\rm imp}$.
Fig.\ref{n-dep-D} also shows that the change-point for small ${\mathcal
S}$ is on the high temperature side, while the change-point for larger
${\mathcal S}$ is on the low temperature side, mostly.

We next consider the interpretation of the above results.
In regions of $N_{\rm nrg}$ where $\langle n\rangle$ and $\langle
n_{\uparrow}n_{\downarrow}\rangle$ are not $N_{\rm nrg}$-dependent,
the $N_{\rm nrg}$-dependence of $D$ directly reflects the $N_{\rm
nrg}$-dependence of $\langle s_{z}\rangle$.
In the region of low temperatures below $N_{\rm nrg}\simeq 10$, 
this is in fact the case, and 
$D\propto \langle s_{z}\rangle^{2}$.
Therefore, the power law of $D$ is a reflection of the power law of
$\langle s_{z}\rangle$.
The power law of $\langle s_{z}\rangle$ on the high temperature side due
to scale invariance of the free spin regime 
shows the decrease of magnetization due to spin fluctuation
in the free spin state,
while the power law on the low temperature side due to scale invariance
of the Kondo regime shows
the disappearance of magnetization due to the Kondo effect, i.e. the
formation of many-body singlet states.
%
When the value of ${\mathcal S}$ is large, the value of $D_{\mathcal
S,Q}$ is large because one would expect $\langle s_{z}\rangle$ to be
large, and since states with large ${\mathcal S}$ generally correspond to
excited states with higher energies, 
they follow the behavior of lower energy states at lower temperatures, 
as interpreted by Wilson's image of renormalization flow.
 As a result, the change-points for various values of ${\mathcal S}$ are
 widely distributed in the Kondo screening region.
%
%


%
For the above reason, 
 the Kondo screening region can
be detected by looking at the $N_{\rm nrg}$-dependence of the relative
entropy $D_{{\mathcal S}{\mathcal Q}}(N_{\rm nrg})$ for many ${\mathcal
S}\neq 0$.
%

The fact that several magnetic excited states are necessary to capture
the Kondo screening process as described above can be understood from
the fact that the calculation of $S_{\rm imp}$, one of the conventional
quantities to capture the Kondo screening process, requires all state
including excited states in NRG.

These features are also seen when the parameters are varied as shown in
Fig.\ref{ued-dep-D}
(For a system away from the electron-hole symmetry shown below, the equation
$D_{S,Q}(N)=D_{S,-Q}(N)$ does not hold and the inequality
$D_{S,Q}(N)>D_{S,-Q}(N) \ (Q>0)$ holds.).
Fig.\ref{ued-dep-D} shows that the $N_{\rm nrg}$-dependence of the
entangle entropy $S_{\rm E.E.}$ of the ground state for each $N_{\rm
nrg}$,
the relative entropy $D_{{\mathcal S}{\mathcal Q}}(N_{\rm nrg})$ for $({\mathcal
 S},{\mathcal Q})=(1/2,-1)$ and $(5/2,3)$, and 
the impurity entropy $S_{\rm imp}$,  
for a electron-hole symmetric system with smaller value of $U$ and 
a system away from the electron-hole symmetry.
The Fig.\ref{ued-dep-D} shows that 
the transition regions from $\log(4)$ to $\log(2)$ and 
the Kondo screening regions 
change from the previous results due to the change in the value of $U$
and the deviation from the electron-hole symmetry,
and correspondingly, the regions of decreasing entanglement entropy and
the regions where the change-points appear in the relative entropy,
respectively.

From these results, it is found that to capture the Kondo screening process, it
is appropriate to consider relative entropy $D$ for several magnetic
states of total system among the quantum informative quantities
considered in this paper, $S_{\rm E.E.}, \ I$ and $D$.

\begin{figure}[h]
\includegraphics[width=0.5\textwidth]{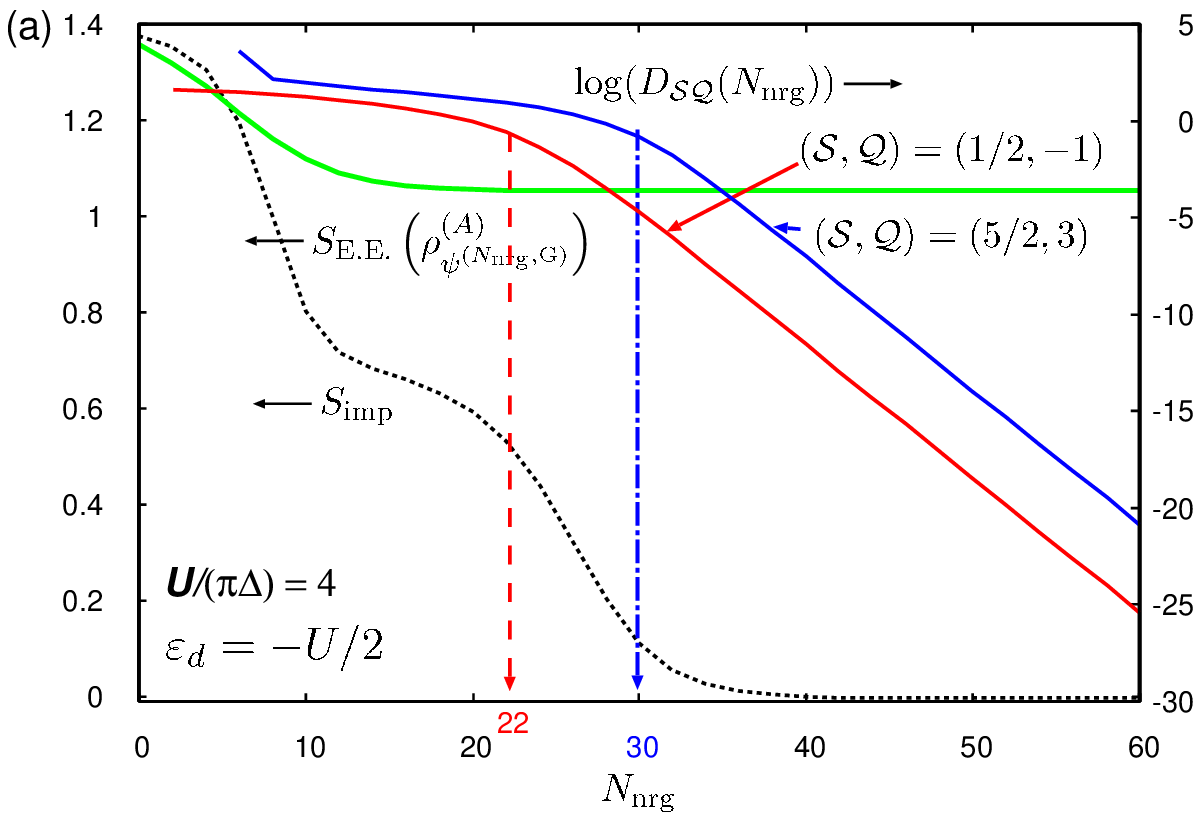}\\
\includegraphics[width=0.5\textwidth]{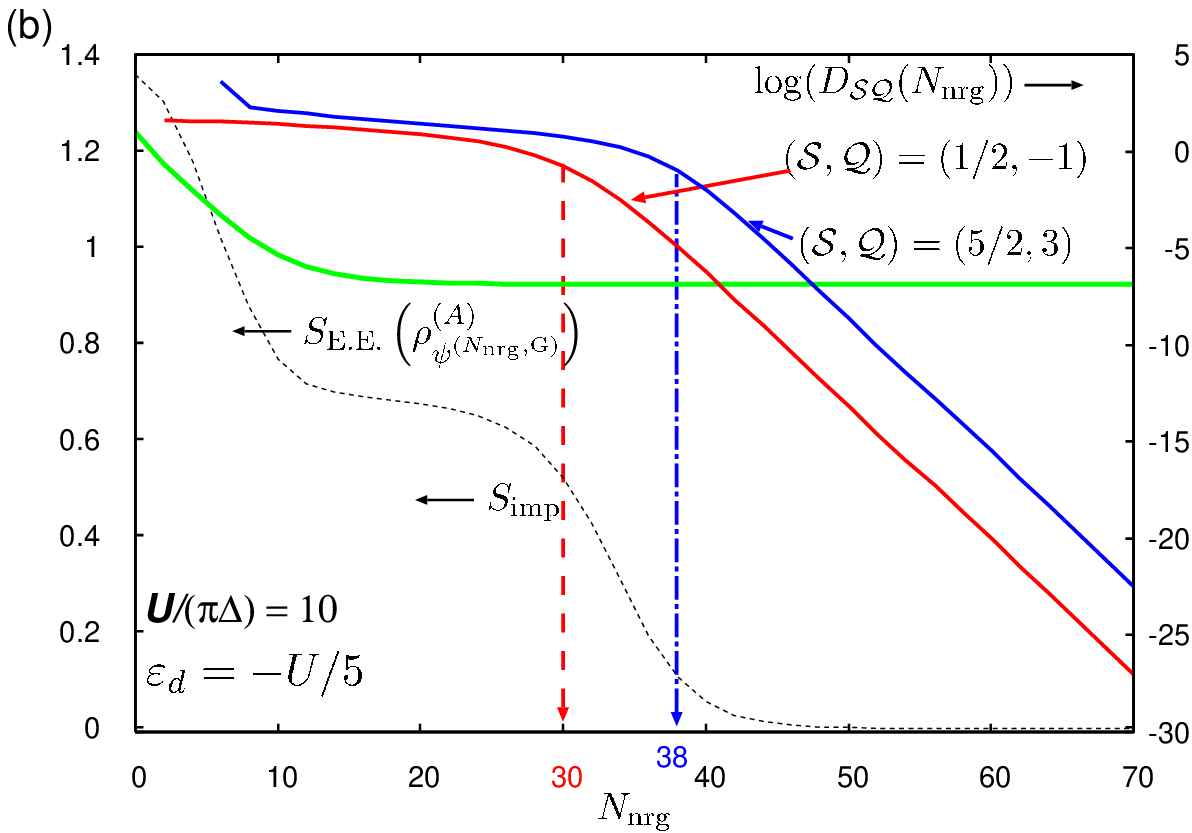}
\caption{(Color on line) $N_{\rm nrg}$-dependence of $S_{\rm imp}$, $S_{\rm
 E.E.}\left(\rho^{(A)}_{\psi^{({\rm G},N_{\rm nrg})}}\right)$, and
 $D_{{\mathcal S}{\mathcal Q}}(N_{\rm nrg})$ for $({\mathcal
 S},{\mathcal Q})=(1/2.-1)$ and $(5/2,3)$. (a)A electron-hole
 symmetric system with smaller value of
 $U$,(b) A system away from the electron-hole symmetry.}\label{ued-dep-D}
\end{figure}

\section{Conclusion}\label{last}
In order to formulate the distribution of quantum entanglement on
quantum impurities and 
quantum entanglement in quantum impurity pair in the system, for states of interest in
quantum impurity systems, 
we have considered quantum entanglement between a subsystem consisting
of one or two quantum impurities arbitrarily selected from the system
and its complement, the environmental system.
For this purpose, the pure state of interest has been reduced to the
subsystem consisting of selected quantum impurities to obtain the
density operator, and quantum informative quantities such as entanglement
entropy have been calculated.
The pure states of interest in quantum impurity systems are often
simultaneous eigenstates of physical quantities such as 
total particle number $Q$, 
total spin $S_{z}$ ($\Delta Q$), etc. (and often also energy eigenstates
of the system).
It has been shown that the density operator is diagonal or almost
diagonal in 
 several cases.
%
As a result, we have shown that entanglement entropy, mutual
information, and relative entropy are given by relatively simple
formulas, and that these quantities can be expressed in terms of
expectation values of physical quantities on the selected quantum
impurities and their correlation functions which can be calculated
independently of each other.
As we have demonstrated, these expectation values and correlation functions can be calculated
with high precision using NRG.
It should also be denoted that, apart from their usefulness and possibility of highly accurate
numerical calculations, as can be seen within the derivation
process, the equations derived here can be 
applied to systems other than those presented in this paper(e.g.,
periodic systems
, etc.).
%

Numerical evaluation of these formulas has been performed for 
the dimer with several types of internal interaction, short Hubbard
chains and SIAM
.
In the dimer with several types of internal interaction, we have considered the case of three type of two-body
interactions.
The results have shown that the maximum quantum entanglement cannot be
achieved when the interaction is introduced alone, 
but if Eq.(\ref{max_tqi}) is satisfied, 
the three type of two-body interactions can be introduced while
maintaining maximum quantum entanglement.
Although Eq.(\ref{max_tqi}) itself would be applicable only to dimers, 
it suggests that in systems where several types of two-body interactions
exist,
the entanglement entropy varies diversely due to competition or
cooperation among them.
The result is promising for future studies in systems where several
types of two-body interactions exist.
In the $3-6$ site Hubbard chain, the relationship between its ground
state and the entanglement entropy distribution has been studied.
In these systems, the entanglement entropy distributions are not flat
due to the finite systems with edges.
It would be interesting to see how these distributions changes when
electron reservoirs are connected to these systems.
The behavior of the mutual information in the $4-6$ site Hubbard chains, 
which is not a monotonically decreasing inter-site distance, also
suggests
that other systems deserve further study.

The SIAM has only one Anderson impurity to choose from, but if that one
Anderson impurity system is regarded as a composite system of two
spinless Anderson impurity systems, one can choose two
of them from the system.
This can be regarded as a simplified template for the set up in the subsection \ref{twoI}
that discusses the case where the subsystem consists of two quantum
impurities.
In fact, we have also calculated the mutual information between the up-
and down-spin electron systems from this perspective.
As a result, we have confirmed that the behavior of SIAM from the high
temperature region to the low temperature region including the Kondo
screening region can be captured with the quantum informative quantities.
In this process, we have calculated anew the magnetization $s_{z}$ of the
impurity in non zero $S_{z}$ state in SIAM under zero magnetic field
with SU(2) symmetry, a calculation not previously reported.
(The quantity $s_{z}$ belongs to the representation of $L=1$ in the
representation theory of SU(2), so the computational complexity in NRG
is larger than that of the quantity belonging to
$L=0$($n,n_{\uparrow}n_{\downarrow}$ etc.) or $L=1/2$( the creation
operator of the discretized conduction electron in NRG etc.), which has
been repeatedly calculated hitherto).
We have also analyzed separately all excited states with high $S$
retained in the NRG calculation and have decomposed the Kondo screening
process along these states.
These are new findings for SIAM itself.

In addition, our method is a very localized method, and 
it is important to relate it to related previous research methods that
have different approaches from our method.
However, it is difficult at this time to make clear the relevance of our
method to related previous studies, since they used different models, 
different settings, and significantly different approaches.
It is expected that studies using both methods will be developed and the
relationship between them will be discussed in the future.

In summary, the method proposed in this paper is expected to elucidate
the quantum entanglement of states of various multiple quantum impurity
systems beyond 
 systems presented in this paper.

%
Using the method described in the subsection \ref{twoI}, 
it is possible to analyze the entanglement between the two selected
quantum impurities. This is expected to provide a better understanding
of the quantum entanglement in the state of interest than the analysis for a single choice of
quantum impurity(even without the assumption that the state of interest
is an eigenstate of $\Delta Q$, the number of independent physical
quantities to be calculated is 25 for systems without any symmetry
which, while computationally
time-consuming is not impractical).
In particular, when the state of interest is further an eigenstate of $\Delta
Q$, the method is computationally more accessible. One of the future
task is to apply these to investigations of the duality between itinerant and localized
nature in the new (topological) local Fermi
liquids.

\begin{acknowledgements}
Numerical computation was partly carried out in Yukawa Institute
Computer Facility.
\end{acknowledgements}

\appendix

\section{Matrix elements of the representation matrix of $\rho^{(ab)}_{\psi}$}
Here, we show the matrix elements of the representation matrix of
$\rho^{(ab)}_{\psi}$ for a general pure state.
The representation matrix of $\rho^{(ab)}_{\psi}$ is no longer a block
diagonal matrix consisting of matrices labeled by $N_{Q}$.
There exist non-zero non-diagonal block matrices labeled by $(N_{Q},
N_{Q}^{\prime})$.
Furthermore, the block diagonal matrices labeled by $N_{Q}$ are not generally block
diagonal matrices either.

The matrix elements of the block matrices labeled by
$(N_{Q},N_{Q}^{\prime})=(0,0)$,$(0,1)$,$(0.2)$,$(0.3)$,$(0.4)$ and $(4,4)$
are as follows;
\begin{eqnarray}
    \ev{\ketbra{0}}
    &=&
    \ev{h_{a,\uparrow}h_{a,\downarrow}h_{b,\uparrow}h_{b,\downarrow}},\nonumber\\
    \ev{d^{\dagger}_{i,\sigma}\ketbra{0}}
    &=&
    \ev{d^{\dagger}_{i,\sigma}h_{i,\sigma}h_{i^c,\uparrow}h_{i^c,\downarrow}},\nonumber\\
    \ev{d^{\dagger}_{i,\uparrow}d^{\dagger}_{i,\downarrow}\ketbra{0}}
    &=&
    \ev{d^{\dagger}_{i,\uparrow}d^{\dagger}_{i,\downarrow}h_{i^c,\uparrow}h_{i^c,\downarrow}},\nonumber\\
    \ev{d^{\dagger}_{a,\sigma}d^{\dagger}_{b,\tau}\ketbra{0}}
    &=&
    \ev{d^{\dagger}_{a,\sigma}h_{a,\sigma}d^{\dagger}_{b,\tau}h_{b,\tau}},\nonumber\\
 \ev{d^{\dagger}_{i,\uparrow}d^{\dagger}_{i,\downarrow}d^{\dagger}_{i^c,\sigma}\ketbra{0}}
    &=&
    \ev{d^{\dagger}_{i,\uparrow}d^{\dagger}_{i,\downarrow}d^{\dagger}_{i^c,\sigma}h_{i^c,\sigma}},\nonumber\\
    \ev{d^{\dagger}_{a,\uparrow}d^{\dagger}_{a,\downarrow}d^{\dagger}_{b,\uparrow}d^{\dagger}_{b,\downarrow}\ketbra{0}}
    &=&
    \ev{d^{\dagger}_{a,\uparrow}d^{\dagger}_{a,\downarrow}d^{\dagger}_{b,\uparrow}d^{\dagger}_{b,\downarrow}},\nonumber\\
\ev{d^{\dagger}_{a,\uparrow}d^{\dagger}_{a,\downarrow}d^{\dagger}_{b,\uparrow}d^{\dagger}_{b,\downarrow}\ketbra{0}d_{b,\downarrow}d_{b,\uparrow}d_{a,\downarrow}d_{a,\uparrow}}
    &=&
    \ev{n_{a,\uparrow}n_{a,\downarrow}n_{b,\uparrow}n_{b,\downarrow}}.\nonumber
%
\end{eqnarray}
The matrix elements of block matrix labeled by
$(N_{Q},N_{Q}^{\prime})=(1.1)$ are as follows;
\begin{eqnarray}
    \ev{d^{\dagger}_{i,\sigma}\ketbra{0}d_{i,\sigma}}
    &=&
    \ev{n_{i,\sigma}h_{i,\sigma}h_{i^c,\uparrow}h_{i^c,\downarrow}}.\nonumber\\
 \ev{d^{\dagger}_{i,\sigma}\ketbra{0}d_{i^c,\tau}}
    &=&
    \ev{d^{\dagger}_{i,\sigma}h_{i,\sigma}d_{i^c,\tau}h_{i^c,\tau}},\nonumber\\
    \ev{d^{\dagger}_{i,\sigma}\ketbra{0}d_{i,-\sigma}}
    &=&
    \ev{d^{\dagger}_{i,\sigma}d_{i,-\sigma}h_{i^c,\uparrow}h_{i^c,\downarrow}}.\nonumber
\end{eqnarray}
The matrix elements of block matrix labeled by
$(N_{Q},N_{Q}^{\prime})=(1,2)$ as follows;
\begin{eqnarray}
\ev{d^{\dagger}_{i,\uparrow}d^{\dagger}_{i,\downarrow}\ketbra{0}d_{i,\sigma}}
    &=&(\delta_{\downarrow,\sigma}-\delta_{\uparrow,\sigma})\nonumber\\
&\times&
    \ev{d^{\dagger}_{i,-\sigma}n_{i,\sigma}h_{i^c,\uparrow}h_{i^c,\downarrow}},\nonumber\\
    \ev{d^{\dagger}_{i,\uparrow}d^{\dagger}_{i,\downarrow}\ketbra{0}d_{i^c,\sigma}}
    &=&
    \ev{d^{\dagger}_{i,\uparrow}d^{\dagger}_{i,\downarrow}d_{i^c,\sigma}h_{i^c,\sigma}},\nonumber\\
    \ev{d^{\dagger}_{i,\sigma}d^{\dagger}_{i^c,\tau}\ketbra{0}d_{i,\sigma}}
    &=&-
    \ev{n_{i,\sigma}h_{i,\sigma}d^{\dagger}_{i^c,\tau}h_{i^c,\tau}},\nonumber\\
    \ev{d^{\dagger}_{i,\sigma}d^{\dagger}_{i^c,\tau}\ketbra{0}d_{i,-\sigma}}
    &=&-
    \ev{d^{\dagger}_{i,\sigma}d_{i,-\sigma}d^{\dagger}_{i^c,\tau}h_{i^c,\tau}}.\nonumber
%
\end{eqnarray}
The matrix elements of block matrix labeled by
$(N_{Q},N_{Q}^{\prime})=(1,3)$ as follows;
\begin{eqnarray}
    \ev{d^{\dagger}_{i,\uparrow}d^{\dagger}_{i,\downarrow}d^{\dagger}_{i^c,\sigma}\ketbra{0}d_{i,\tau}}
    &=&(\delta_{\uparrow,\tau}-\delta_{\downarrow,\tau})\nonumber\\
&\times&
    \ev{d^{\dagger}_{i,-\tau}n_{i,\tau}d^{\dagger}_{i^c,\sigma}h_{i^c,\sigma}},\nonumber\\
    \ev{d^{\dagger}_{i,\uparrow}d^{\dagger}_{i,\downarrow}d^{\dagger}_{i^c,\sigma}\ketbra{0}d_{i^c,\sigma}}
    &=&
    \ev{d^{\dagger}_{i,\uparrow}d^{\dagger}_{i,\downarrow}n_{i^c,\sigma}h_{i^c,\sigma}},\nonumber\\
    \ev{d^{\dagger}_{i,\uparrow}d^{\dagger}_{i,\downarrow}d^{\dagger}_{i^c,\sigma}\ketbra{0}d_{i^c,-\sigma}}
    &=&
    \ev{d^{\dagger}_{i,\uparrow}d^{\dagger}_{i,\downarrow}d^{\dagger}_{i^c,\sigma}d_{i^c,-\sigma}}.\nonumber
%
\end{eqnarray}
The matrix elements of block matrix labeled by
$(N_{Q},N_{Q}^{\prime}=(1,4)$ as follows;
\begin{eqnarray}
    \ev{d^{\dagger}_{a,\uparrow}d^{\dagger}_{a,\downarrow}d^{\dagger}_{b,\uparrow}d^{\dagger}_{b,\downarrow}\ketbra{0}d_{i,\sigma}}
    &=&(\delta_{\downarrow,\sigma}-\delta_{\uparrow,\sigma})\nonumber\\
&\times&
    \ev{d^{\dagger}_{i,-\sigma}n_{i,\sigma}d^{\dagger}_{i^c,\uparrow}d^{\dagger}_{i^c,\downarrow}}.\nonumber
%
\end{eqnarray}

The matrix elements of block matrix labeled by
$(N_{Q},N_{Q}^{\prime})=(2,2)$ as follows;
\begin{eqnarray}
    \ev{d^{\dagger}_{i,\uparrow}d^{\dagger}_{i,\downarrow}\ketbra{0}d_{i,\downarrow}d_{i,\uparrow}}
    &=&
    \ev{n_{i,\uparrow}n_{i,\downarrow}h_{i^c,\uparrow}h_{i^c,\downarrow}},\nonumber\\
    \ev{d^{\dagger}_{i,\uparrow}d^{\dagger}_{i,\downarrow}\ketbra{0}d_{i^c,\sigma}d_{i,\tau}}
    &=&(\delta_{\uparrow,\tau}-\delta_{\downarrow,\tau})\nonumber\\
&\times&
    \ev{d^{\dagger}_{i,-\tau}n_{i,\tau}d_{i^c,\sigma}h_{i^c,\sigma}},\nonumber\\
    \ev{d^{\dagger}_{i,\uparrow}d^{\dagger}_{i,\downarrow}\ketbra{0}d_{i^c,\downarrow}d_{i^c,\uparrow}}
    &=&
    \ev{d^{\dagger}_{i,\uparrow}d^{\dagger}_{i,\downarrow}d_{i^c,\downarrow}d_{i^c,\uparrow}},\nonumber\\
    \ev{d^{\dagger}_{a,\sigma}d^{\dagger}_{b,\tau}\ketbra{0}d_{b,\tau}d_{a,\sigma}}
    &=&
    \ev{n_{a,\sigma}h_{a,\sigma}n_{b,\tau}h_{b,\tau}},\nonumber\\
    \ev{d^{\dagger}_{i,\sigma}d^{\dagger}_{i^c,\tau}\ketbra{0}d_{i^c,-\tau}d_{i,\sigma}}
    &=&
    \ev{n_{i,\sigma}h_{i,\sigma}d^{\dagger}_{i^c,\tau}d_{i^c,-\tau}},\nonumber\\
    \ev{d^{\dagger}_{a,\sigma}d^{\dagger}_{b,\tau}\ketbra{0}d_{b,-\tau}d_{a,-\sigma}}
    &=&
    \ev{d^{\dagger}_{a,\sigma}d_{a,-\sigma}d^{\dagger}_{b,\tau}d_{b,-\tau}}.\nonumber
%
\end{eqnarray}

The matrix elements of block matrix labeled by
$(N_{Q},N_{Q}^{\prime})=(2,3)$ as follows;
\begin{eqnarray}
    \ev{d^{\dagger}_{i,\uparrow}d^{\dagger}_{i,\downarrow}d^{\dagger}_{i^c,\sigma}\ketbra{0}d_{i,\downarrow}d_{i,\uparrow}}
    &=&
    \ev{n_{i,\uparrow}n_{i_\downarrow}d^{\dagger}_{i^c,\sigma}h_{i^c,\sigma}},\nonumber\\
    \ev{d^{\dagger}_{i,\uparrow}d^{\dagger}_{i,\downarrow}d^{\dagger}_{i^c,\sigma}\ketbra{0}d_{i^c,\sigma}d_{i,\tau}}
    &=&(\delta_{\downarrow,\tau}-\delta_{\uparrow,\tau})\nonumber\\
&\times&
    \ev{d^{\dagger}_{i,-\tau}n_{i_\tau}n_{i^c,\sigma}h_{i^c,\sigma}},\nonumber\\
    \ev{d^{\dagger}_{i,\uparrow}d^{\dagger}_{i,\downarrow}d^{\dagger}_{i^c,\sigma}\ketbra{0}d_{i^c,-\sigma}d_{i,\tau}}
    &=&(\delta_{\downarrow,\tau}-\delta_{\uparrow,\tau})\nonumber\\
&\times&
    \ev{d^{\dagger}_{i,-\tau}n_{i_\tau}d^{\dagger}_{i^c,\sigma}d_{i^c,-\sigma}},\nonumber\\
    \ev{d^{\dagger}_{i,\uparrow}d^{\dagger}_{i,\downarrow}d^{\dagger}_{i^c,\sigma}\ketbra{0}d_{i^c,\downarrow}d_{i^c,\uparrow}}
    &=&(\delta_{\downarrow,\sigma}-\delta_{\uparrow,\sigma})\nonumber\\
&\times&
    \ev{d^{\dagger}_{i,\uparrow}d^{\dagger}_{i,\downarrow}d_{i^c,-\sigma}n_{i^c,\sigma}}.\nonumber
%
\end{eqnarray}
The matrix elements of block matrix labeled by
$(N_{Q},N_{Q}^{\prime})=(2,4)$ as follows;
\begin{eqnarray}
    \ev{d^{\dagger}_{a,\uparrow}d^{\dagger}_{a,\downarrow}d^{\dagger}_{b,\uparrow}d^{\dagger}_{b,\downarrow}\ketbra{0}d_{i,\downarrow}d_{i,\uparrow}}
    &=&
    \ev{n_{i,\uparrow}n_{i,\downarrow}d^{\dagger}_{i^c,\uparrow}d^{\dagger}_{i^c,\downarrow}},\nonumber\\
    \ev{d^{\dagger}_{a,\uparrow}d^{\dagger}_{a,\downarrow}d^{\dagger}_{b,\uparrow}d^{\dagger}_{b,\downarrow}\ketbra{0}d_{b,\sigma}d_{a,\tau}}
    &=&(\delta_{\sigma,-\tau}-\delta_{\sigma,\tau})\nonumber\\
&\times&
    \ev{d^{\dagger}_{a,-\tau}n_{a,\tau}d^{\dagger}_{b,-\sigma}n_{b,\sigma}}.\nonumber
%
\end{eqnarray}
The matrix elements of block matrix labeled by
$(N_{Q},N_{Q}^{\prime})=(3,3)$ as follows;
\begin{eqnarray}
    \ev{d^{\dagger}_{i,\uparrow}d^{\dagger}_{i,\downarrow}d^{\dagger}_{i^c,\sigma}\ketbra{0}d_{i^c,\sigma}d_{i,\downarrow}d_{i,\uparrow}}
    &=&
    \ev{n_{i,\uparrow}n_{i,\downarrow}n_{i^c,\sigma}h_{i^c,\sigma}},\nonumber\\
    \ev{d^{\dagger}_{i,\uparrow}d^{\dagger}_{i,\downarrow}d^{\dagger}_{i^c,\sigma}\ketbra{0}d_{i^c,-\sigma}d_{i,\downarrow}d_{i,\uparrow}}
    &=&
    \ev{n_{i,\uparrow}n_{i,\downarrow}d^{\dagger}_{i^c,\sigma}d_{i^c,-\sigma}},\nonumber\\
    \ev{d^{\dagger}_{i,\uparrow}d^{\dagger}_{i,\downarrow}d^{\dagger}_{i^c,\sigma}\ketbra{0}d_{i^c,\downarrow}d_{i^c,\uparrow}d_{i,\tau}}
    &=&(\delta_{\sigma,-\tau}-\delta_{\sigma,\tau})\nonumber\\
&\times&
    \ev{d^{\dagger}_{i,-\tau}n_{i,\tau}d_{i^c,-\sigma}n_{i^c,\sigma}}.\nonumber
\end{eqnarray}
The matrix elements of block matrix labeled by
$(N_{Q},N_{Q}^{\prime})=(3,4)$ as follows;
\begin{eqnarray}
    \ev{d^{\dagger}_{a,\uparrow}d^{\dagger}_{a,\downarrow}d^{\dagger}_{b,\uparrow}d^{\dagger}_{b,\downarrow}\ketbra{0}d_{i^c,\sigma}d_{i,\downarrow}d_{i,\uparrow}}
    &=&(\delta_{\downarrow,\sigma}-\delta_{\uparrow,\sigma})\nonumber\\
&\times&
    \ev{n_{i,\uparrow}n_{i,\downarrow}d^{\dagger}_{i^c,-\sigma}n_{i,\sigma}}.\nonumber
%
\end{eqnarray}

\section{Stationary problem of $S_{\rm E.E.}^{(ab)}$}

As we did in subsection \ref{oneI},
we consider the stationary problem of $S_{\rm E.E.}^{(ab)}$ and the
linear response of $S_{\rm E.E.}^{(ab)}$ to 16 variables.
The results of differentiating $S_{\rm E.E.}^{(ab)}$ by these 16
variables is as follows;
\begin{eqnarray}
\frac{\partial S^{(ab)}_{\rm E.E.}}{\partial c^{(1)}_{i\sigma}}
&=&
\log\left(\frac{\lambda^{(0)}}{\lambda^{(1)}_{i\sigma}}\right),\\
\frac{\partial S^{(ab)}_{\rm E.E.}}{\partial c^{(2)}_{i\uparrow,i\downarrow}}
&=&
\log\left(\frac{\lambda^{(1)}_{i\uparrow}\lambda^{(1)}_{i\downarrow}}{\lambda^{(0)}\lambda^{(2)}_{i\uparrow,i\downarrow}}\right),\\
\frac{\partial S^{(ab)}_{\rm E.E.}}{\partial c^{(2)}_{a\sigma,b\sigma}}
&=&
\log\left(\frac{\lambda^{(1)}_{a\sigma}\lambda^{(1)}_{b\sigma}}{\lambda^{(0)}\lambda^{(2)}_{a\sigma,b\sigma}}\right),\nonumber\\
\frac{\partial S^{(ab)}_{\rm E.E.}}{\partial c^{(2)}_{a\sigma,b-\sigma}}
&=&
\log\left(\frac{\lambda^{(1)}_{a\sigma}\lambda^{(1)}_{b-\sigma}}{\lambda^{(0)}\left(\lambda^{(2)}_{+}\right)^{\delta_{\sigma}}\left(\lambda^{(2)}_{-}\right)^{\delta_{-\sigma}}}\right),\\
\frac{\partial S^{(ab)}_{\rm E.E.}}{\partial c^{(3)}_{a\uparrow,a\downarrow,b\sigma}}
&=&
\log
\left(\frac{
\lambda^{(0)}
\lambda^{(2)}_{a\uparrow,a\downarrow}
\lambda^{(2)}_{a\sigma,b\sigma}
\left(
\lambda^{(2)}_{+}
\right)^{\delta_{-\sigma}}
\left(
\lambda^{(2)}_{-}
\right)^{\delta_{\sigma}}
}
{
\lambda^{(1)}_{a\uparrow}\lambda^{(1)}_{a\downarrow}\lambda^{(1)}_{b\sigma}
\lambda^{(3)}_{a\uparrow,a\downarrow,b\sigma}
}\right),\nonumber\\
\end{eqnarray}
\begin{eqnarray}
\frac{\partial S^{(ab)}_{\rm E.E.}}{\partial c^{(3)}_{a\sigma,b\uparrow,b\downarrow}}
&=&
\log\left(
\frac{
\lambda^{(0)}
\lambda^{(2)}_{b\uparrow,b\downarrow}
\lambda^{(2)}_{a\sigma,b\sigma}
\left(
\lambda^{(2)}_{+}
\right)^{\delta_{\sigma}}
\left(
\lambda^{(2)}_{-}
\right)^{\delta_{-\sigma}}
}
{
\lambda^{(1)}_{a\sigma}
\lambda^{(1)}_{b\uparrow}
\lambda^{(1)}_{b\downarrow}
\lambda^{(3)}_{a\sigma,b\uparrow,b\downarrow}
}
\right),\nonumber\\
\end{eqnarray}
\begin{eqnarray}
\frac{\partial S^{(ab)}_{\rm E.E.}}{\partial c^{(4)}}
&=&
\log\left(
\frac
{
\underset{i=a,b,\sigma=\uparrow,\downarrow}{\prod}
\left(
\lambda^{(1)}_{i\sigma}
\right)
\underset{\sigma=\uparrow,\downarrow}{\prod}
\left(
\lambda^{(3)}_{a\uparrow,a\downarrow,b\sigma}
\lambda^{(3)}_{a\sigma,b\uparrow,b\downarrow}
\right)
}
{
\lambda^{(0)}
\underset{i=a,b}{\prod}
\left(
\lambda^{(2)}_{i\uparrow,i\downarrow}
\right)
\underset{\sigma=\uparrow,\downarrow}{\prod}
\left(
\lambda^{(2)}_{a\sigma,b\sigma}
\right)
\lambda^{(2)}_{+}\lambda^{(2)}_{-}
\lambda^{(4)}
}
\right),\nonumber\\
\end{eqnarray}
\begin{eqnarray}
\frac{\partial S^{(ab)}_{\rm E.E.}}{\partial |C|}
&=&
\log\left(\frac{\lambda^{(2)}_{-}}{\lambda^{(2)}_{+}}\right)^{\Delta},
\end{eqnarray}
where 
$\delta_{\pm}\equiv\frac{1}{2}\left(1\pm\frac{\lambda^{(2)}_{a\uparrow,b\downarrow}
-
\lambda^{(2)}_{a\downarrow,b\uparrow}}{\sqrt{
(
\lambda^{(2)}_{a\uparrow,b\downarrow}
-
\lambda^{(2)}_{a\downarrow,b\uparrow}
)^{2}
+
4|C|^{2}
}}\right)$
and
$\Delta\equiv
\frac{2|C|}{\sqrt{
(
\lambda^{(2)}_{a\uparrow,b\downarrow}
-
\lambda^{(2)}_{a\downarrow,b\uparrow}
)^{2}
+
4|C|^{2}
}}$.
Similar to the results obtained in subsection \ref{oneI},
the stationary value condition of $S_{\rm E.E.}^{(ab)}$ 
for each variables gives the relationship among several correlation
functions and 
the logarithm of the displacement of the relation gives the linear
response of $S_{\rm E.E.}^{(ab)}$ to each variable.
The equal probability conditions obtained from the stationary value
conditions of $S_{\rm E.E.}^{(ab)}$ for all variables, written in 
terms of correlation functions, are
$c^{(1)}_{\cdots}=\frac{1}{2}$,
$c^{(2)}_{\cdots}=\frac{1}{4}\left(=(\frac{1}{2})^2\right)$,
$c^{(3)}_{\cdots}=\frac{1}{8}\left(=(\frac{1}{2})^{3}\right)$, 
$c^{(4)}=\frac{1}{16}\left(=(\frac{1}{2})^{4}\right)$, and $C=0$
, where the maximum value of $S_{\rm E.E.}^{(ab)}$ is $\log(16)$.
Again, to have maximum entanglement with $B$, the environmental system
of $A$, there must be no correlations inside $A$,
and the monogamy of quantum entanglement can be seen.


The trace distance between $\rho_{\psi}^{(a)}\otimes\rho_{\psi}^{(b)}$
and $\rho_{\psi}^{(ab)}$
is
\begin{eqnarray}
& &\mbox{}{\rm Tr}\left|
\rho_{\psi}^{(a)}\otimes\rho_{\psi}^{(b)}
-
\rho_{\psi}^{(ab)}
\right|\nonumber\\
&=&
\left|\delta_{ab}\lambda^{(0)}\right|
+\sum_{i=a,b,\sigma=\uparrow,\downarrow}
\left|\delta_{ab}\lambda^{(1)}_{i\sigma}\right|\nonumber\\
& &\mbox{}
+\sum_{i=a,b}
\left|\delta_{ab}\lambda^{(2)}_{i\uparrow,i\downarrow}\right|
+\sum_{\sigma=\uparrow,\downarrow}
\left|\delta_{ab}\lambda^{(2)}_{a\sigma,b\sigma}\right|\nonumber\\
& &\mbox{}
+\sum_{j=\pm}
\left|\delta_{ab}\lambda^{(2)}_{j}\right|
+\sum_{\sigma=\uparrow,\downarrow}
\left|\delta_{ab}\lambda^{(3)}_{a\uparrow,a\downarrow,b\sigma}\right|\nonumber\\
& &\mbox{}
+\sum_{\sigma=\uparrow,\downarrow}
\left|\delta_{ab}\lambda^{(3)}_{a\sigma,b\uparrow,b\downarrow}\right|
+
\left|\delta_{ab}\lambda^{(4)}\right|,
\end{eqnarray}
which directly measures the correlation between selected Anderson
impurity $a$ and $b$.
Here,
$\delta_{ab}\lambda^{0}\equiv\lambda^{(0)}
-
\langle
h_{a\uparrow}h_{a\downarrow}\rangle\langle
h_{b\uparrow}h_{b\downarrow}\rangle,\cdots$
and 
$\delta_{ab}\lambda^{(2)}_{\pm}
\equiv
\frac{
\delta_{ab}\lambda^{(2)}_{a\uparrow,b\downarrow}
+
\delta_{ab}\lambda^{(2)}_{a\downarrow,b\uparrow}
\pm
\sqrt{
(
\delta_{ab}\lambda^{(2)}_{a\uparrow,b\downarrow}
-
\delta_{ab}\lambda^{(2)}_{a\downarrow,b\uparrow}
)^{2}
+
4|C|^{2}
}
}
{2}$.

\bibliography{
bibtex
}

\begin{thebibliography}{52}
\bibitem[{odd()}]{oddN}
\bibinfo{note}{We describe what Fig.\ref{n-dep-sIsimp-G} would look like if we
  consider odd numbers of $N_{\rm nrg}$. In the half-filling case, the ground
  state for odd $N_{\rm nrg}$ is a doublet state.At high temperatures, $S_{\rm
  E.E.}$ takes a value close to zero, and at low temperatures it converges to
  the same value as in the case of even $N_{\rm nrg}$. In other words, the
  system goes from a disentangled state to an entangled state, opposite to the
  case of even $N_{\rm nrg}$. The reason for this is that at high temperature,
  the quantum impurity takes on the doublet state of the total system, and the
  state reduced to the quantum impurity is almost equal to
  $|\uparrow\rangle\langle\uparrow |$ and becomes a pure state, and thus does
  not entangle with its environmental system. As the temperature decreases, the
  value of $\langle s_{z}\rangle$ decreases while at the same time $\langle
  n_{\uparrow}n_{\downarrow}\rangle$ has a value, and the value of $S_{\rm
  E.E.}$ increases.(The ground state considered here adopts the state with
  ${\mathcal S}_{z}=1/2$.) The temperature region where the entanglement
  entropy converges does not change depending on the even-oddness of $N_{\rm
  nrg}$. This is also true for quantum relative entropy. Since it is more
  natural to regard the spin-doublet ground state as an artificial state due to
  the discretization of NRG, we discuss only the results for even $N_{\rm nrg}$
  in the main text.}
\end{thebibliography}
%
%
\bibliographystyle{apsrev}


\end{document}